\documentclass[final,twocolumn,5p,times]{elsarticle}

\usepackage{amssymb}

\usepackage{graphicx}
\usepackage{subcaption}

\usepackage{url}
\usepackage{color}

\usepackage{tabularx}\newcolumntype{Y}{>{\centering\arraybackslash}X}
\usepackage{multirow}
\usepackage{booktabs}

\usepackage{comment}

\usepackage{amsmath}

\usepackage{algorithm}
\usepackage{algorithmic}

\usepackage{setspace}

\usepackage[font=small,skip=4pt]{caption}

\newif\ifremark
\long\def\remark#1{
\ifremark%
        \begingroup%
        \dimen0=\columnwidth
        \advance\dimen0 by -1in%
        \setbox0=\hbox{\parbox[b]{\dimen0}{\protect\em #1}}
        \dimen1=\ht0\advance\dimen1 by 2pt%
        \dimen2=\dp0\advance\dimen2 by 2pt%
        \vskip 0.25pt%
        \hbox to \columnwidth{%
                \vrule height\dimen1 width 3pt depth\dimen2%
                \hss\copy0\hss%
                \vrule height\dimen1 width 3pt depth\dimen2%
        }%
        \endgroup%
\fi}

\remarkfalse

\journal{Journal of Parallel and Distributed Computing}

\begin{document}

\begin{frontmatter}


\title{Understanding Cloud Workloads Performance in a Production like Environment}


\author[disca]{Luc\'ia~Pons}
\ead{lupones@disca.upv.es}
\author[disca]{Josu\'e~Feliu}
\ead{jofepre@gap.upv.es}
\author[disca]{Jos\'e~Puche}
\ead{jopucla@gap.upv.es}

\author[huawei]{Chaoyi~Huang}
\ead{joehuang@huawei.com}

\author[disca]{Salvador~Petit}
\ead{spetit@disca.upv.es}
\author[disca]{Julio~Pons}
\ead{jpons@disca.upv.es}
\author[disca]{Mar\'ia~E.~G\'omez}
\ead{megomez@disca.upv.es}

\author[disca]{Julio~Sahuquillo}
\ead{jsahuqui@disca.upv.es}

\address[disca]{Department of Computer Engineering, Universitat Polit\`ecnica de Val\`encia, Valencia, Spain.}
\address[huawei]{Huawei Technologies CO., LDT.}

\begin{abstract}
Understanding inter-VM interference is of paramount importance to provide a sound knowledge and understand where performance degradation comes from in the current public cloud. With this aim, this paper devises a workload taxonomy that classifies applications according to how the major system resources affect their performance (e.g., tail latency) as a function of the level of load (e.g., QPS). After that, we present three main studies addressing three major concerns to improve the cloud performance: impact of the level of load on performance, impact of hyper-threading on performance, and impact of limiting the major system resources (e.g., last level cache) on performance.
In all these studies we identified important findings that we hope help cloud providers improve their system utilization.
\end{abstract}


\end{frontmatter}

\section{Introduction}\label{sec:intro}

Public cloud systems share the system resources among the tenant applications typically hosted by virtual machines (VMs).
All the major system resources such as the CPU, the memory bandwidth, the remote storage server, in addition to the network connecting to the server machines, are shared.
Typically, the load of the system is rather low (i.e., around 15\% processor utilization). Therefore, the cloud administrator can allocate on the same physical machine different VMs with certain guarantees that quality of service (QoS) will not be violated. However, since VMs from different tenants compete among them for shared resources, it may happen that the peak consumption periods of the workload of a given VM severely damage the QoS of another tenant.
It is the job of the cloud system provider to avoid such undesirable adverse effect on performance by properly sizing the required resources to each tenant VM.

This work characterizes the behavior of cloud applications from a system resource consumption perspective, and studies how the overall performance is affected by each major system resource.
Since nowadays public cloud systems execute a wide variety of workloads, the first step is to select the applications to be studied.
Special interest is taken in latency-critical workloads. Many interactive services such as MongoDB \cite{mongodb} and NGINX \cite{nginx} are examples of  latency-critical workloads. An important characteristic of these workloads is that \textit{tail latency} is the key performance metric instead of average latency. Tail latency values are normally much lower. For instance, as experimental results will show, some applications like \texttt{specjbb} or \texttt{silo} must provide a $95^{th}$ tail latency of less than one millisecond. Once selected the target applications, a workload characterization study is required in order to analyze similarities and differences among applications.
This study is aimed at helping cloud researchers choose the workloads of interests for performance analysis. The results of this study conclude presenting a workload taxonomy, where applications are classified in four main categories depending on how the major system resources impact on their performance.

The main objective of this work is to provide some highlights that help cloud providers to understand the reasons why the overall system performance can be affected 
and what situations can make it to drop. This is a challenging task since performance variations depend on many factors, both from the client side, such as the level of load of the application (e.g., queries per second or QPS), and from the server side, such as the processor utilization level or disk bandwidth.
Moreover, tail latency is extremely sensitive to small performance degradation. While a small and short interference in other workloads can be averaged along their execution time and result in minor performance degradation, interference immediately impacts on the tail latency and thus in the user-perceived QoS. 

This paper analyzes three major concerns regarding cloud performance:
i) how performance (e.g., tail latency) is affected varying the level of load and number of threads, ii) how hyper-threading impacts on performance, and iii) how constraining the major system resources affects the performance. 

Workload variation allows us to evaluate the effect in the studied metrics and tail latency. 
One of the most critical metrics in public clouds today is the CPU utilization.
to study the effect on performance in a wide range of the CPU utilization, which grows from around 10\% up to saturating, . A key goal for cloud providers is to find out the CPU utilization that is able to satisfy tenant applications without compromising the QoS. Since this is a challenging task, cloud providers simplify this issue by merely establishing an upper threshold (e.g., a CPU utilization by 50\%). 
In this paper we study the effect on performance in a wide range of the CPU utilization (from around 10\% up to saturating) and we found that the server can support a CPU utilization higher than 50\% in some applications (e.g., \emph{img-dnn}) while still satisfying the client demands. This means that other metrics should be used in addition to the CPU utilization in order properly size the supported workload. In other words, further research is required to model performance degradation taking into account the interference at other system components. This research should identify the CPU utilization level that makes each application to violate QoS. Addressing this issue is required to further improve resource utilization.

Hyper-threading affects differently the performance of multi-threaded tenant applications.
To study in-depth this issue, we tested if two threads of the same application running on the same physical core perform better (i.e., the core is entirely occupied by threads of the same application) than if each thread of the same application was pinned to a different core, that is, the physical core is occupied by threads of different applications. 
Results reveal that not always having the threads in different physical cores is the configuration that achieves the highest performance. Instead, some applications achieve similar performance when running their two threads in the same physical core. This means that efforts must be put on new thread-to-core allocation strategies for cloud workloads.

Finally, in a second study we analyzed how 
limiting a major system shared resource (e.g., the LLC cache or main memory bandwidth) can be useful to mimic real public cloud situations where a VM has only a fraction of the resource since the remaining fraction is being used by contending VMs.
Results shown that the performance (i.e., tail latency) of some applications like \emph{img-dnn} is very sensitive to the occupied cache space, and therefore the number of supported QPS, while the performance of important applications like \emph{media-streaming} show little sensitiveness. This observation emphasizes the need of LLC partitioning approaches focusing on cloud applications.

In summary, this paper covers three major concerns of current cloud providers: impact on performance of the load level, hyper-threading, and resource sharing. We discuss and analyze the major finding of each study which makes the contributions of this paper. We hope these findings provide insights and help cloud providers to improve their system performance. In order to provide representativeness to the obtained results, all experiments were launched on an experimental hardware/software platform that closely resembles to a typical platform of a real cloud system.

The remainder of this paper is organized as follows. 
Section \ref{sec:related} discusses the related work.
Section \ref{sec:system} presents the hardware/software platform.
Section \ref{sec:workloads}  discusses key characteristics of latency critical applications.
Section \ref{sec:characterization}  presents the workload characterization study.
Section \ref{sec:taxonomy} introduces the devised workload taxonomy.
Section \ref{sec:constrain_results} analyzes the effect on performance of constraining the major system resources.
Finally, Section \ref{sec:conclus} presents some concluding remarks.

\section{Related Work}\label{sec:related}

This section summarizes previous research on inter-VM interference in the public cloud. 
Due to the nature and fast evolution of this research topic and industry problems, much research work has emerged in the last few years. 
Previous works have used distinct methodologies that present important differences regarding the virtualization level (e.g., VMs, containers or no virtualization), the studied workloads (e.g., HPC, Cloud, etc.) and the target interference resource (e.g., LLC, memory bandwidth, I/O, etc.). 

From a cloud provider perspective, the virtualization level is the most critical since it establishes the main features of the target platform, thus limiting 
the techniques that can be used to address the inter-VM interference problem.
In other words, the challenges dealing with interference widely differ depending on the virtualization level that is supported by the system.

First we focus on those works that do not consider virtualization.
With the aim of improving system fairness, CoPart~\cite{copart} leverages Intel's CAT and MBA technologies and characterizes the behavior of each application according to the number of LLC misses and the consumed memory bandwidth.
High-performance workloads and a single latency-critical application are studied in this work.
A similar methodology is followed in HyPart~\cite{hypart}, which is a hybrid partitioning approach that combines Intel MBA with two additional techniques.
In the experimental evaluation, mostly high performance benchmarks are employed.
Finally, in ~\cite{quasar}, the Quasar cluster management system is presented. The main aim of Quasar is to increase resource utilization.

Next, we move to works that consider just LXC virtualization~\cite{lxc} (i.e., Linux containers).
In~\cite{heracles}, D. Lo et al. aim to increase server utilization by tolerating some interference of Best-Effort (BE) tasks, provided that the service level objectives (SLO) of a single running latency-critical (LC) application is met.
To do this, the approach first characterizes 3 different LC Google workloads, detects the situations when guaranteeing SLO becomes problematic, and then limits the amount of resources assigned to the BE workloads.
Pursuing the same target, PARTIES~\cite{parties} monitors tail latency applications, memory capacity and network bandwidth usage.
Upon SLO violation detection, this approach initiates the allocation of one or more resources to the LC service whose latency is growing.
In the evaluation, six open-source LC applications (two of them from TailBench) are used as well as a multi-threaded BE job running in a separate container.

Finally, we analyze those approaches that consider VM virtualization.
HCloud~\cite{hcloud}, is a resource provisioning system that assigns resources (e.g., VMs, container instances) and instance sizes (e.g., number of vCPUs) depending on the QoS constrains of the jobs.
The approach is evaluated on Google Compute Engine (GCE) using three representative workload scenarios (from low to high load variability) that run a mix of batch workloads and LC workloads.
However, authors do not consider reducing unpredictability through resource partitioning.
AVMMC~\cite{avmmc}, on the other hand, determines dynamically the best VM to core mapping on the studied system.
In this approach, authors pursue to improve the capabilities of the hypervisor to monitor and manage co-running VMs. To do so, they classify each VM as compute or memory intensive to group together VMs with different behaviors.
The evaluated workloads are just mixes composed of SPEC CPU2006 benchmarks.
Finally, DeepDive~\cite{deepdive} pursues to identify and manage performance interference VMs in IaaS clouds.
To do so, the system collects about a dozen low-level metrics including performance counters in two (interference and no-interference) multi-dimensional cluster.
An important weakness of this approach is that cloning and running VMs in isolation may present
important problems in production systems, since isolation based methodologies may become costly and prohibitive.

\section{Experimental Hardware/Software System}\label{sec:system}

This section presents a general overview of the experimental platform - both hardware and software - used in this characterization study.
Besides, the devised VM infrastructure is also explained.

\subsection{Hardware/Software System Specifications}\label{sec:system:subsec:specs}

\begin{figure}[tb] 
\begin{center}
\includegraphics[width=1.0\columnwidth]{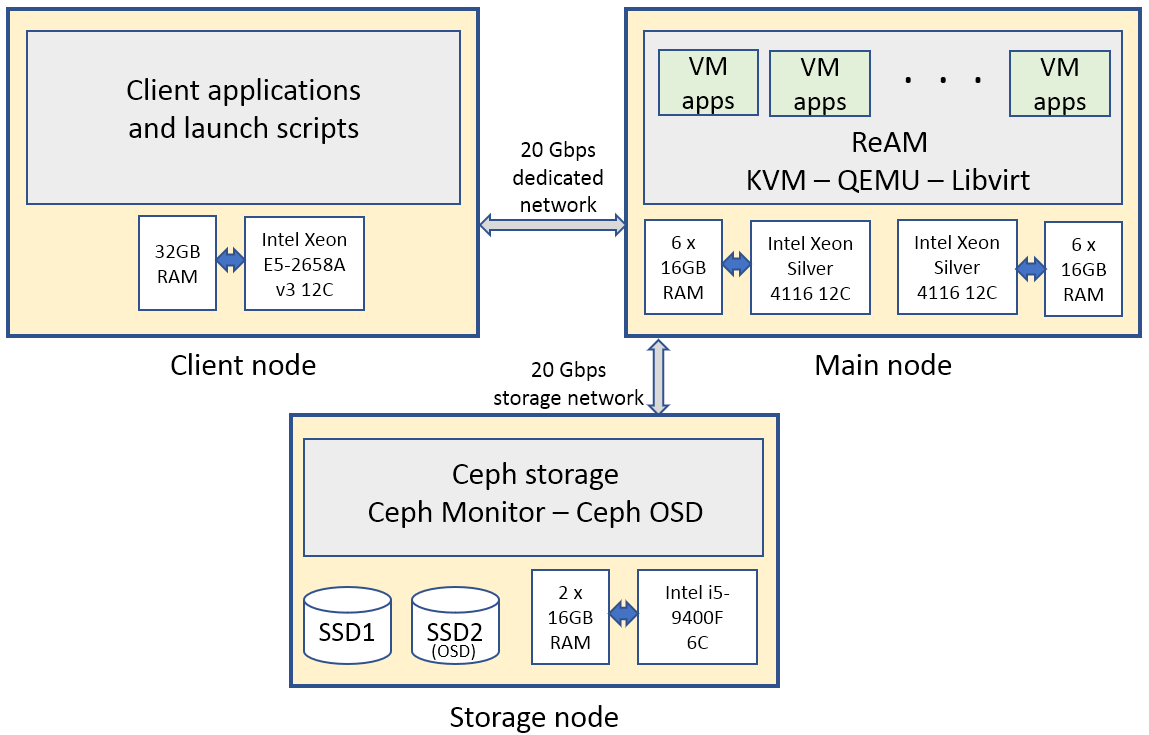}
\caption{Overview of the complete experimental framework.}
\label{fig:experimental_setup}
\end{center}
\end{figure}

The experimental framework is made up of three physical servers (main, client and storage nodes) interconnected with two 20 Gbps dedicated links.
Figure~\ref{fig:experimental_setup} shows a block diagram of the experimental platform, including the three nodes: main node, client node and storage node, and the installed packages and libraries. Below these nodes are described.

The main node is where the developed framework, Resource and Application Manager software (ReAM), is installed. This node runs the VMs hosting the corresponding applications or benchmarks to be studied. The node has two Intel Xeon Silver 4116 processors (i.e., two sockets), with six memory channels holding 16GB each, which amounts to 96GB (6x16GB) DRAM and supports a bandwidth of 107.3 GB/s. 
This processor is deployed with 12 cores with a 16.5GB 11-way LLC. 
It supports Intel Cache Allocation Technology (CAT) and Memory Bandwidth Allocation (MBA)~\cite{intel_rdt}, allowing to perform cache and memory bandwidth partitioning studies.

The client node is only used to run the clients for client-server benchmarks, e.g., the applications from the TailBench benchmark suite. In this type of applications, the main node starts the client applications on the client node with an ssh command, which executes a script in this node to launch the client-side of the application.

The storage node implements a remote storage system with Ceph~\cite{ceph}. This node has a SATA SSD hard drive devoted to the remote storage. It runs a Ceph OSD daemon and a Ceph monitor.

Regarding the system software, the main node has a virtual switch configured to connect the physical network interfaces with the VMs. 
The virtual switch is configured using Open vSwitch~\cite{ovSwitch} (OvS) and Data Plane Development Kit~\cite{dpdk} (DPDK). 
OvS and DPDK enable direct transfer of packets between the user space and physical interface, bypassing the kernel network stack. 
This setup boosts network performance compared to the default packet forwarding mechanism implemented in the Linux kernel. 
Additionally, KVM is installed as hypervisor, QEMU as virtualizer and Libvirt as virtualization manager in both client and server nodes. 
This configuration is similar to that used by OpenStack on its compute nodes~\cite{openstack}.

\subsection{VM Infrastructure}\label{sec:system:subsec:vms}

To replicate a common infrastructure that cloud service providers deploy on their servers, the framework presented in this paper works with VMs.
More precisely, the framework can allocate resource partitions to each individual VM (e.g., number of CPUs, main memory space, LLC space, and main memory, network and disk bandwidth), and monitor its consumption.

The installed operating system in the nodes is Ubuntu 18.04 LTS. 
The VMs have three main benchmark suites installed: TailBench, SPEC CPU2006 and SPEC CPU2017 (see Section~\ref{sec:workloads}).
The client node has installed the clients of the Tailbench and \emph{media-streaming} (from CloudSuite) workloads.

To reduce the start-up overhead, the designed framework makes use of the snap-shots feature of \emph{libvirt}~\cite{libvirt}.
For each VM, we have taken a snapshot of the state where the OS boot process is already performed and it is ready to receive the command to launch the target benchmark.

\subsection{Intel Hyper-Threading}

The processor of the experimental platform implements the Hyper-Threading~\cite{hyper-threading} technology, that is, Intel's implementation of the simultaneous multi-threading (SMT) paradigm~\cite{smt}. The key characteristic of the SMT architecture is its ability to issue several instructions from multiple threads each cycle, which allows increasing the utilization of the functional units and thus the processor throughput.

Hyper-Threading processors support the concurrent execution of two threads. From the operating system perspective, a given physical core is seen as two logical cores or two different CPUs; however, both are on the same physical core.

In a multi-core processor with Hyper-Threading cores, it is very important to discern between physical cores and logical cores when talking about resource sharing. Two threads running on two different cores share the inter-core shared resources of the processor which are mainly the LLC and main memory. Two threads running on the logical cores of the same physical core, in addition, compete for  intra-core shared resources. This set of components is critical for performance and includes, among others, the ROB, load queue, store queue, function units, as well as, L1 and L2 caches.

\section{Latency Critical Applications: Load Levels and QoS}\label{sec:workloads}

Latency-critical applications are being widely used in cloud systems. To support these applications, the cloud server typically implements online interactive services (e.g., web search) and must respond to the input requests within specific latency bounds to guarantee QoS and provide a satisfactory user experience.

As a representative set of latency-critical applications, we use the benchmarks of the TailBench benchmark suite~\cite{tailbench}. This suite includes eight representative applications of today's latency-critical applications. For the sake of completeness, we briefly describe the studied applications. 
The main characteristics of the studied applications are:

\begin{itemize}

\item \texttt{\textbf{img-dnn}} is a handwriting recognition application based on OpenCV. The application uses randomly-chosen samples from the MNIST database.

\item \texttt{\textbf{masstree}} is a fast and scalable in-memory key-value store written in C++. Each user request often involves many tens or hundreds of requests to the key-value store; this type of applications, therefore, have very short latency requirements.

\item \texttt{\textbf{moses}} is a statistical machine translation system written in C++. We drive \textit{moses} using randomly-chosen dialogue snippets from the opensubtitles.org English-Spanish corpus.

\item \texttt{\textbf{shore}} is a transactional on-disk database. Database and logs are both stored in a solid state drive.

\item \texttt{\textbf{silo}} is an in-memory transactional database. \textit{Silo} is designed to scale well on modern multicores, and uses TPC-C, an industry-standard OLTP benchmark. \textit{Silo} and \textit{shore} differ significantly in how they store and access data.

\item \texttt{\textbf{specjbb}} is an industry-standard Java middleware benchmark. Java middleware is widely used in business services and must often satisfy strict latency constraints. 

\item \texttt{\textbf{sphinx}} is a compute-intensive speech recognition system written in C++. Speech recognition systems are an important component of speech-based interfaces and applications such as Apple Siri, Google Now, and IBM Speech to Text.

\item \texttt{\textbf{xapian}} is a search engine written in C++ that is widely used both in
software frameworks (e.g., Catalyst) and popular websites (e.g. the Debian wiki).

\end{itemize}
In addition, in this work we also analyze the \emph{media-streaming} workload from CloudSuite~\cite{cloudsuite}. 
This application, based on the NGINX web server, is a streaming server for hosted synthetic videos of various lengths and qualities.
The server is accessed by a client based on the \emph{httperf's wsesslog} session generator, which performs a set of requests per session for the videos stored in the server. 
For this application, we collect the average \emph{response} and \emph{transfer} times across all requests, which are the metric that have been chosen as the most suitable performance metrics.
This popular application in today's datacenters allows to broaden even more the spectrum of behaviors, with the objective to have a wide range of behaviors.

\subsection{Setting Representative Load Levels}\label{sec:workloads:subsec:load}

Unlike other benchmark suites like SPEC CPU, TailBench introduces the Queries Per Second (QPS) parameter that allows generating a wide range of load levels. Therefore, there is a need to tune this parameter in order to match the desired workload level. Below, we discuss the approach followed to obtain a valid and representative range of QPS for each of the studied applications.

In order to simulate real clinet-server behavior, clients of the TailBench applications issue requests to the server following the Zipfian distribution~\cite{baeza05,feitelson15}.
To do so, the workload clients use a \emph{request generator} to indicate the times when the requests must be issued to satisfy the demanded QPS and the Zipfian distribution. Unfortunately, it may happen that when a large QPS is demanded, the client cannot generate and send the requests fast enough. 
In this scenario, the client might still reach the desired QPS on average but it probably breaks the Zipfian distribution, which compromises the representativeness of the experiment.

To address this issue and guarantee the representativeness of the experiments, we defined a new metric called \emph{timely requests ratio}, which accounts for the percentage of requests that the clients are able to issue fulfilling the request times obtained following the Zipfian distribution. 
A request is defined as \emph{non-timely} when the request time, provided by the request generator, is earlier than the current time.
Notice that clients can break the distribution and still meet the requested QPS.
For instance, we experimentally saw that although a single client can generate up to thousands of requests per second, usually it can only generate between 400 and 600 without breaking the distribution.
In the experimental results we present in this study, we always make sure that the target QPS is achieved while guaranteeing that, at least, 97.5\% of the requests are timely.
It is also important to take into account that, to prevent performance from being dominated by huge queuing delays, latency-critical workloads are typically run at low CPU loads~\cite{Barroso07}.
To make our experiments representative, we make sure that, for each application, we evaluate a QPS range that covers, at least, from 20\% to 50\% CPU utilization.
Note that TailBench workloads cover a wide range of response times (e.g., from microseconds to seconds), which affects the QPS range required to evaluate the target loads range for the CPU utilization.

Meeting the discussed criteria, Table~\ref{tab:qps} presents the QPS range defined for each workload and the number of clients used to generate the requests at the client side. 
As observed, QPS widely varies among applications; this metric ranges from 0.2 to 2.0 for \emph{sphinx} and from 250 to 6000 for \emph{silo}. 
For each workload, we have set the minimum number of clients that guarantees 97.5\% of timely requests and keeps the QPS within its range.

\begin{table}[]
\centering
\begin{tabular}{|l|c|c|}
\hline
\textbf{Workload}   & \textbf{QPS range}    & \textbf{Number of clients} \\ \hline
img-dnn             & 100 -- 2000           & 12          \\
masstree            & 250 -- 2500           & 12          \\
moses               & 10 -- 500             & 6          \\
shore               & 10 -- 300            & 12          \\
silo                & 250 -- 6000          & 18          \\
specjbb             & 250 -- 7000          & 18          \\
sphinx              & 0.2 -- 2.0           & 2          \\
xapian              & 100 -- 1100            & 4          \\\hline
\end{tabular}

\vspace{0.2cm}

\caption{QPS range and number of clients used for each workload to guarantee that the ratio of timely requests is above $97.5\%$.}
\label{tab:qps}
\end{table}

Unlike the TailBench applications, the load of \emph{media-streaming} is not computed as the QPS rate but as the number of \emph{sessions}. That is, any number of clients can be launched to complete the target number of sessions, which marks the final amount of requests the server receives. In this work we have explored the client load up to a maximum of 24 sessions and 24 clients (i.e., one session per client), since we found out that a higher number of sessions than clients yields saturated results.

\section{Workload Characterization}\label{sec:characterization}

This section characterizes the behavior of the evaluated applications considering different scenarios.
First. we describe the three scenarios that are considered in the experiments.
Next, we characterize each TailBench application and 
discuss the main observations taken from the experimental results from a latency perspective.

\subsection{Studied Scenarios}\label{sec:charac:subsec:scenarios}

\begin{figure}[tb] 
\begin{center}
\vspace{0.2cm}
\includegraphics[width=1.0\columnwidth]{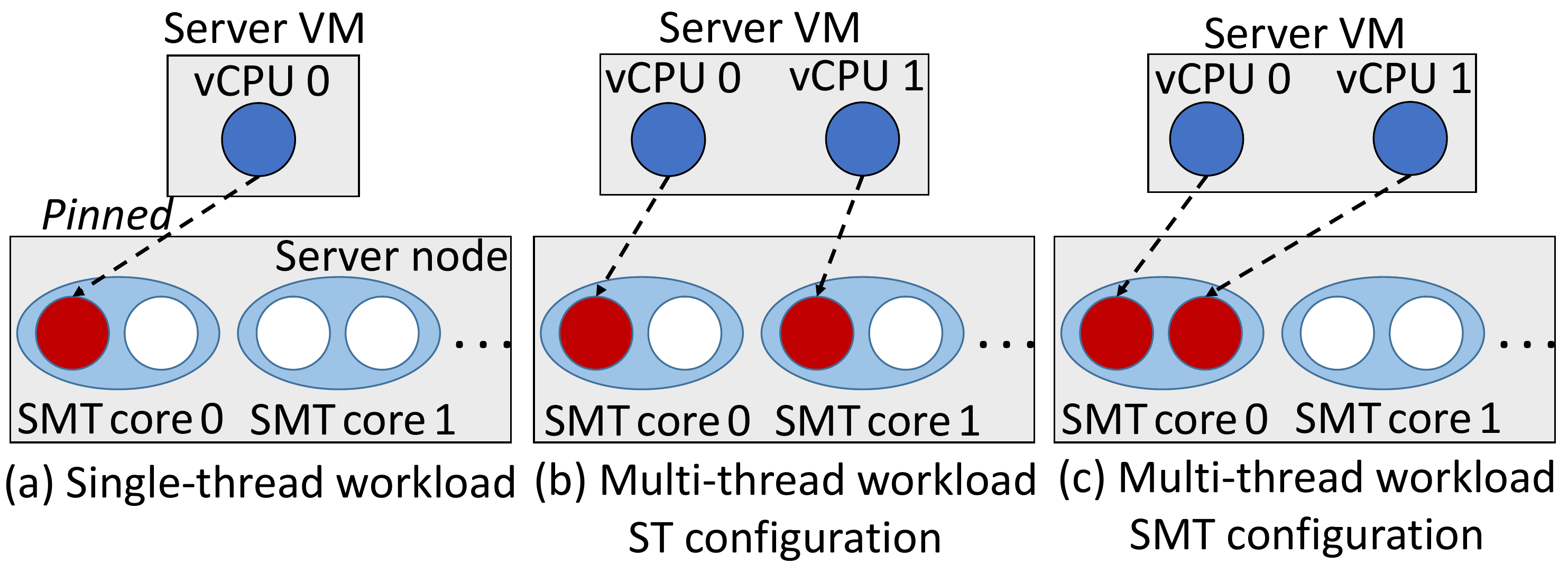}
\caption{Sever VM core configurations studied.}
\label{fig:scenarios}
\end{center}
\end{figure}

In order to study the effect of hyper-threading in the server, we have characterized the studied TailBench applications in the three server scenarios presented in Figure~\ref{fig:scenarios}.
The difference among the studied scenarios lies on the number of threads spawned by the server application and the logical core where each thread is pinned. 
In scenario (a), the server workload runs single-threaded. 
In scenario (b), the server application runs with two threads (vCPU cores), each one pinned to a logical core of different (SMT) physical cores. 
Finally, scenario (c) differs from scenario (b) in that the two threads (vCPU cores) are pinned to logical cores of the same SMT physical core, which runs both of them simultaneously.

Notice that the results of the analysis remain valid regardless of the number of threads being executed. It is a matter of how threads are pinned to cores rather than the number of co-running threads. Therefore, to simplify the analysis only a maximum of two threads has been used in this study.

\subsection{Workload Resource Consumption}
\label{sec:charac:subsec:results_individual}

\begin{figure*}[t!]
    \centering
    \subfloat[$95^{th}$ tail latency (ms)]{%
       \includegraphics[width=0.322\textwidth]{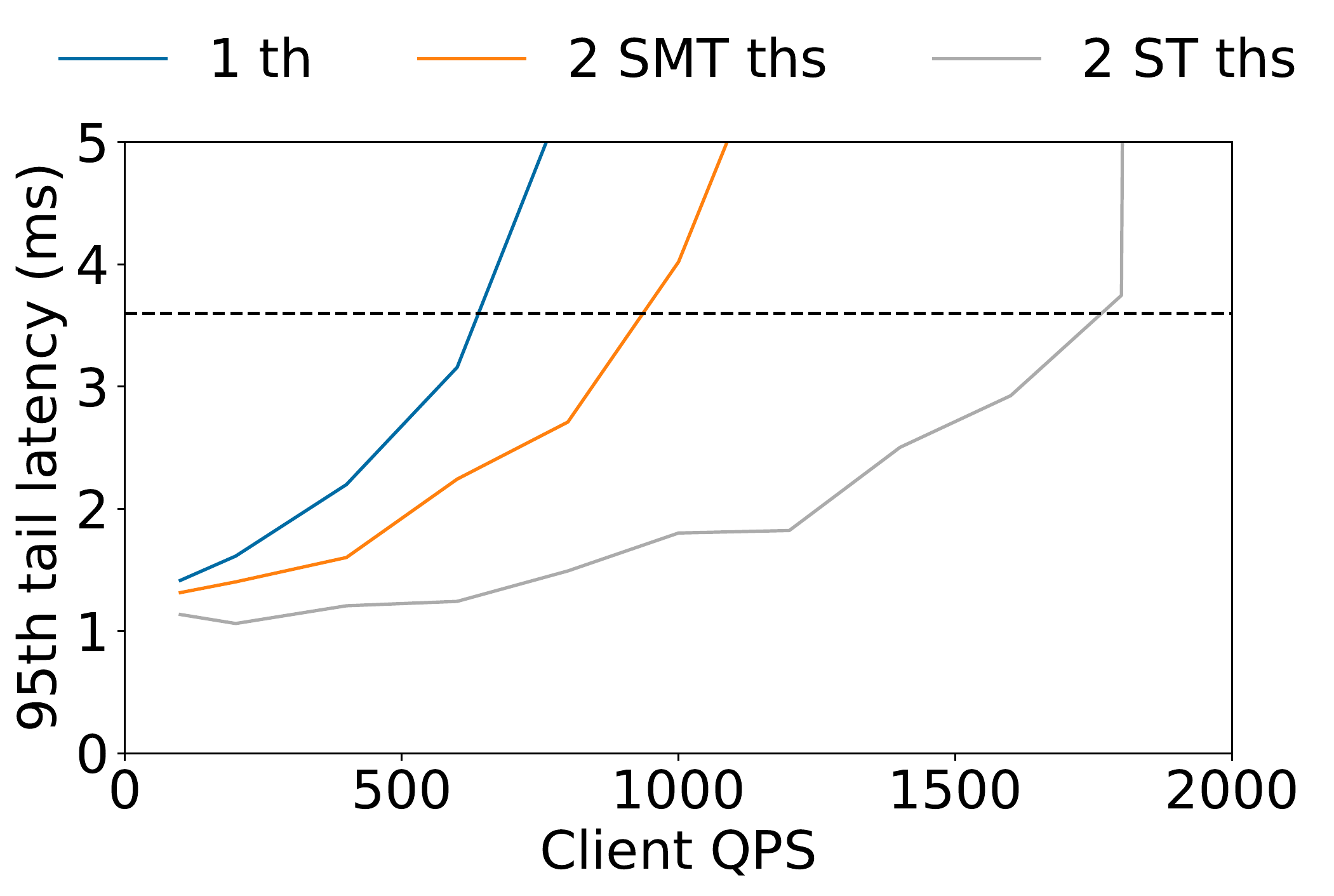}
       \label{fig:img_lat}}
    \subfloat[CPU utilization]{%
       \includegraphics[width=0.322\textwidth]{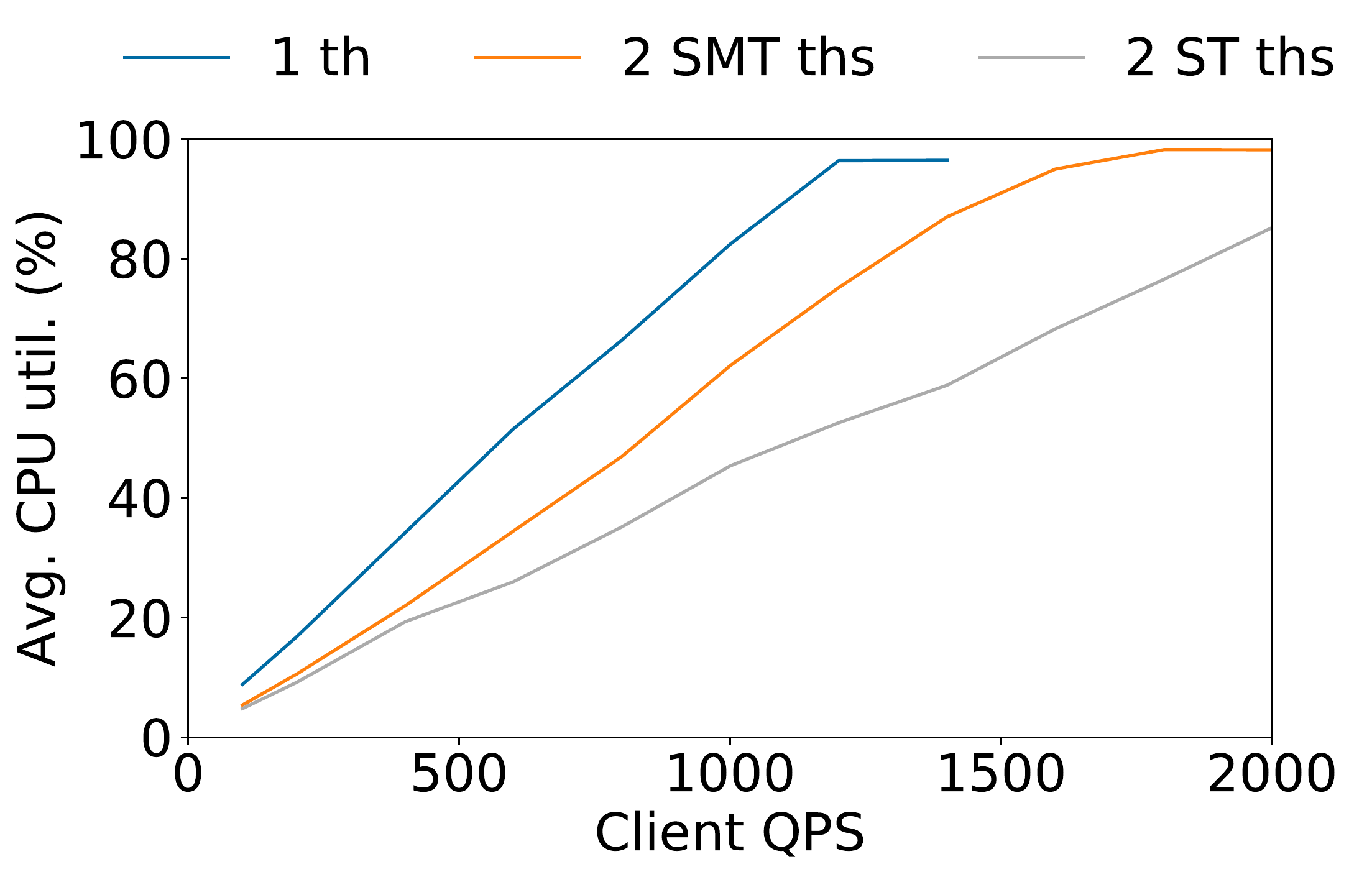}
       \label{fig:img_util}}
    \subfloat[Network transmit bandwidth]{%
       \includegraphics[width=0.322\textwidth]{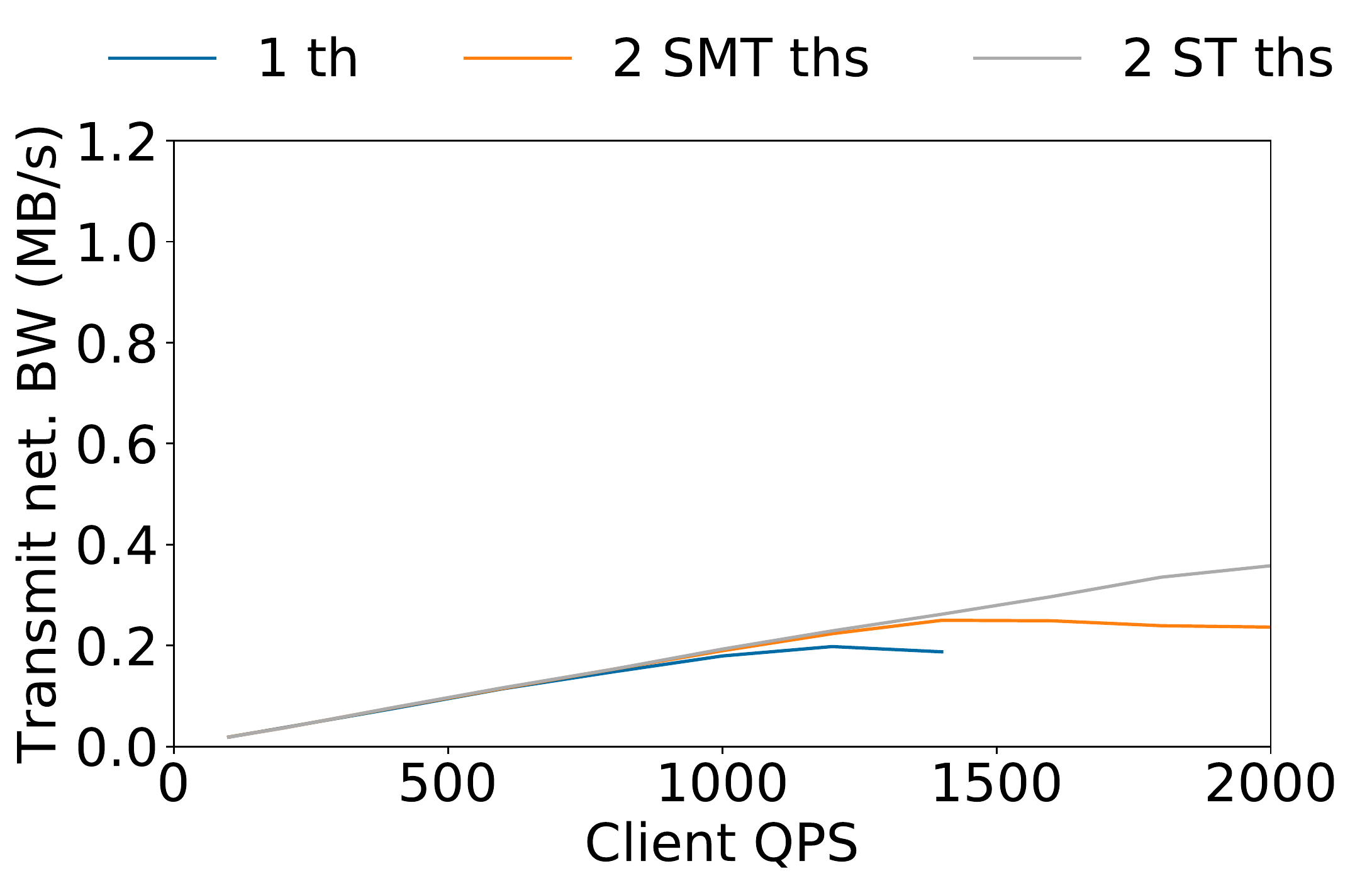}
       \label{fig:img_rx}}
       \\
       \vspace{0.2cm}
    \subfloat[Disk bandwidth]{%
       \includegraphics[width=0.322\textwidth]{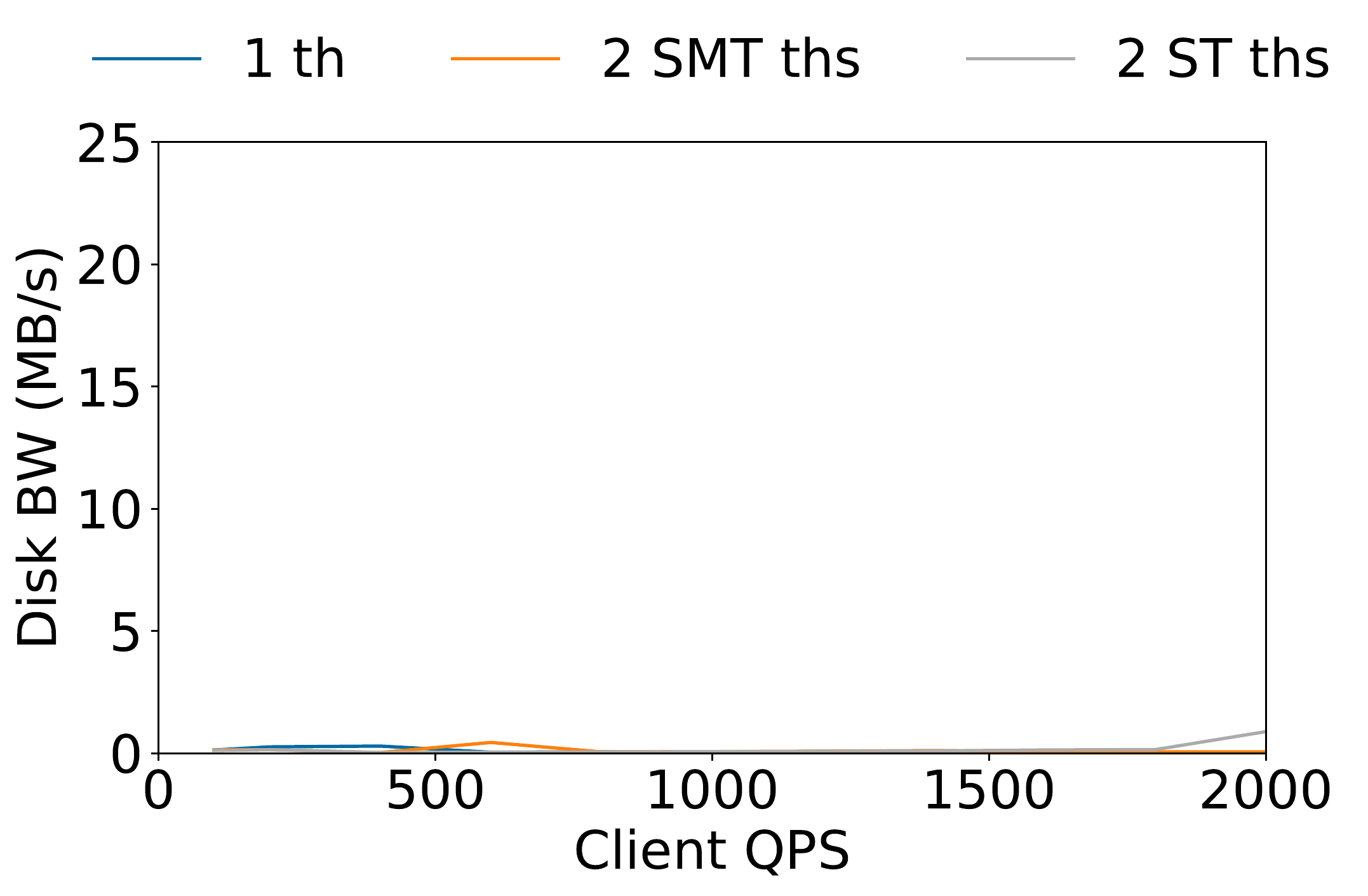}
       \label{fig:img_disk}}
    \subfloat[Main memory bandwidth]{%
       \includegraphics[width=0.322\textwidth]{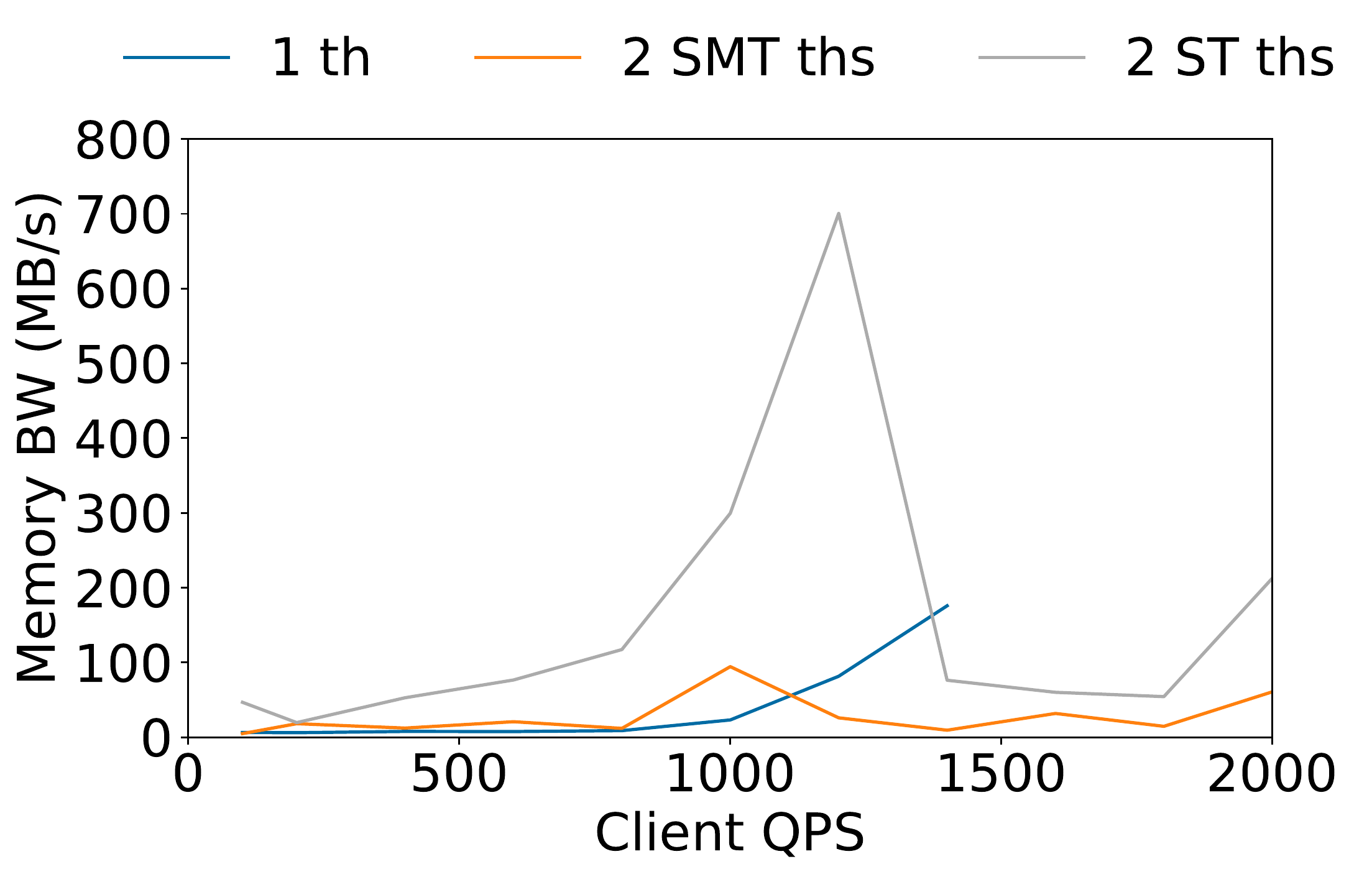}
       \label{fig:img_mem}}
    \subfloat[LLC occupancy]{%
       \includegraphics[width=0.322\textwidth]{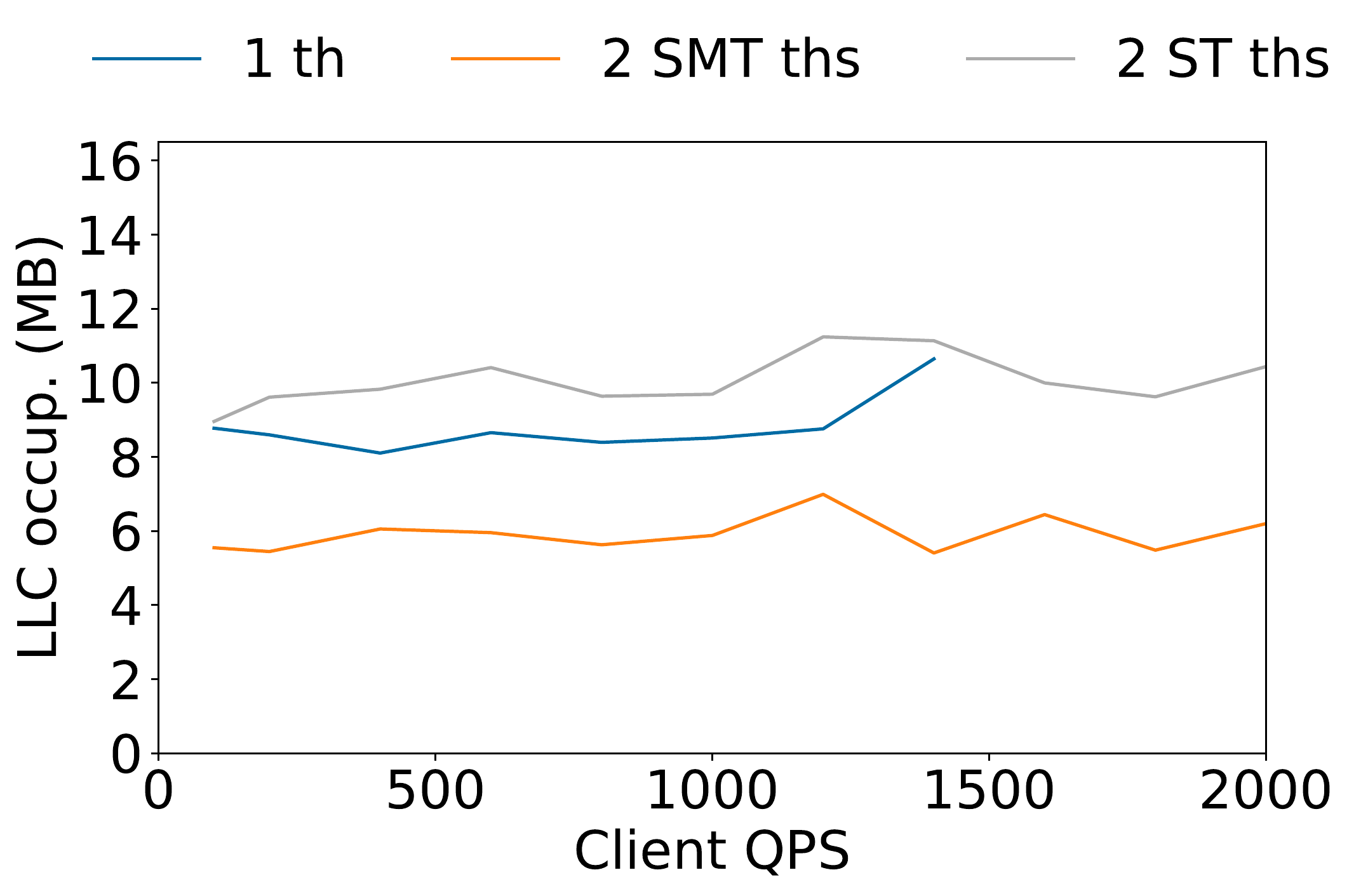}
       \label{fig:img_llc}}
    \caption{Img-dnn characterization.}
    \label{fig:img_noconstrains} 
\end{figure*}

\begin{figure*}[t!]
    \centering
    \subfloat[$95^{th}$ tail latency (ms)]{%
       \includegraphics[width=0.32\textwidth]{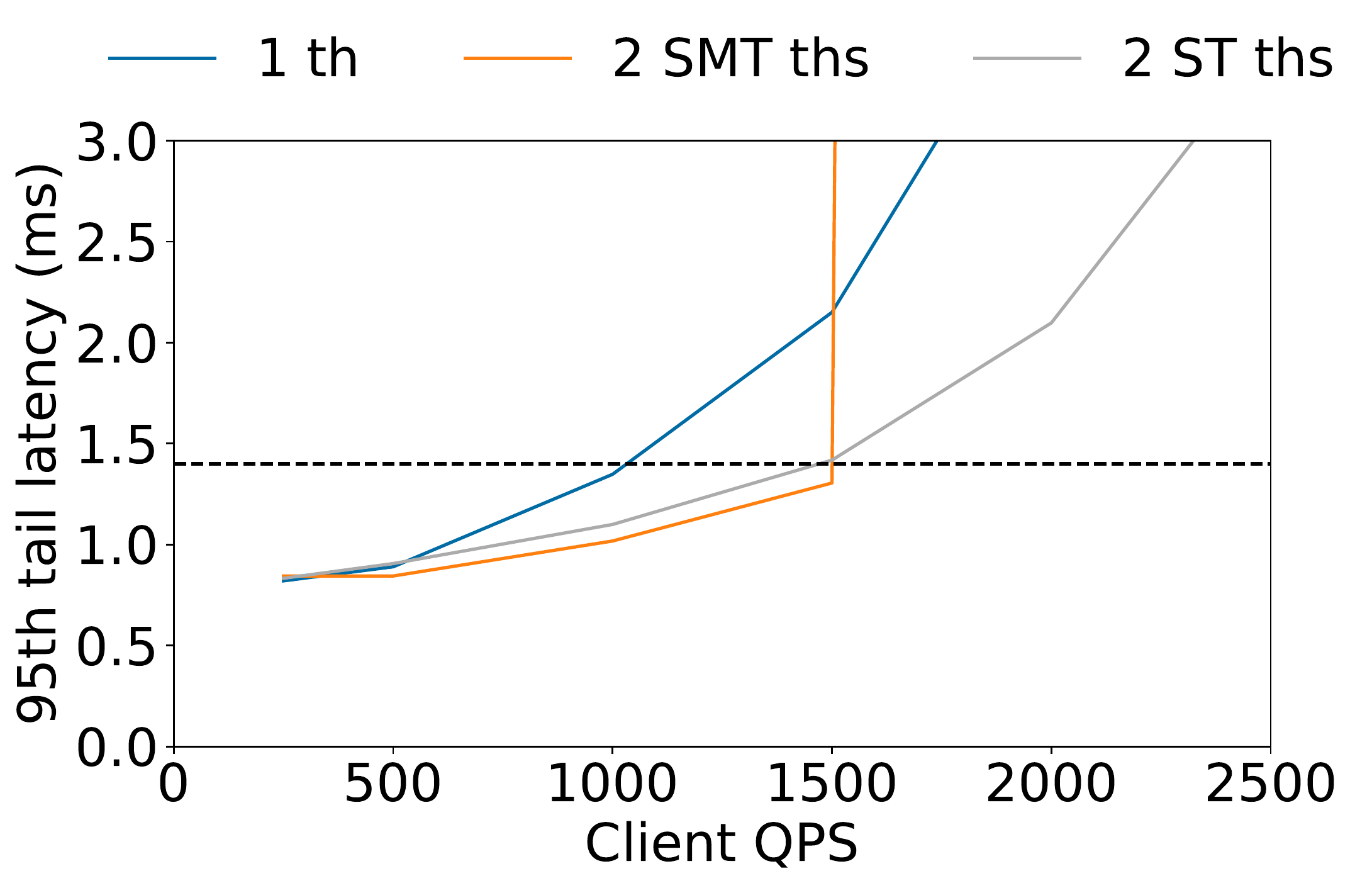}
       \label{fig:masstree_lat}}
    \subfloat[CPU utilization]{%
       \includegraphics[width=0.32\textwidth]{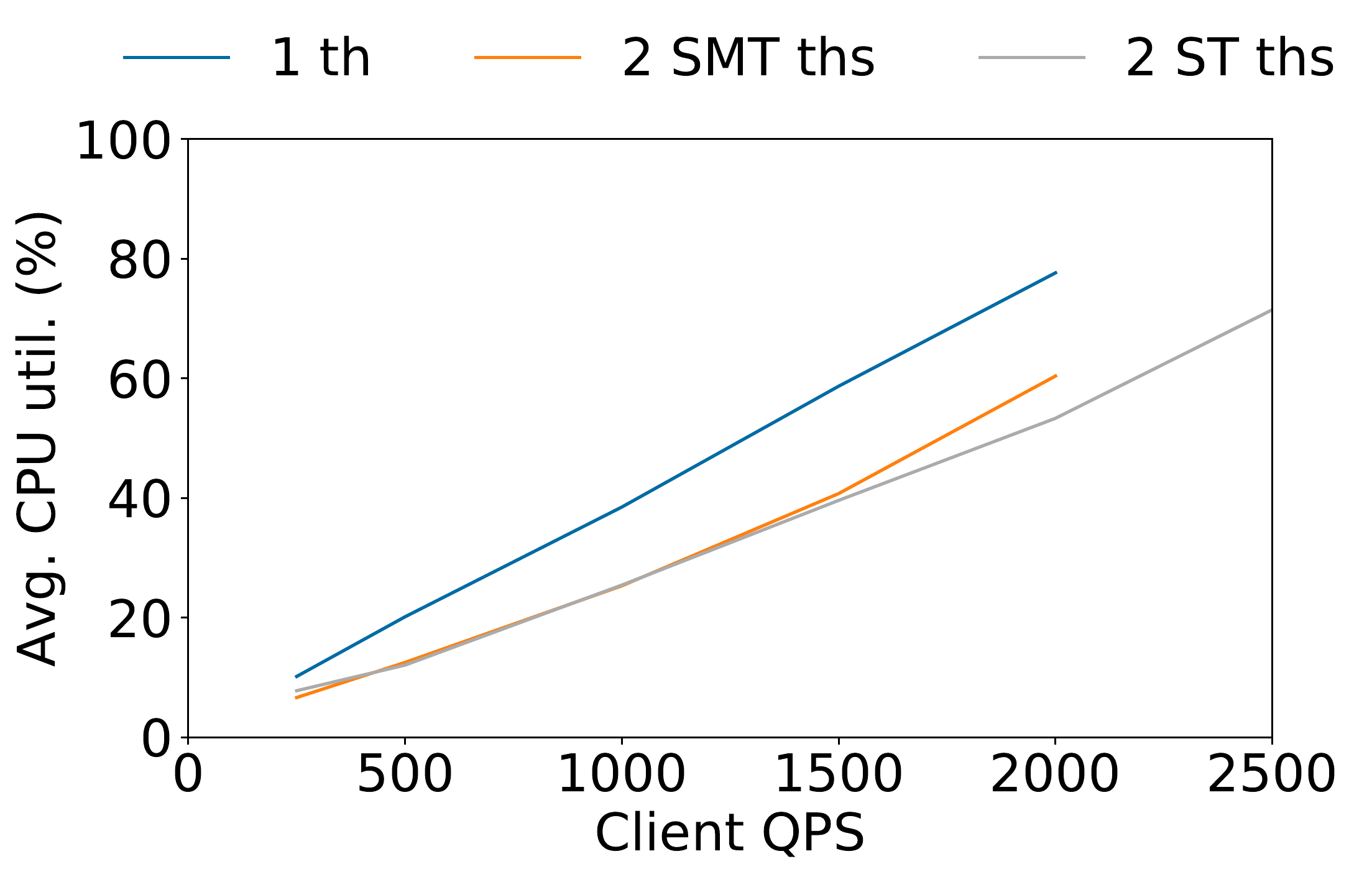}
       \label{fig:masstree_util}}
    \subfloat[Network transmit bandwidth]{%
       \includegraphics[width=0.32\textwidth]{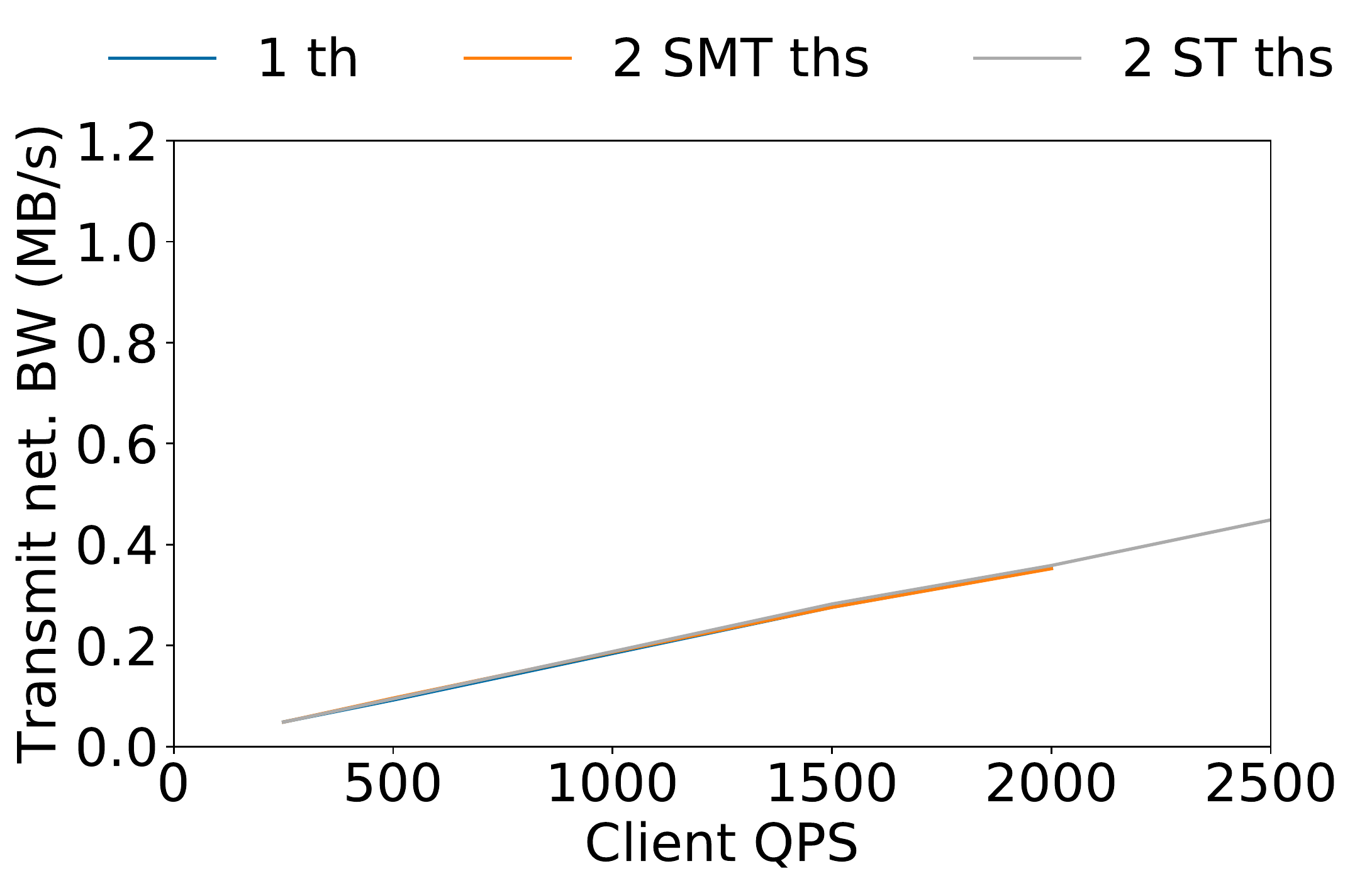}
       \label{fig:masstree_rx}}
       \\
       \vspace{0.2cm}
    \subfloat[Disk bandwidth]{%
       \includegraphics[width=0.32\textwidth]{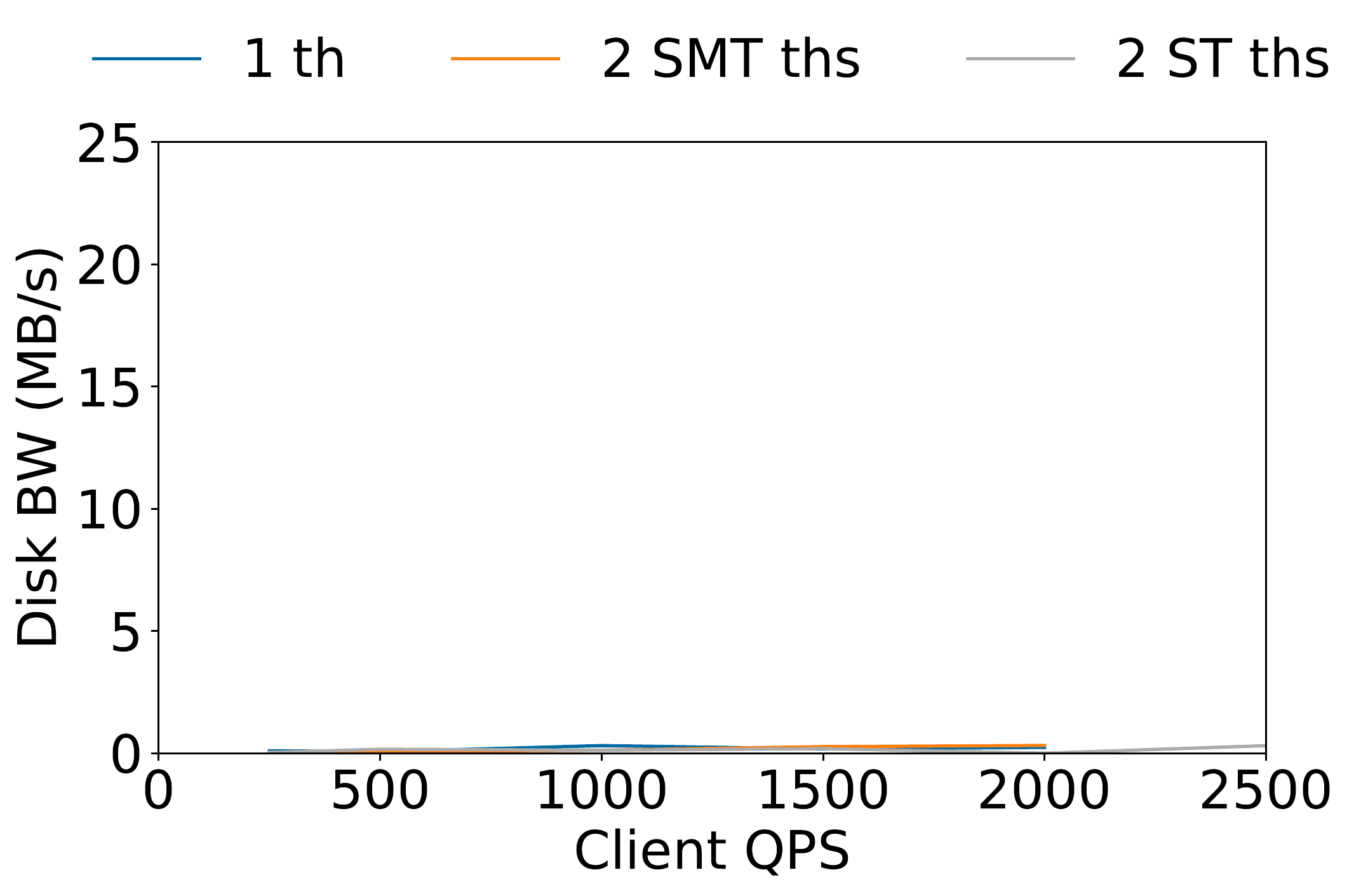}
       \label{fig:masstree_disk}}
    \subfloat[Main memory bandwidth]{%
       \includegraphics[width=0.32\textwidth]{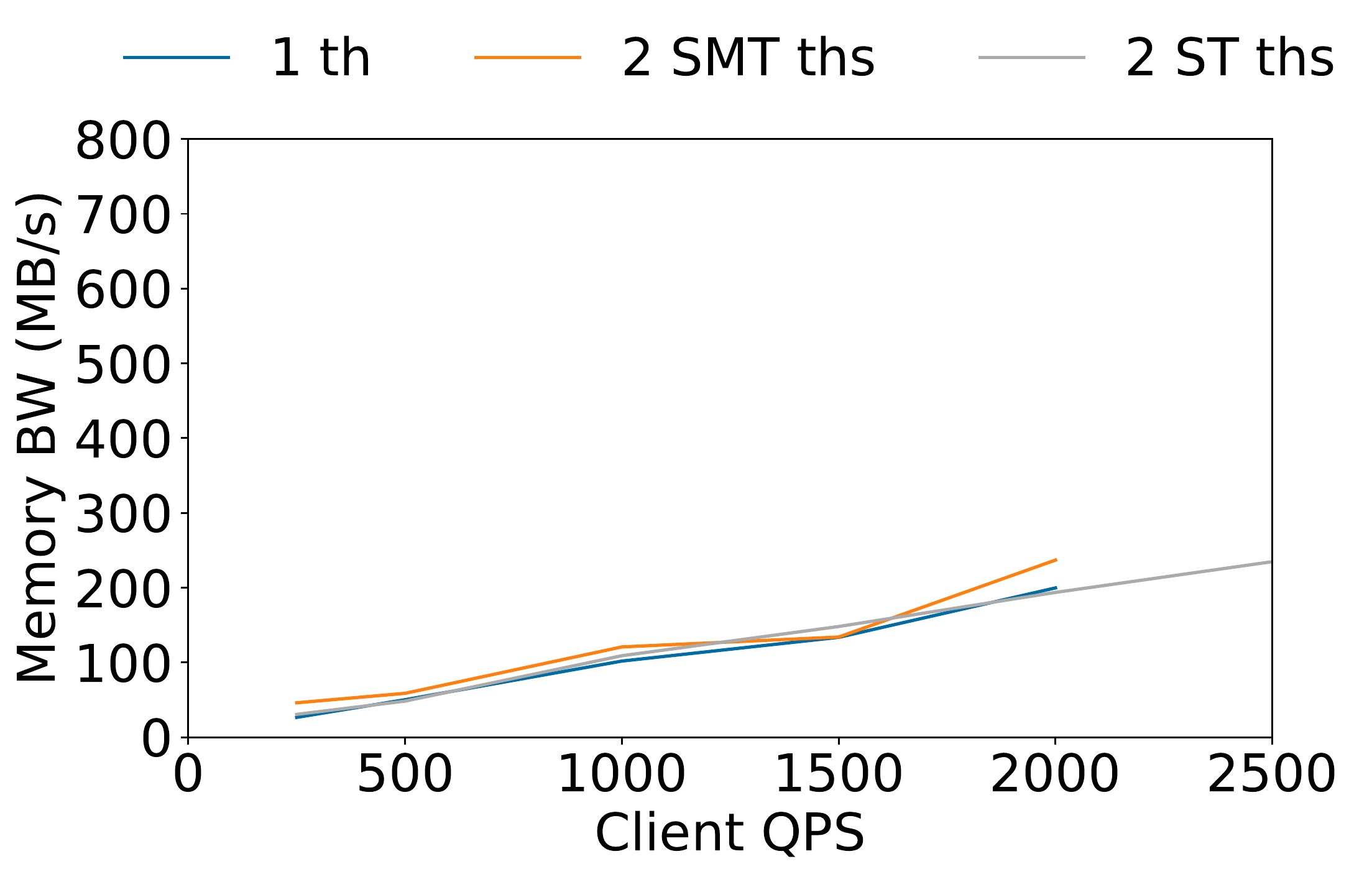}
       \label{fig:masstree_mem}}
    \subfloat[LLC occupancy]{%
       \includegraphics[width=0.32\textwidth]{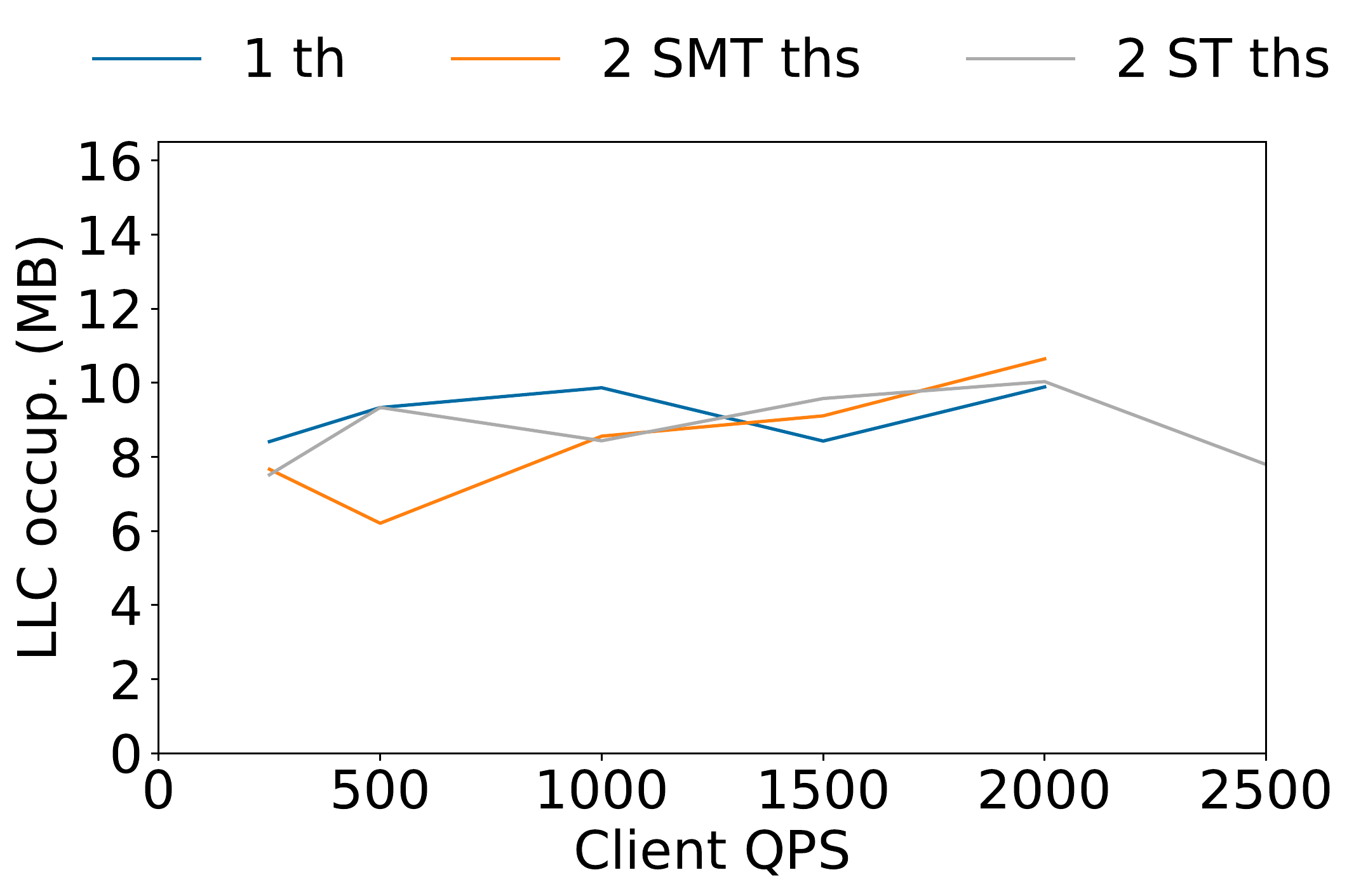}
       \label{fig:masstree_llc}}
    \caption{Masstree characterization.}
    \label{fig:masstree_noconstrains} 
\end{figure*}

\begin{figure*}[t!]
    \centering
    \subfloat[$95^{th}$ tail latency (ms)]{%
       \includegraphics[width=0.32\textwidth]{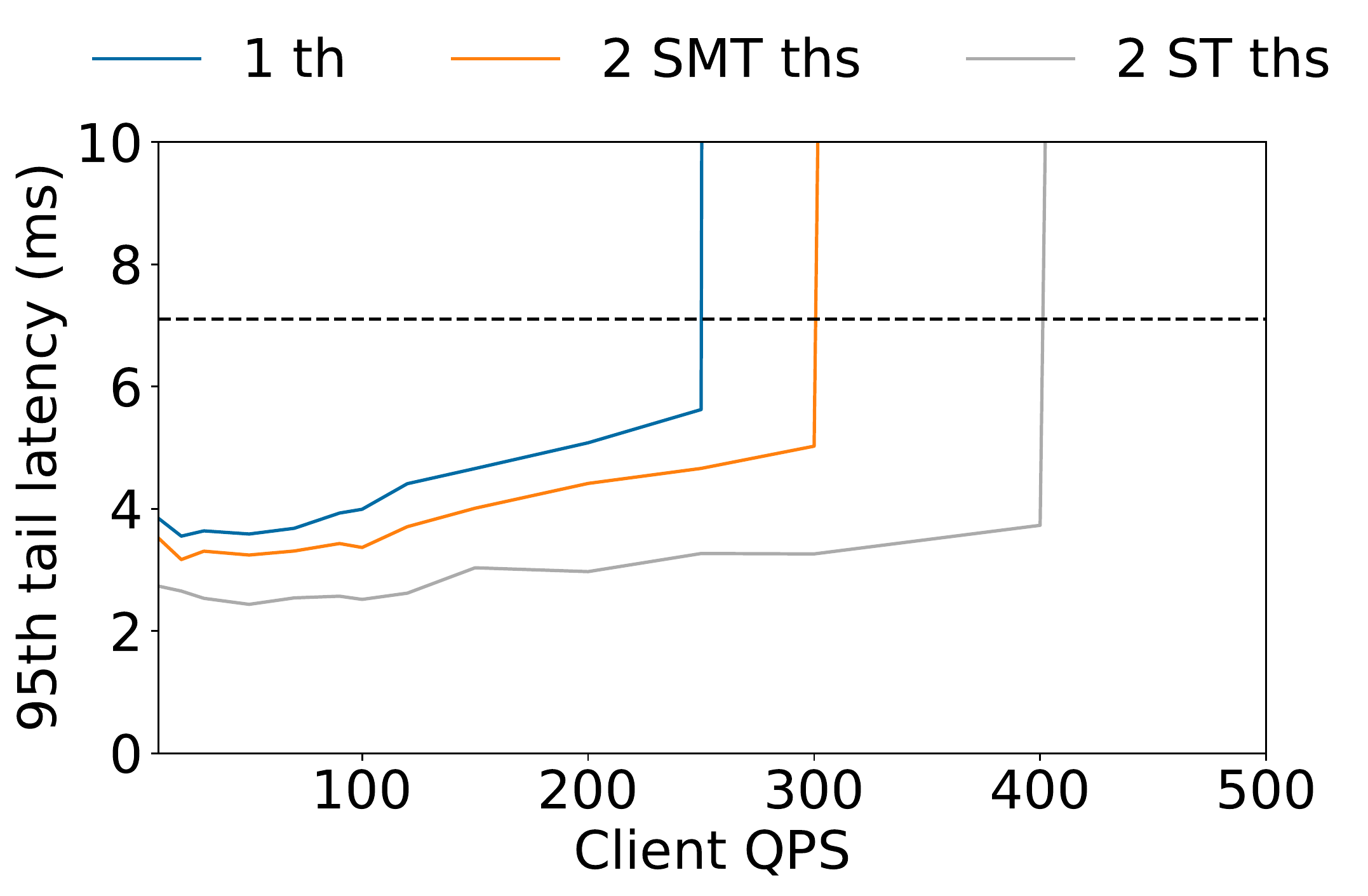}
       \label{fig:moses_lat}}
    \subfloat[CPU utilization]{%
       \includegraphics[width=0.32\textwidth]{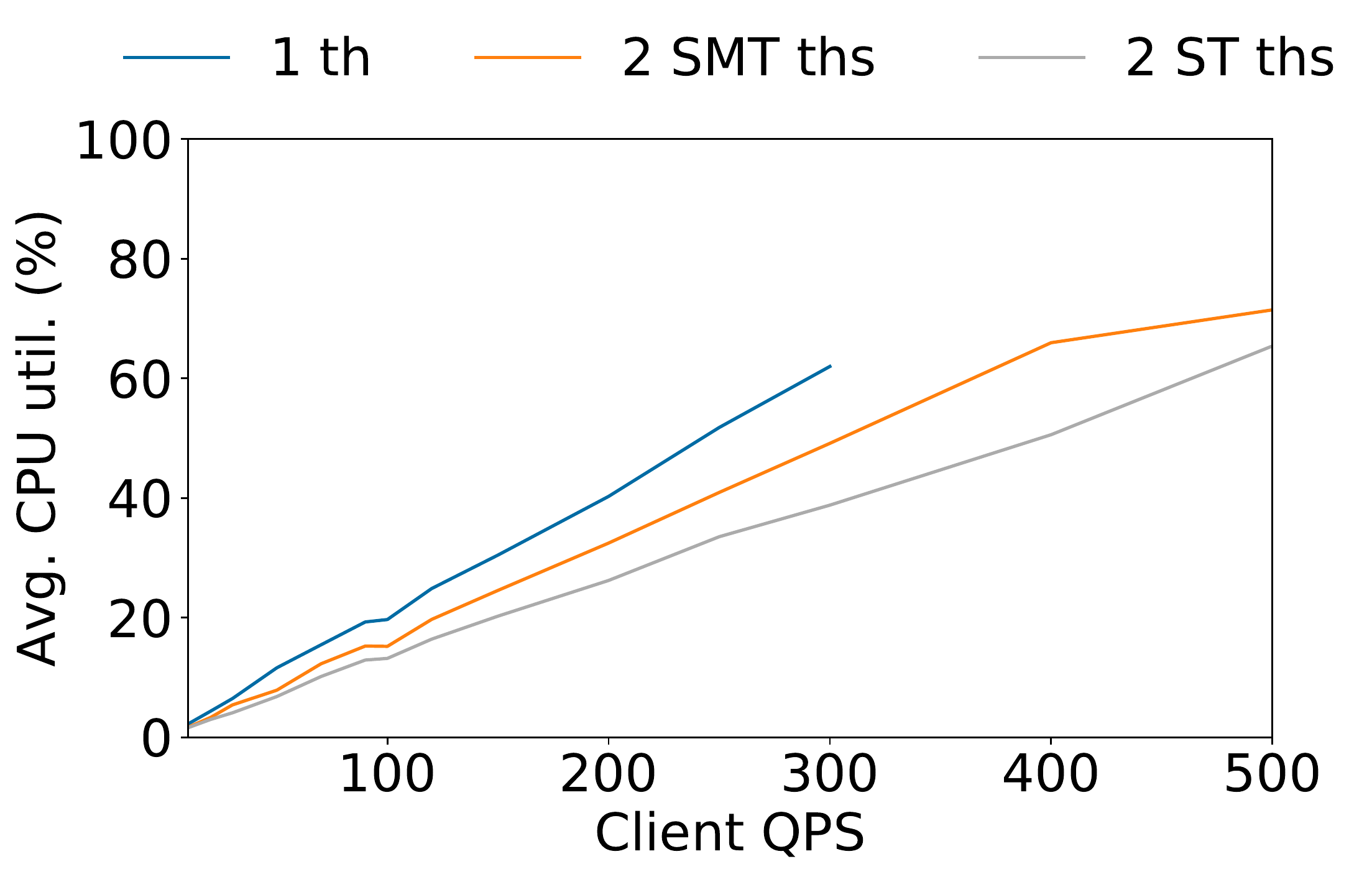}
       \label{fig:moses_util}}
    \subfloat[Network transmit bandwidth]{%
       \includegraphics[width=0.32\textwidth]{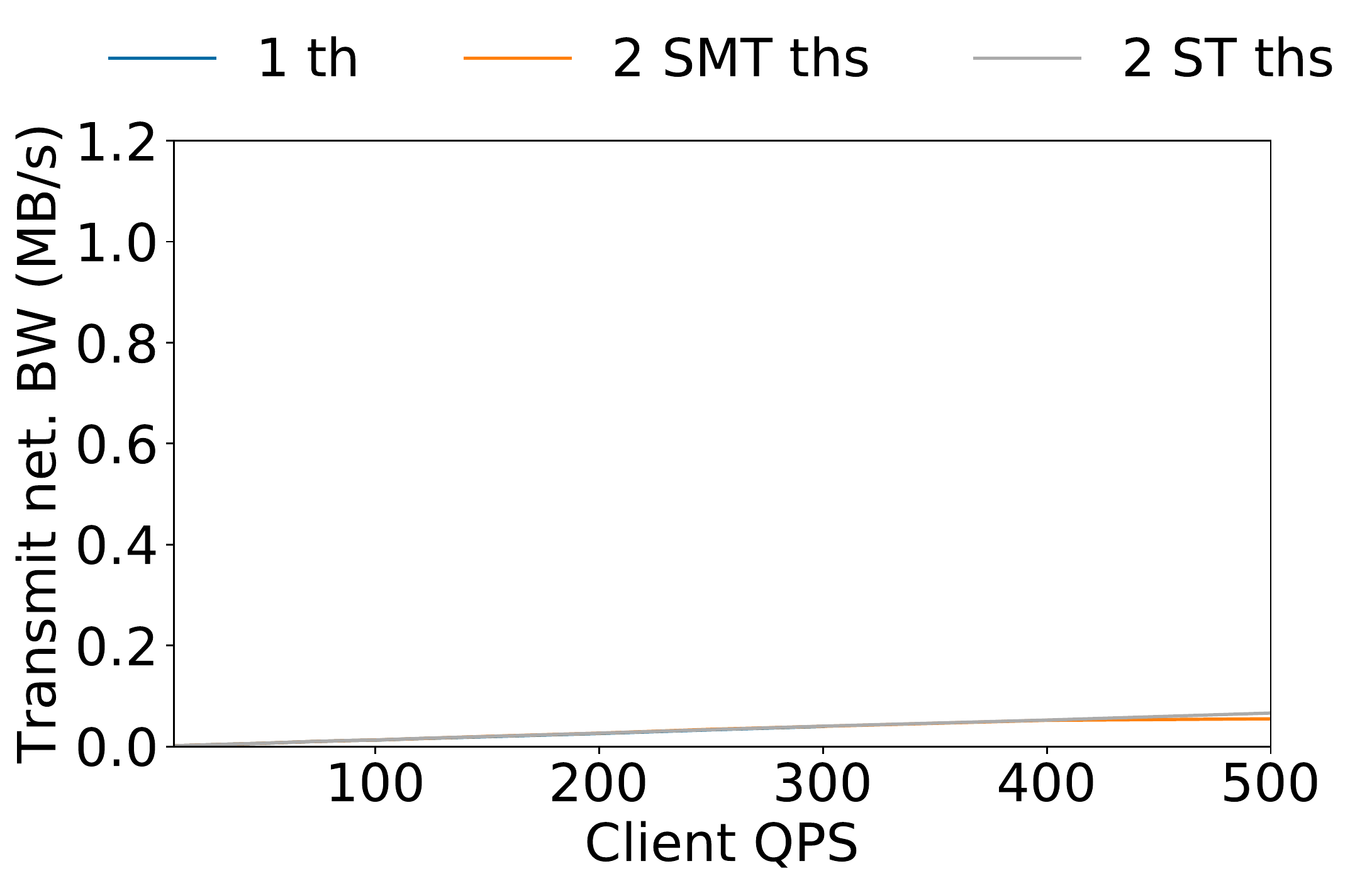}
       \label{fig:moses_tx}}
       \\
       \vspace{0.2cm}
    \subfloat[Disk bandwidth]{%
       \includegraphics[width=0.32\textwidth]{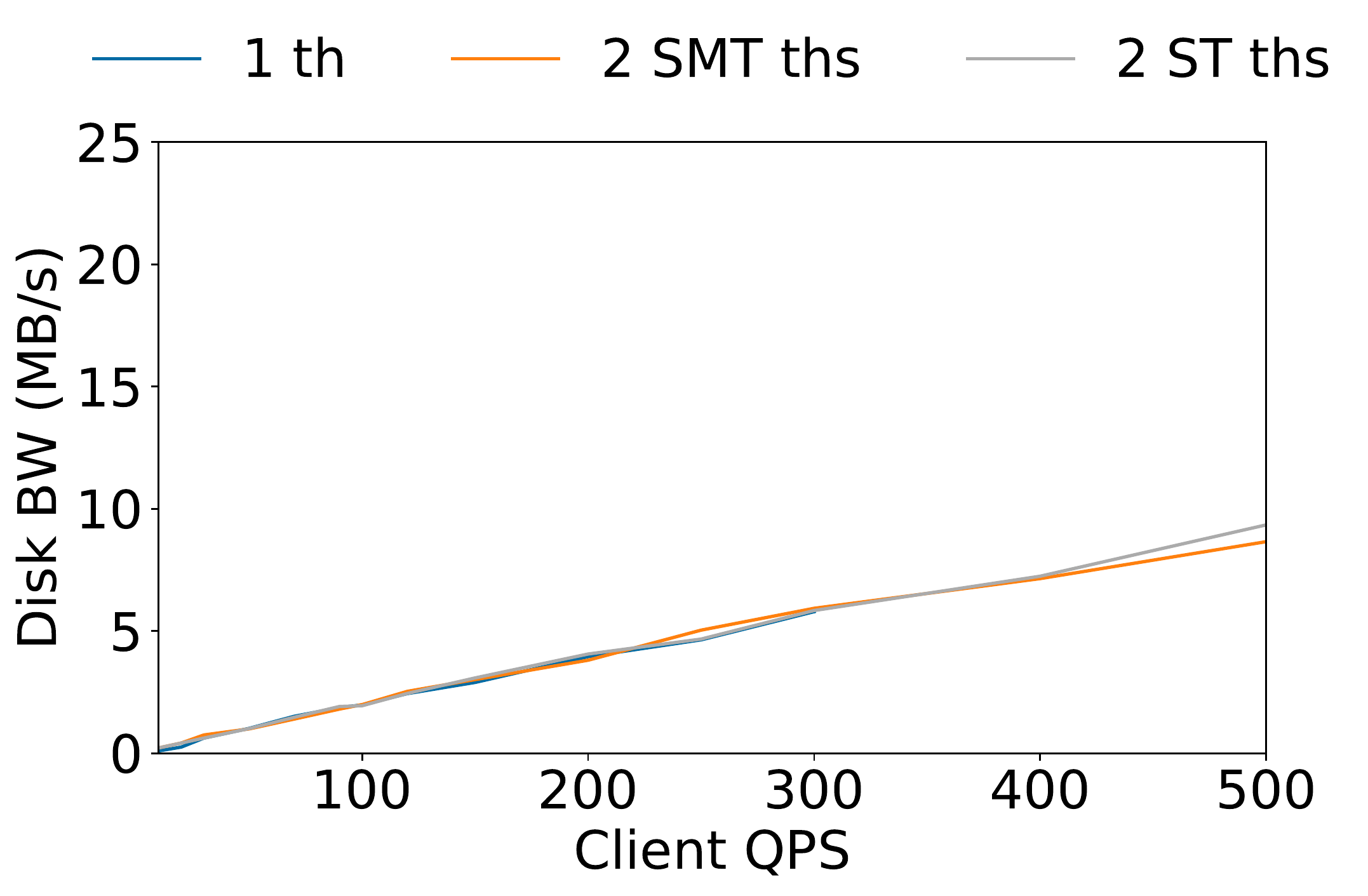}
       \label{fig:moses_disk}}
    \subfloat[Main memory bandwidth]{%
       \includegraphics[width=0.32\textwidth]{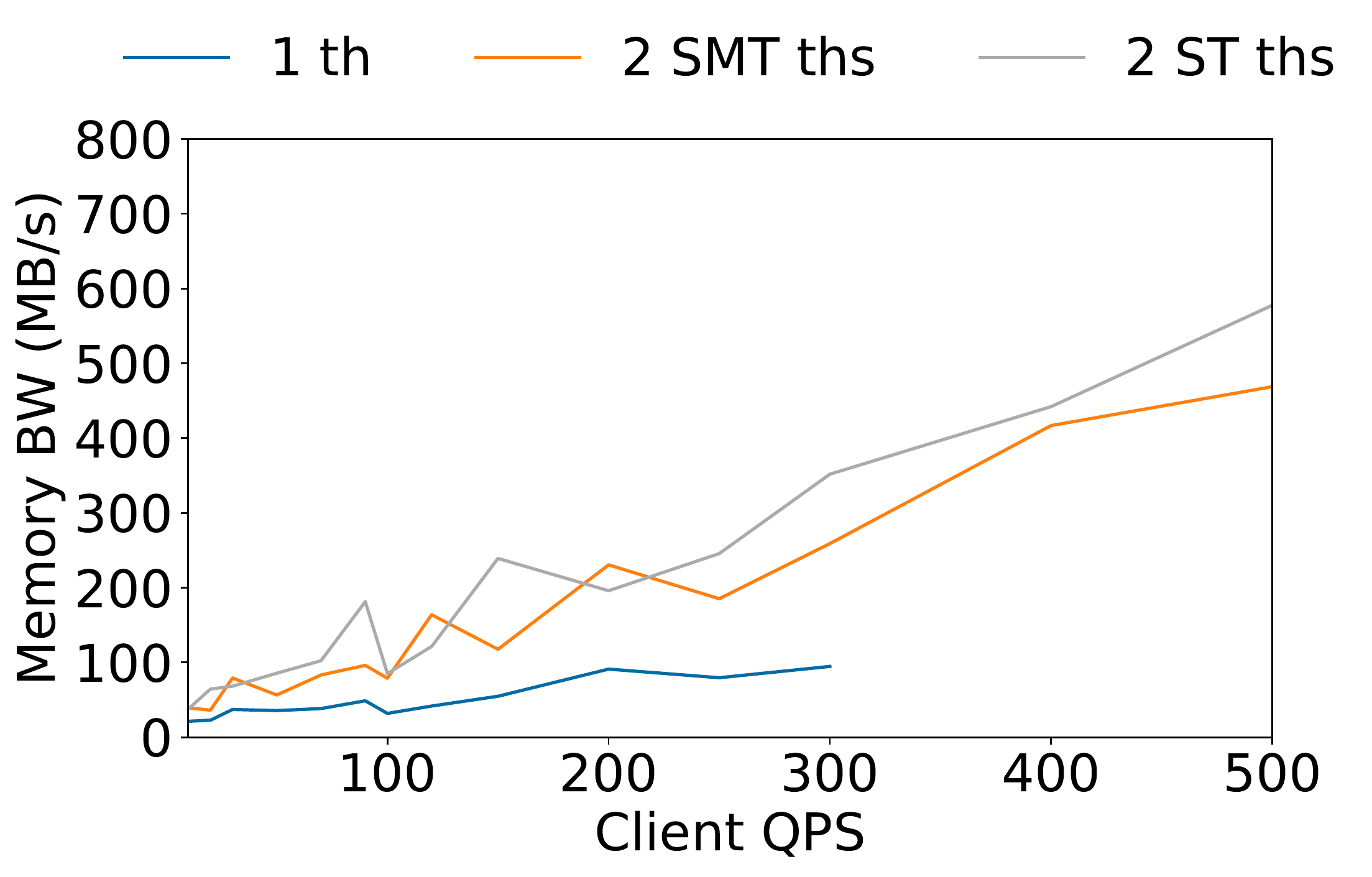}
       \label{fig:moses_mem}}
    \subfloat[LLC occupancy]{%
       \includegraphics[width=0.32\textwidth]{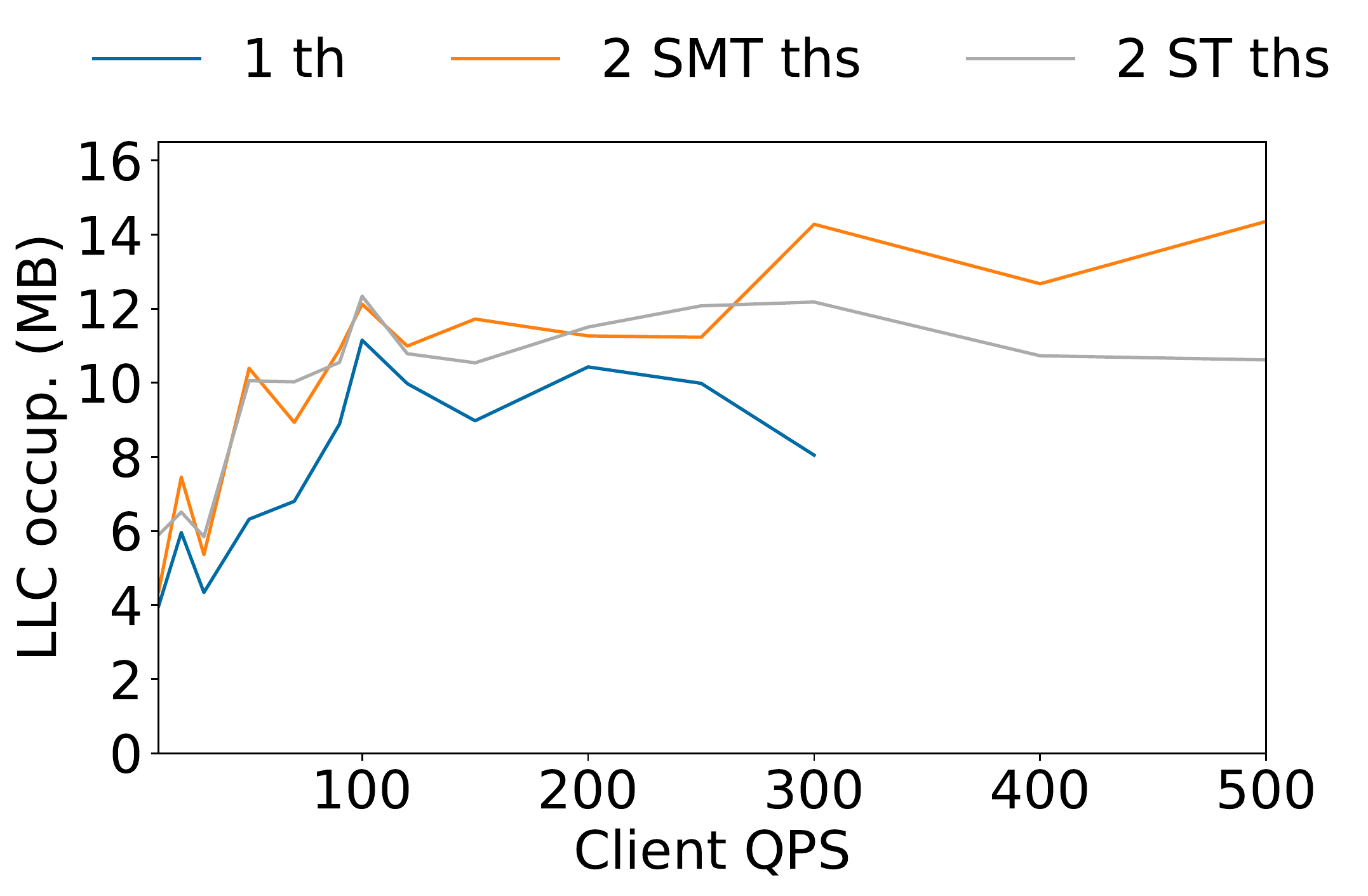}
       \label{fig:moses_llc}}
    \caption{Moses characterization.}
    \label{fig:moses_noconstrains} 
\end{figure*}

\begin{figure*}[t!]
    \centering
    \subfloat[$95^{th}$ tail latency (ms)]{%
       \includegraphics[width=0.32\textwidth]{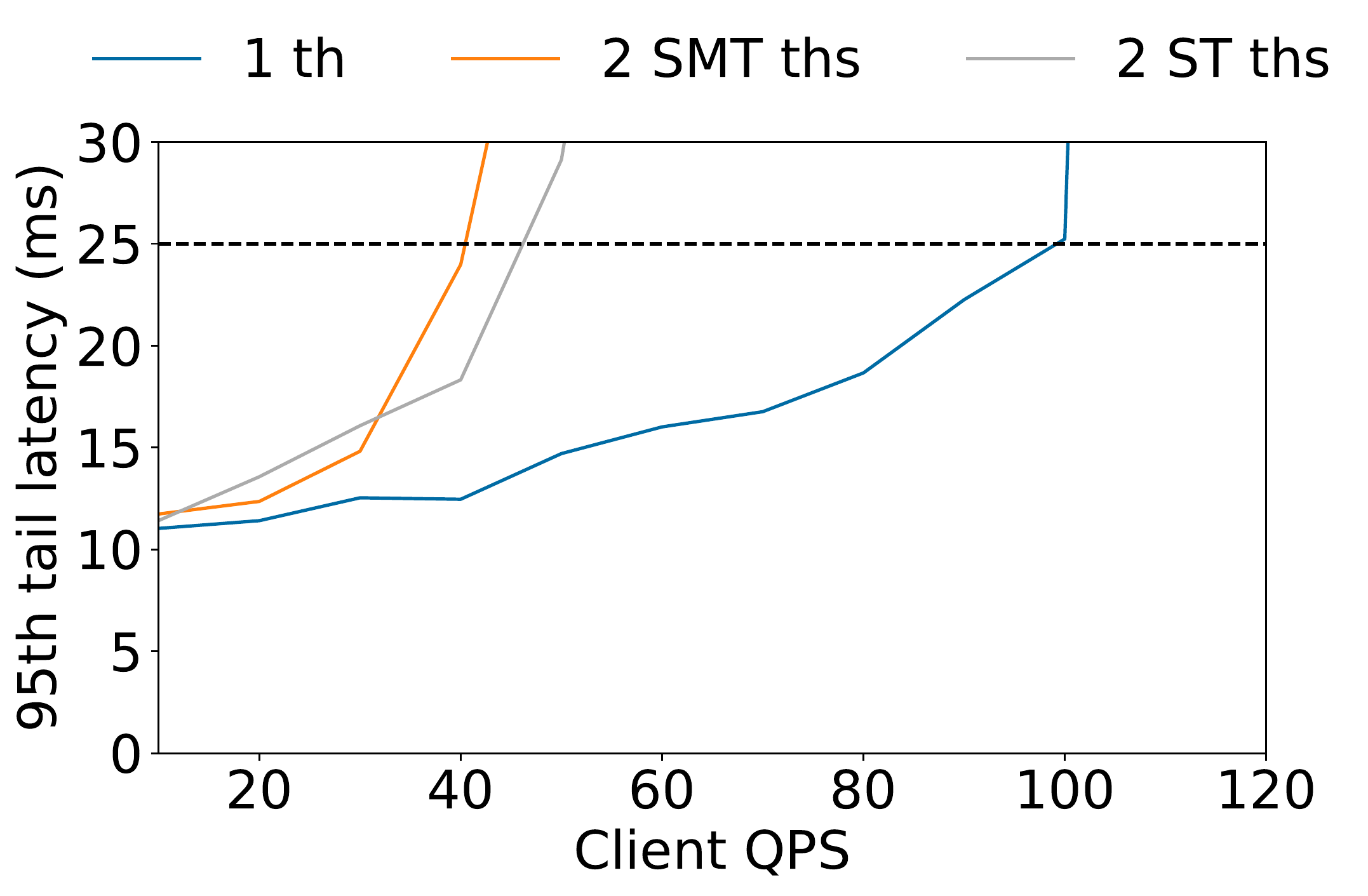}
       \label{fig:shore_lat}}
    \subfloat[CPU utilization]{%
       \includegraphics[width=0.32\textwidth]{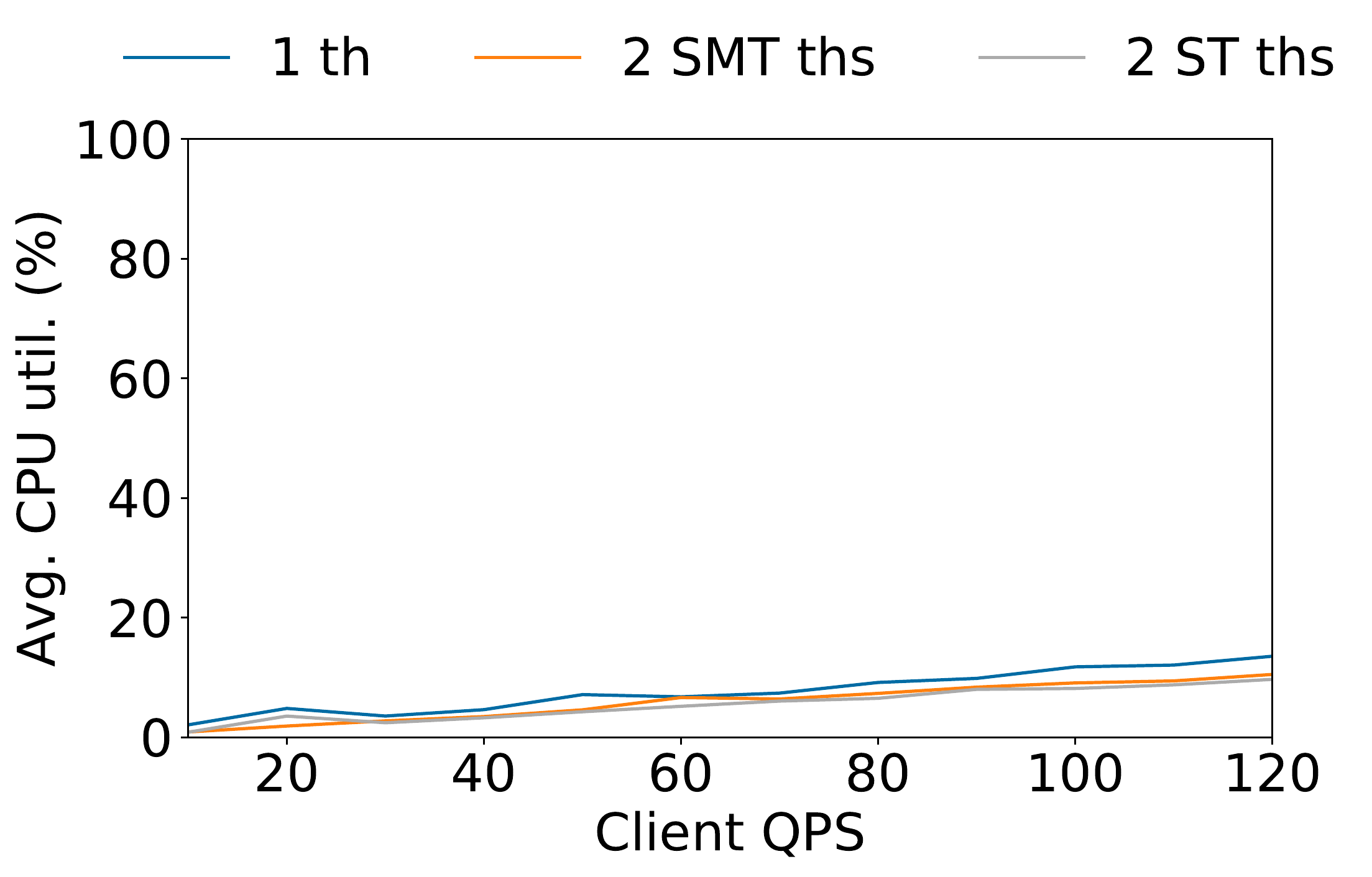}
       \label{fig:shore_util}}
    \subfloat[Network transmit bandwidth]{%
       \includegraphics[width=0.32\textwidth]{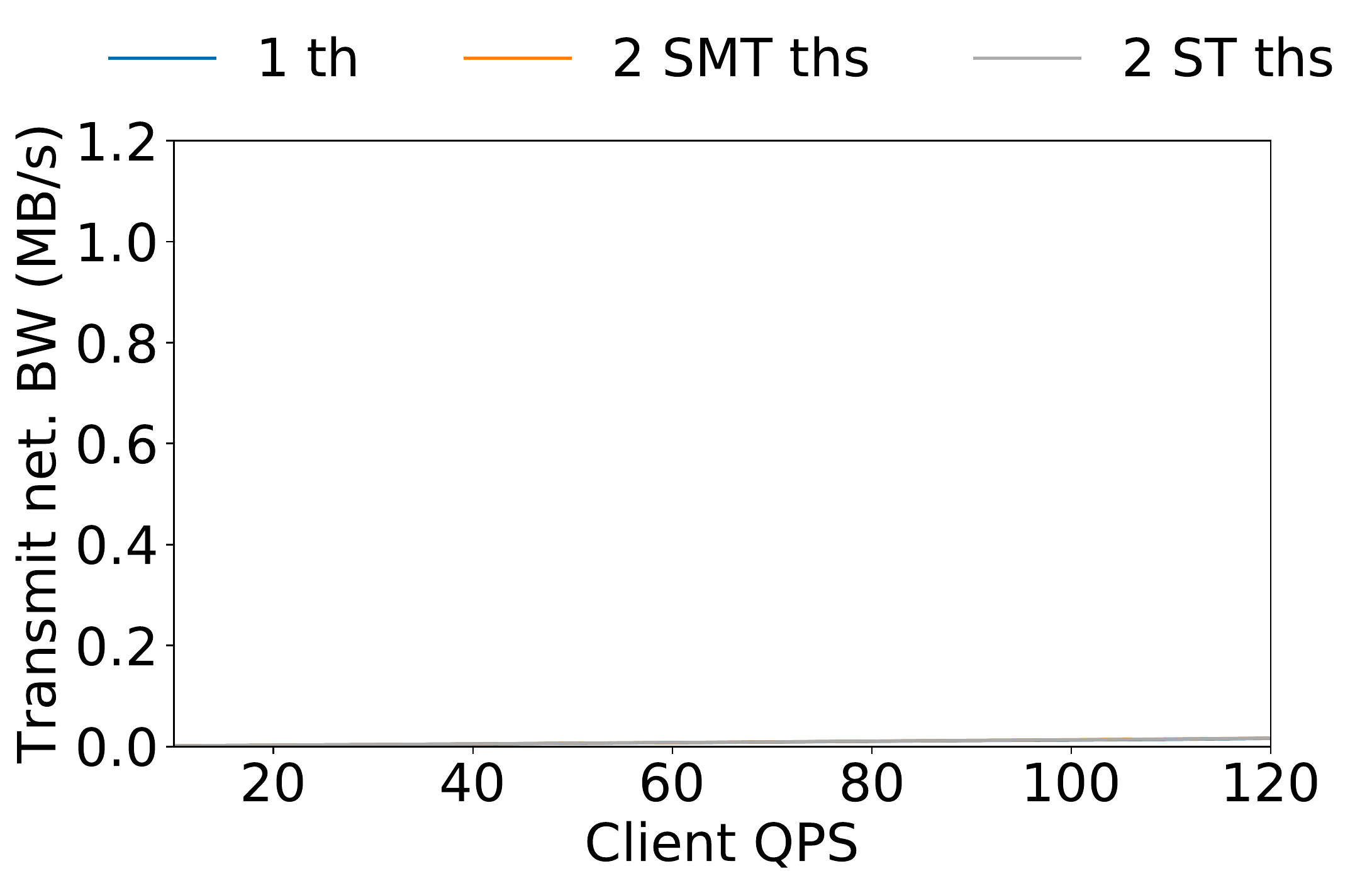}
       \label{fig:shore_tx}}
       \\
       \vspace{0.2cm}
    \subfloat[Disk bandwidth]{%
       \includegraphics[width=0.32\textwidth]{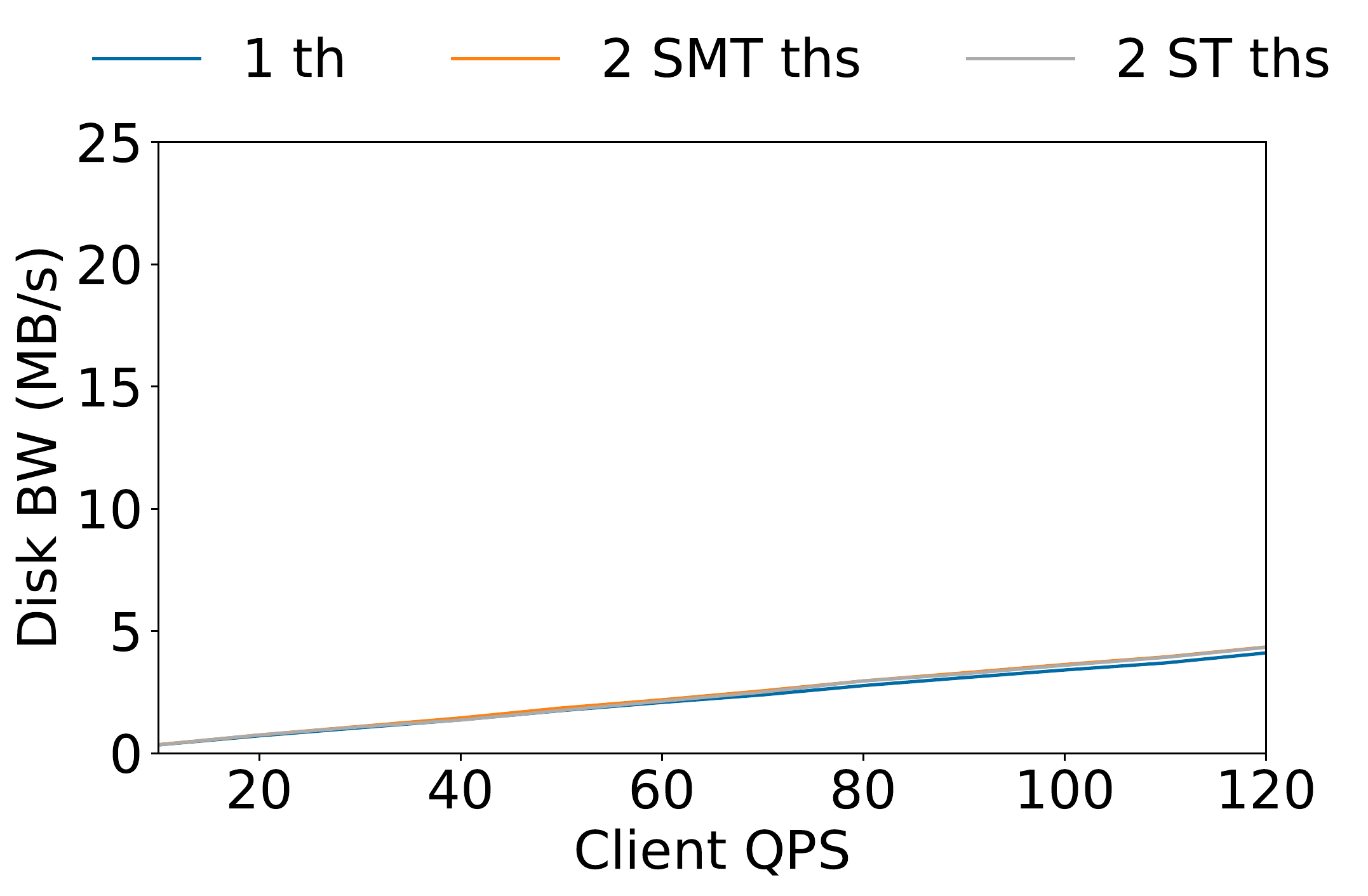}
       \label{fig:shore_disk}}
    \subfloat[Main memory bandwidth]{%
       \includegraphics[width=0.32\textwidth]{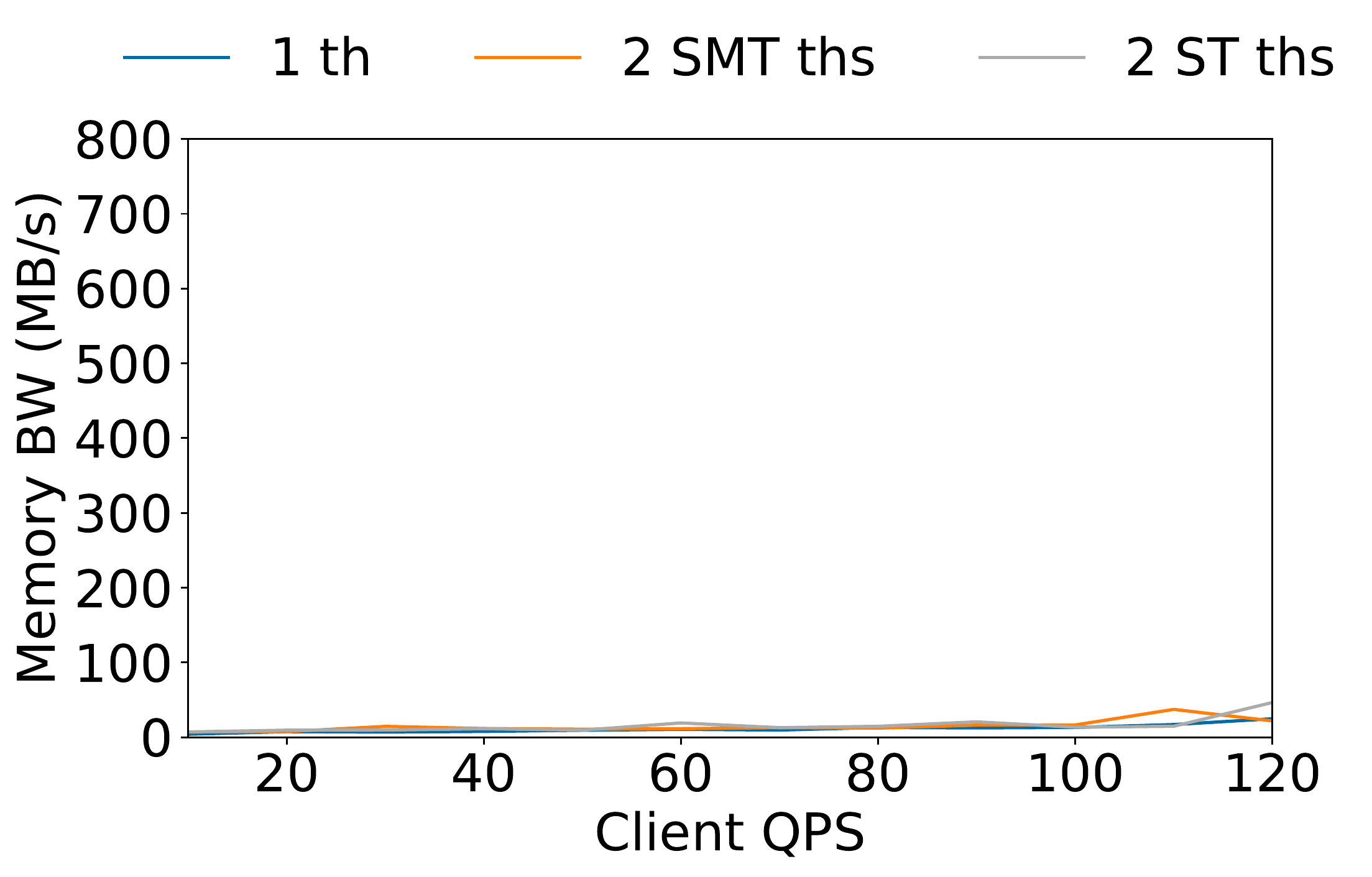}
       \label{fig:shore_mem}}
    \subfloat[LLC occupancy]{%
       \includegraphics[width=0.32\textwidth]{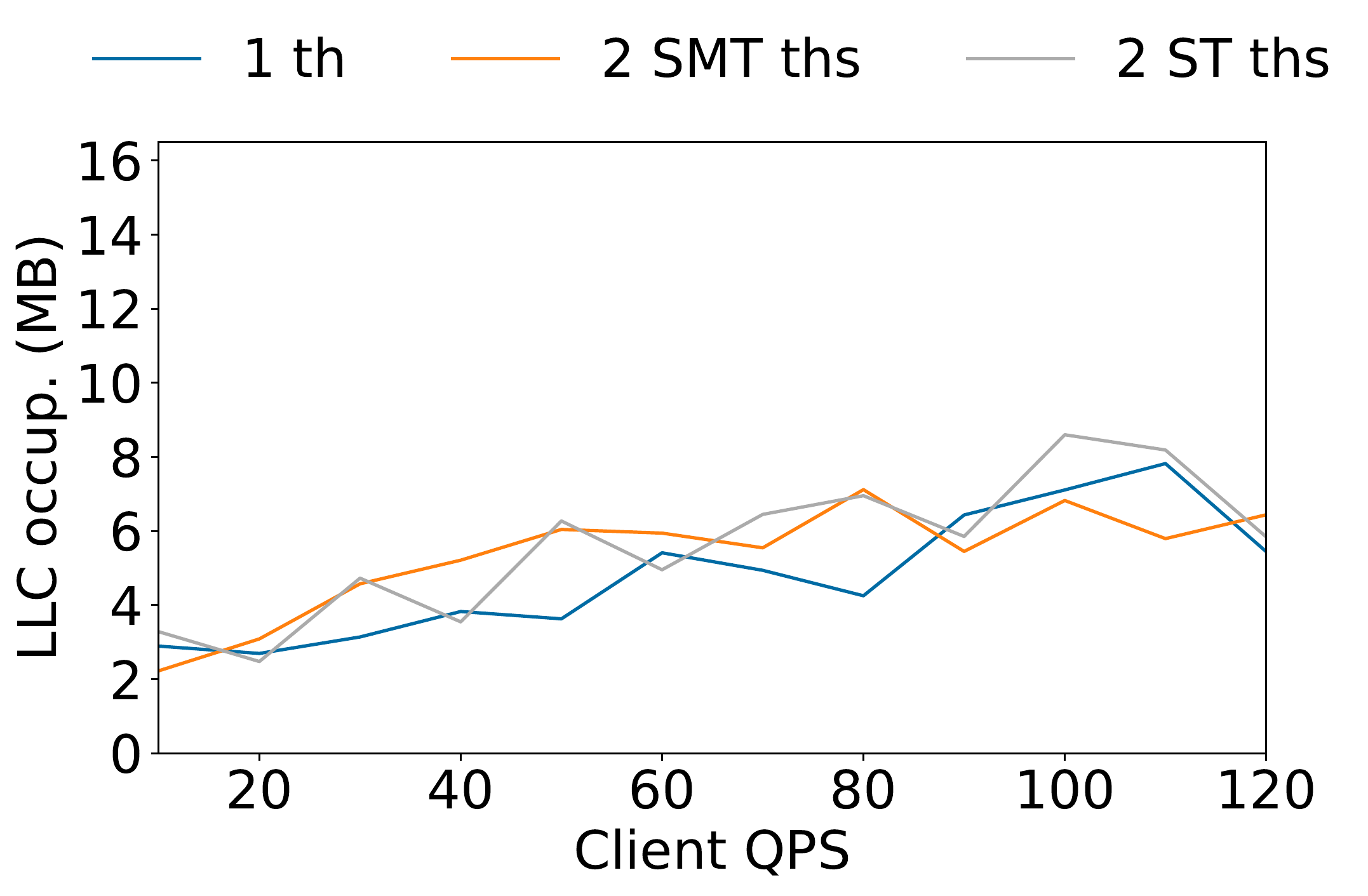}
       \label{fig:shore_llc}}
    \caption{Shore characterization.}
    \label{fig:shore_noconstrains} 
\end{figure*}

\begin{figure*}[t!]
    \centering
    \subfloat[$95^{th}$ tail latency (ms)]{%
       \includegraphics[width=0.32\textwidth]{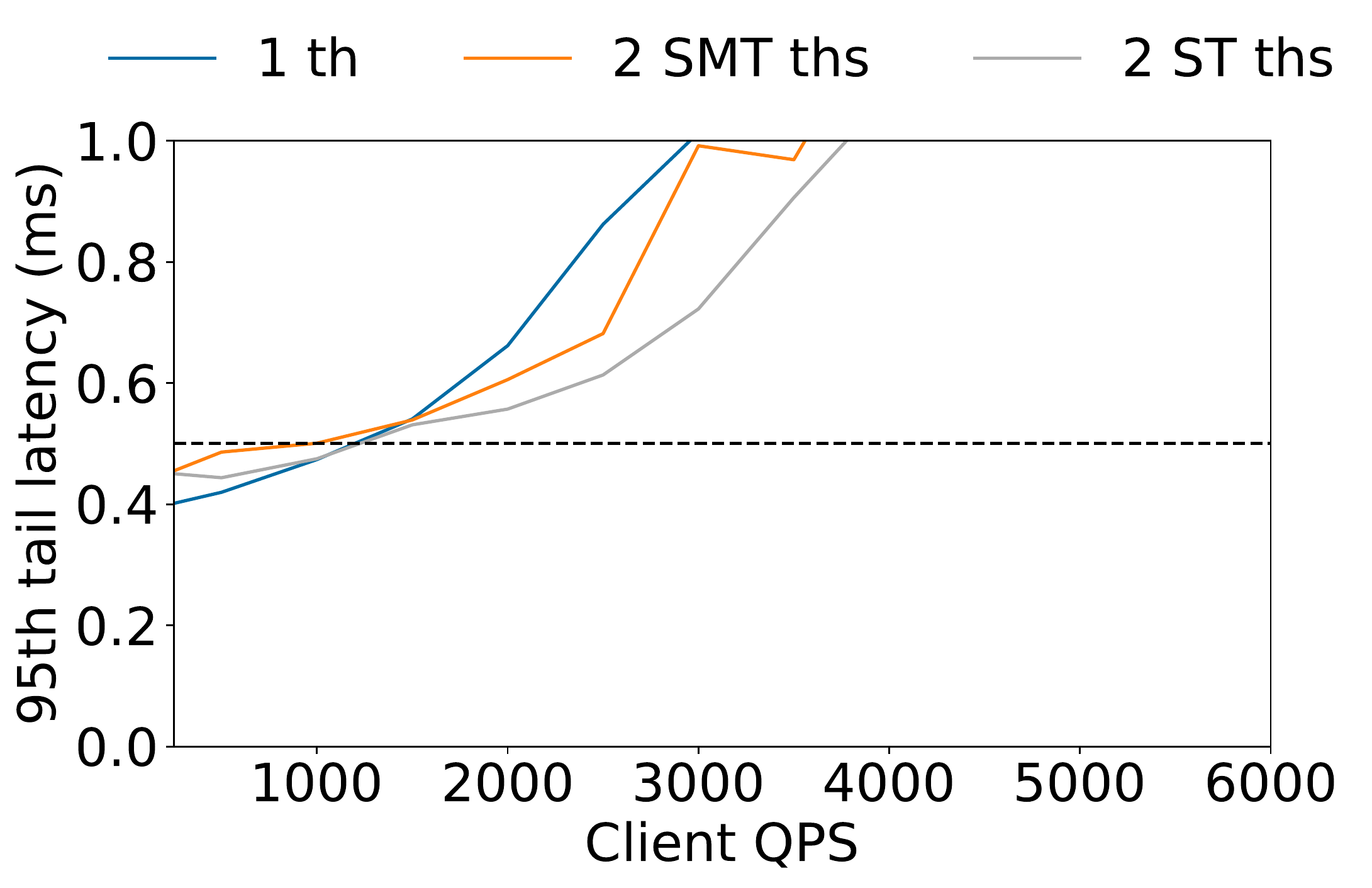}
       \label{fig:silo_lat}}
    \subfloat[CPU utilization]{%
       \includegraphics[width=0.32\textwidth]{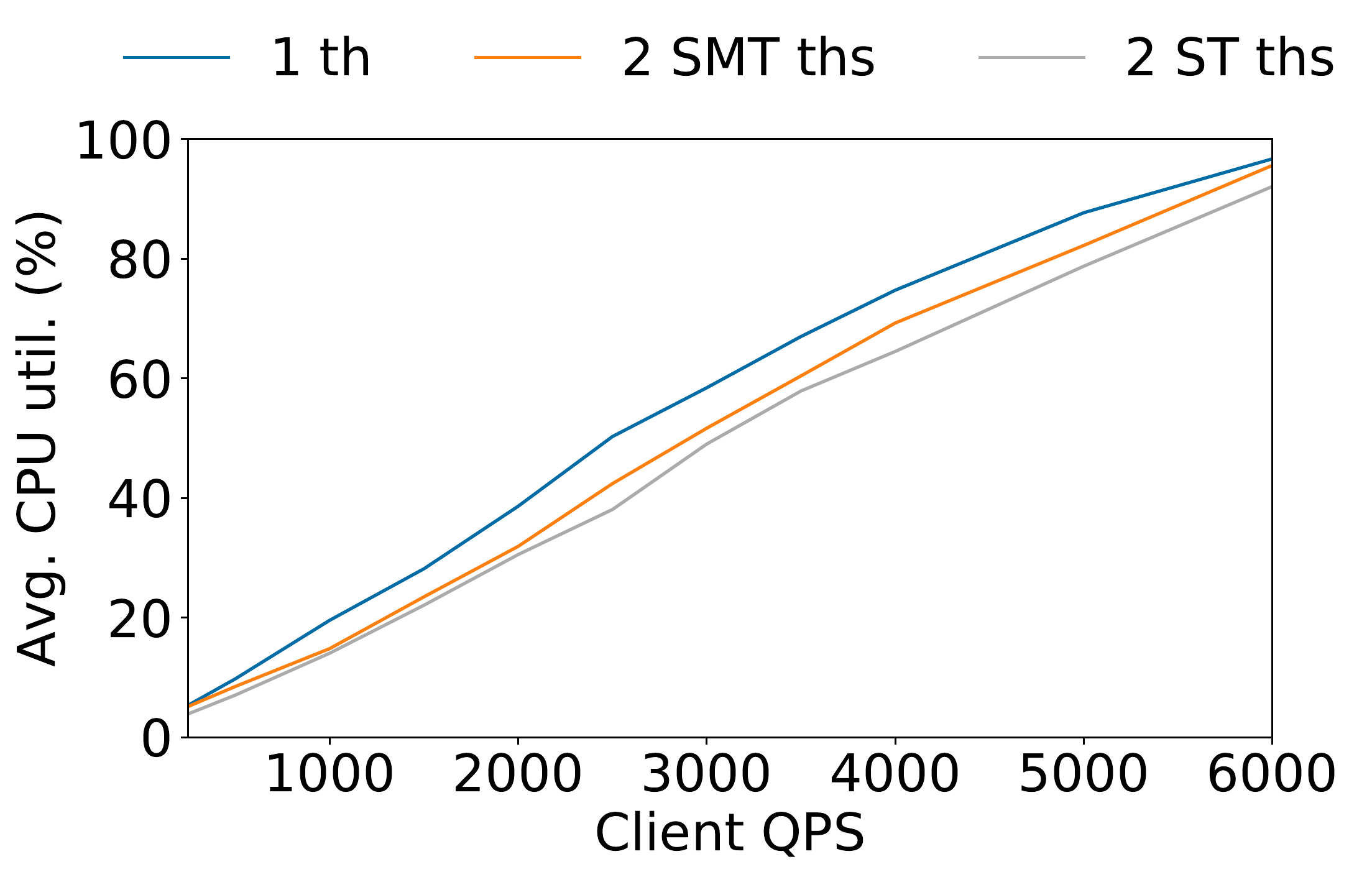}
       \label{fig:silo_util}}
    \subfloat[Network transmit bandwidth]{
       \includegraphics[width=0.32\textwidth]{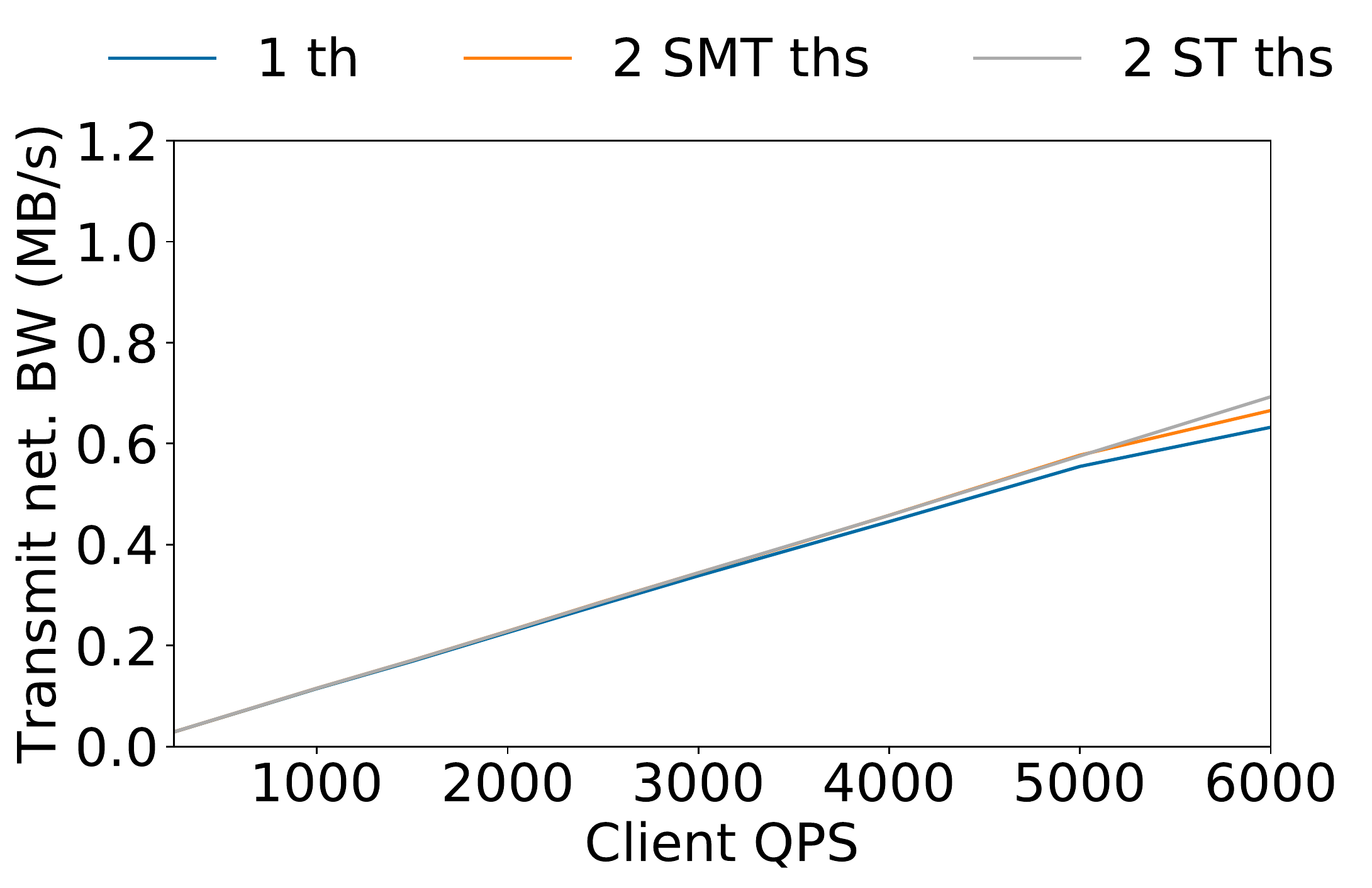}
       \label{fig:silo_tx}}
       \\
       \vspace{0.2cm}
    \subfloat[Disk bandwidth]{%
       \includegraphics[width=0.32\textwidth]{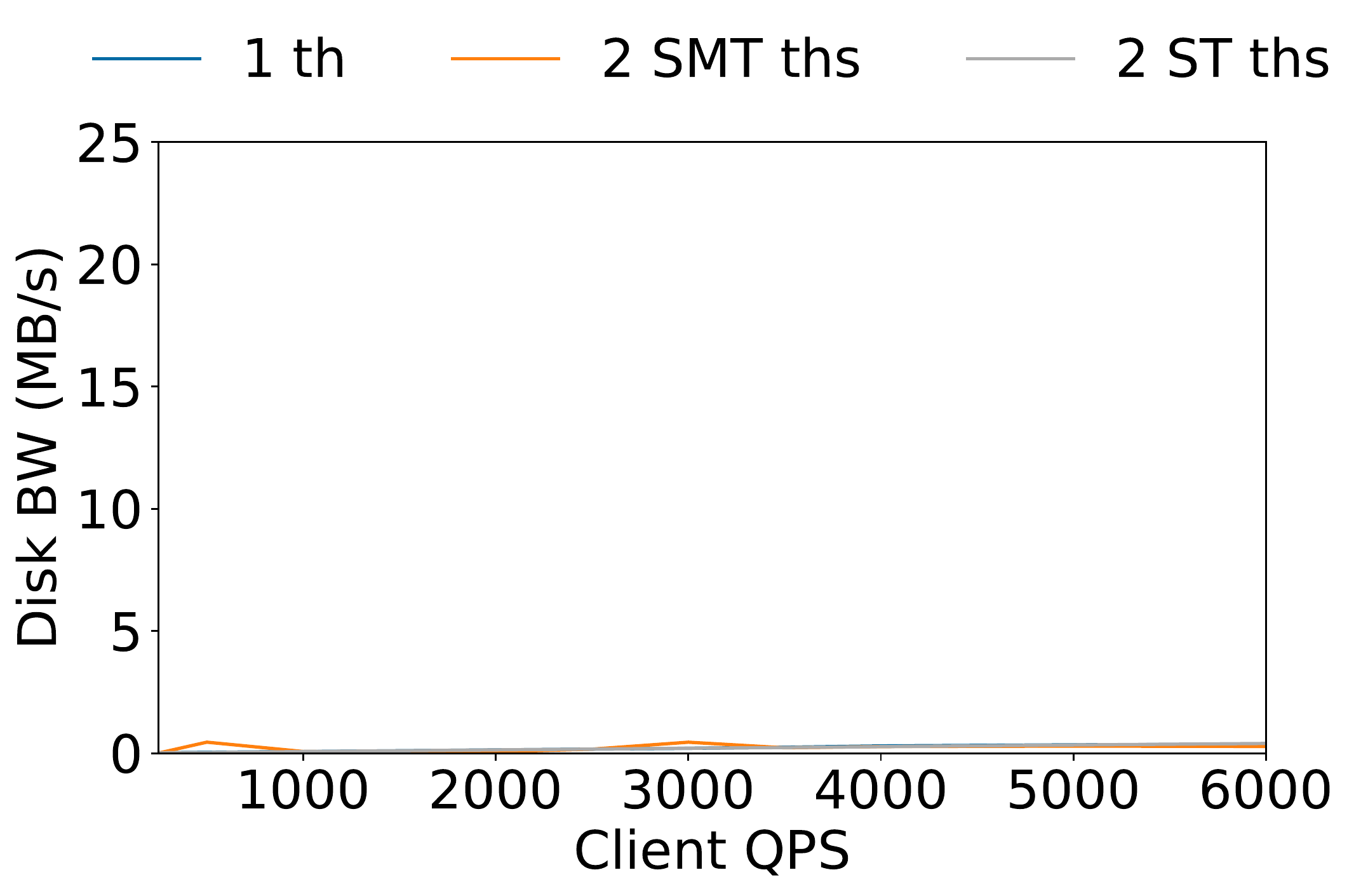}
       \label{fig:silo_disk}}
    \subfloat[Main memory bandwidth]{%
       \includegraphics[width=0.32\textwidth]{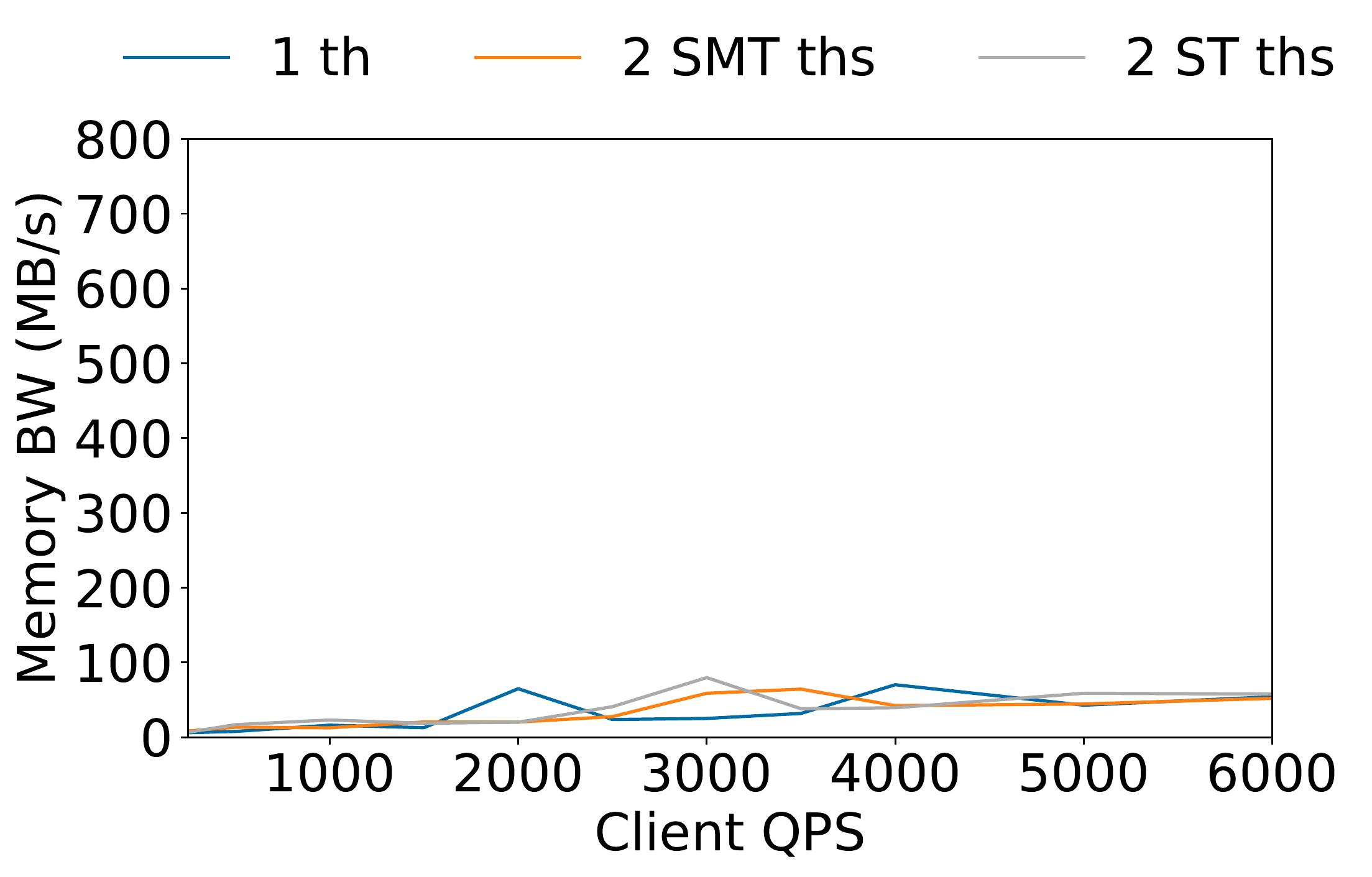}
       \label{fig:silo_mem}}
    \subfloat[LLC occupancy]{%
       \includegraphics[width=0.32\textwidth]{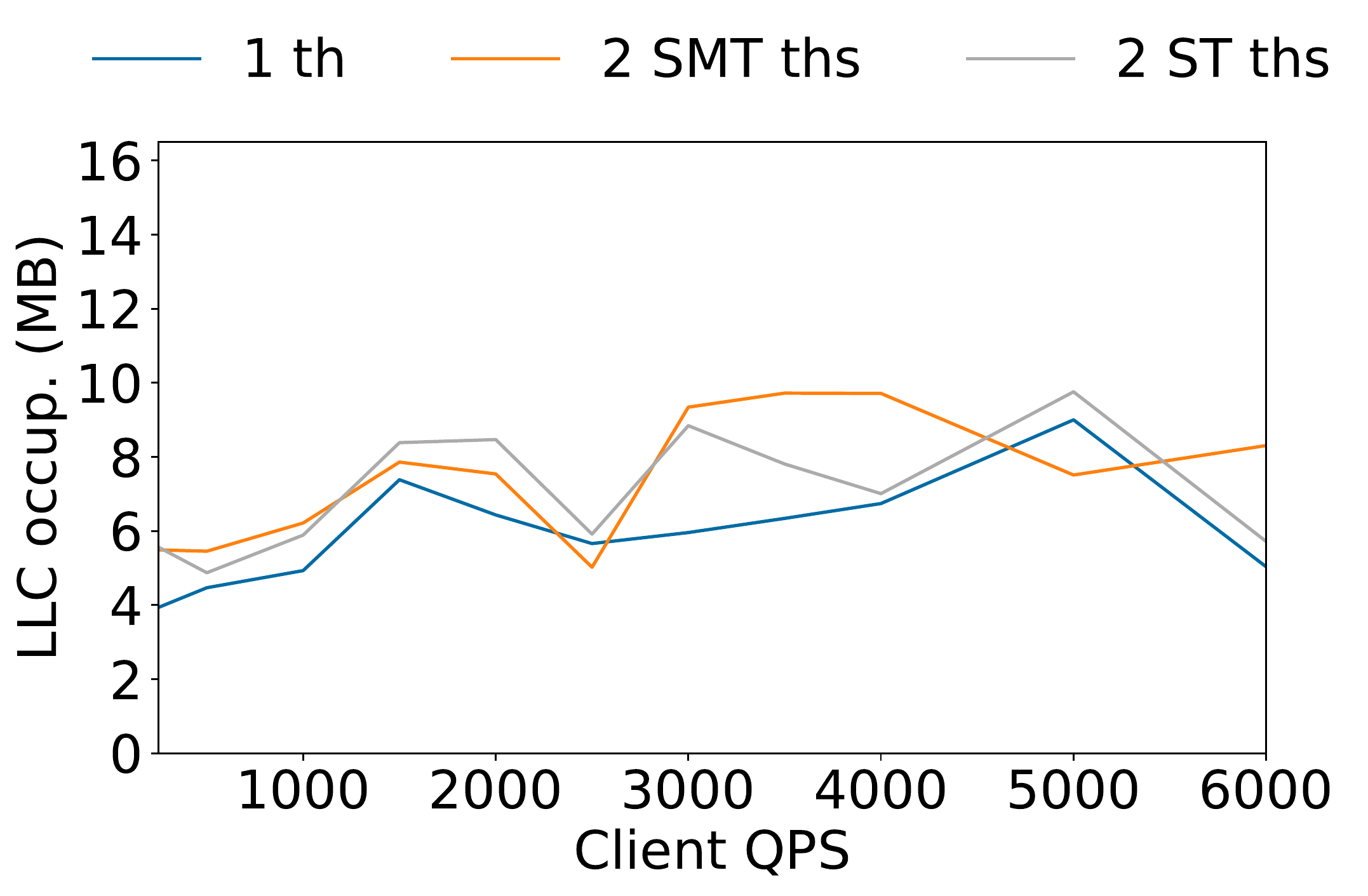}
       \label{fig:silo_llc}}
    \caption{Silo characterization.}
    \label{fig:silo_noconstrains} 
\end{figure*}

\begin{figure*}[t!]
    \centering
    \subfloat[$95^{th}$ tail latency (ms)]{%
       \includegraphics[width=0.32\textwidth]{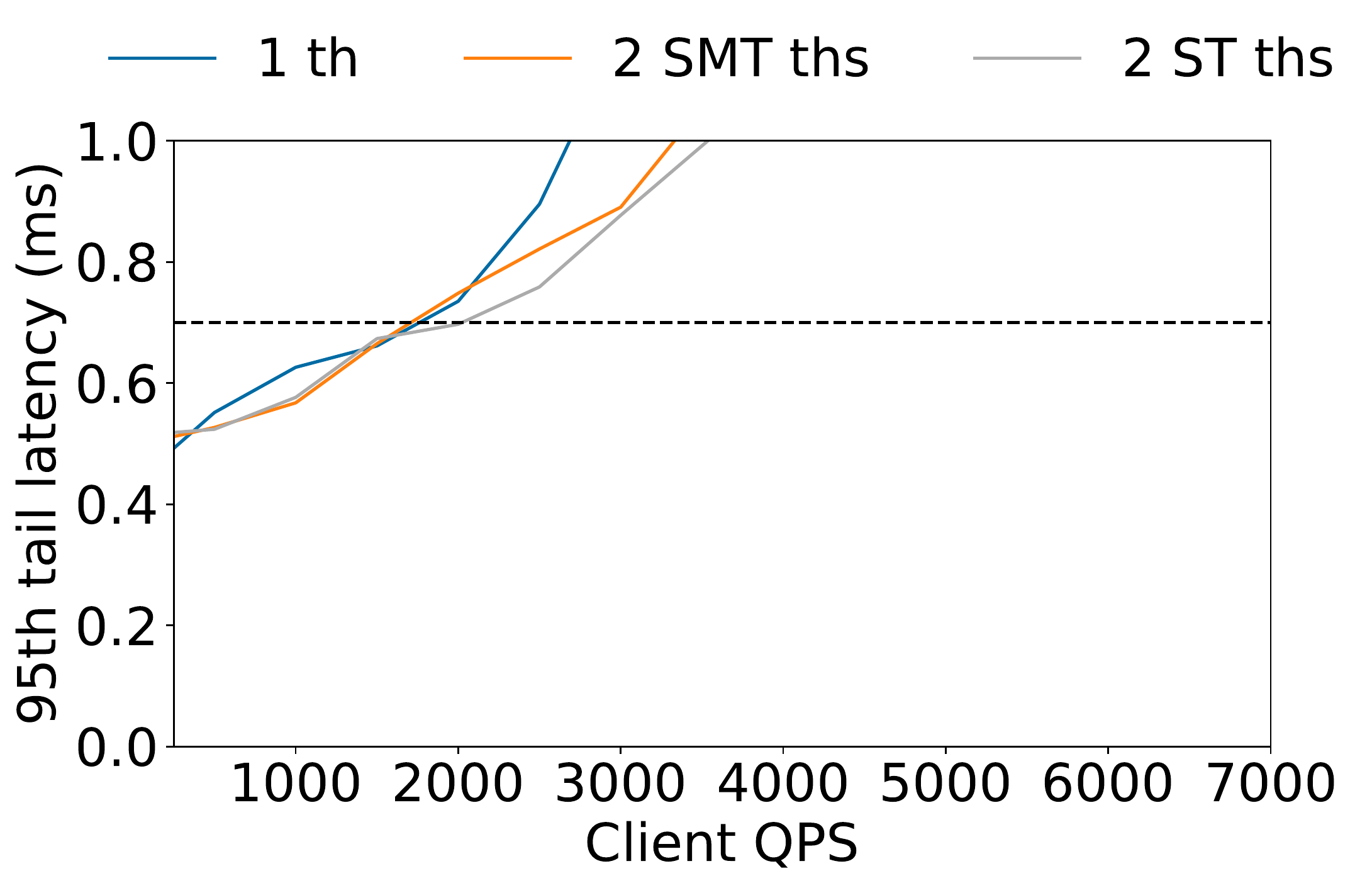}
       \label{fig:specjbb_lat}}
    \subfloat[CPU utilization]{%
       \includegraphics[width=0.32\textwidth]{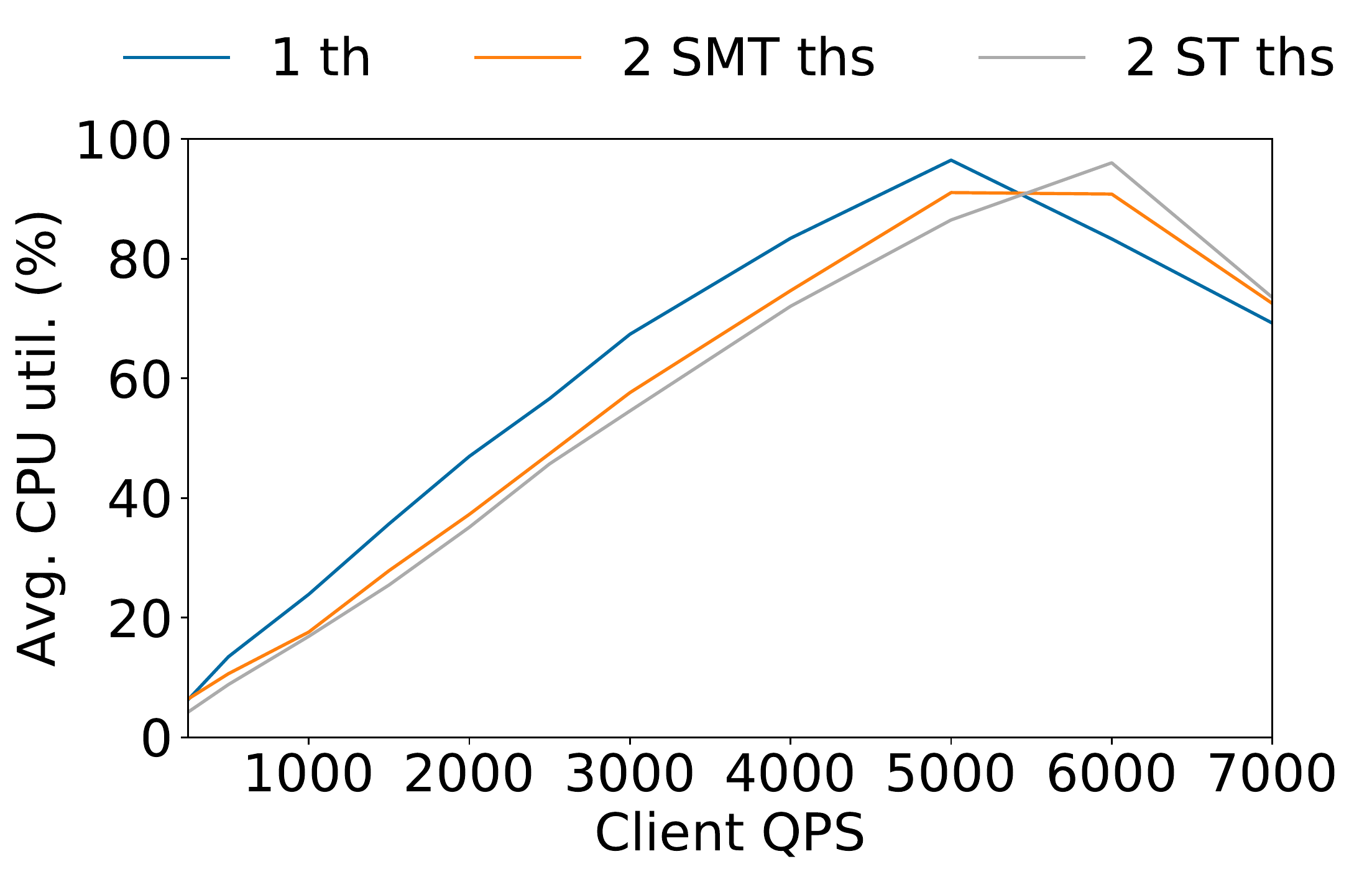}
       \label{fig:specjbb_util}}
    \subfloat[Network transmit bandwidth]{%
       \includegraphics[width=0.32\textwidth]{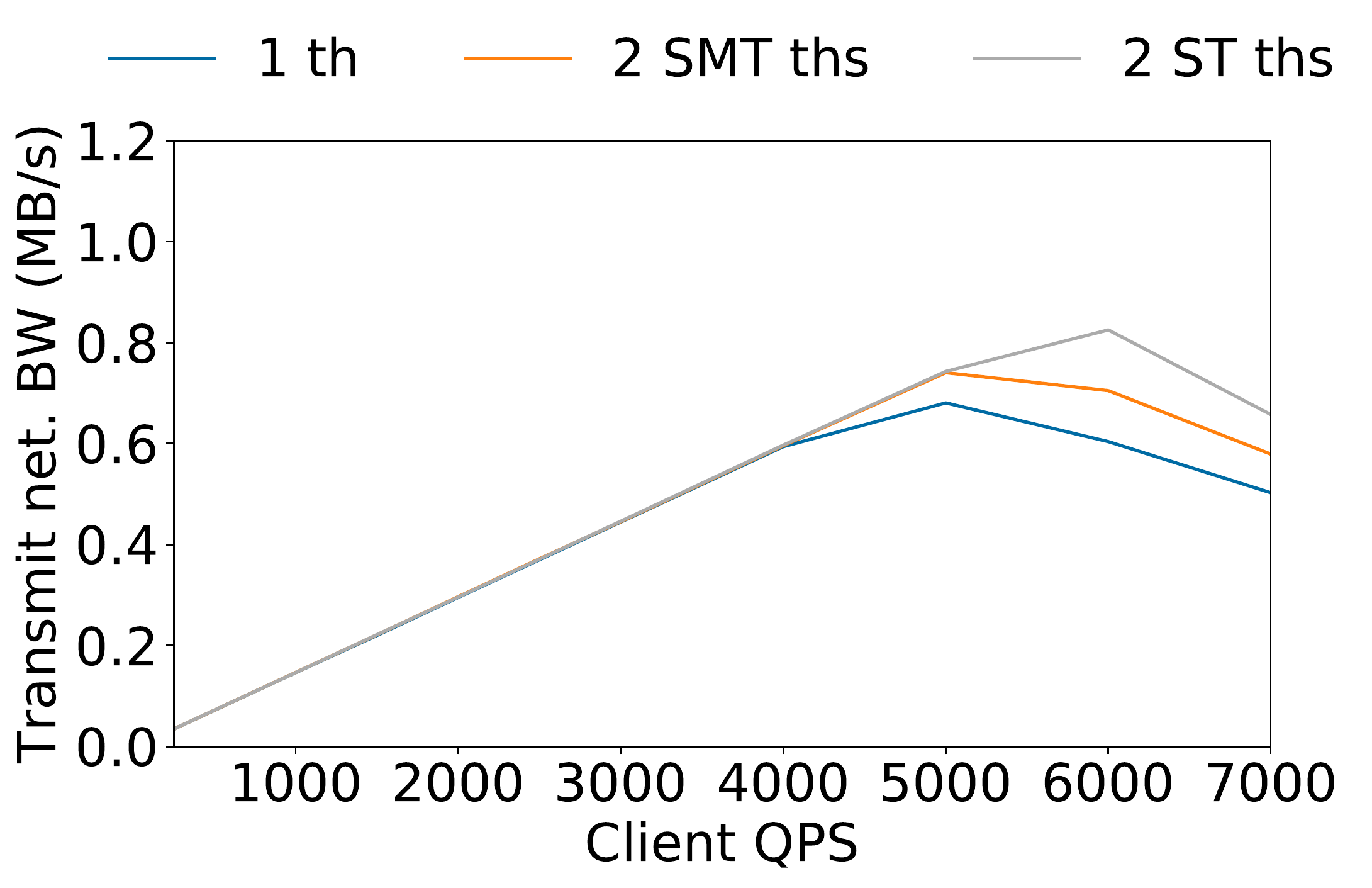}
       \label{fig:specjbb_tx}}
       \\
       \vspace{0.2cm}
    \subfloat[Disk bandwidth]{%
       \includegraphics[width=0.32\textwidth]{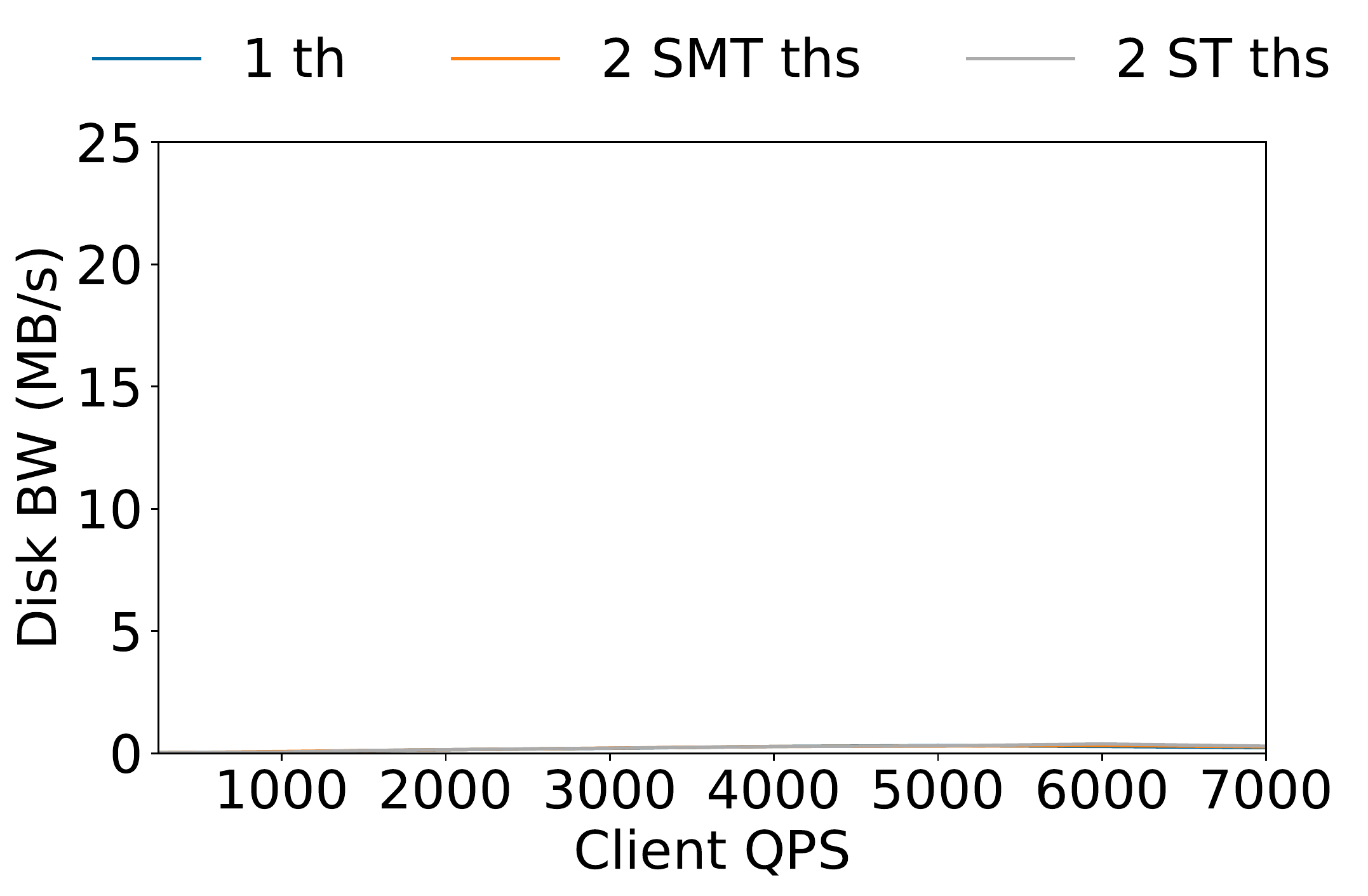}
       \label{fig:specjbb_disk}}
    \subfloat[Main memory bandwidth]{%
       \includegraphics[width=0.32\textwidth]{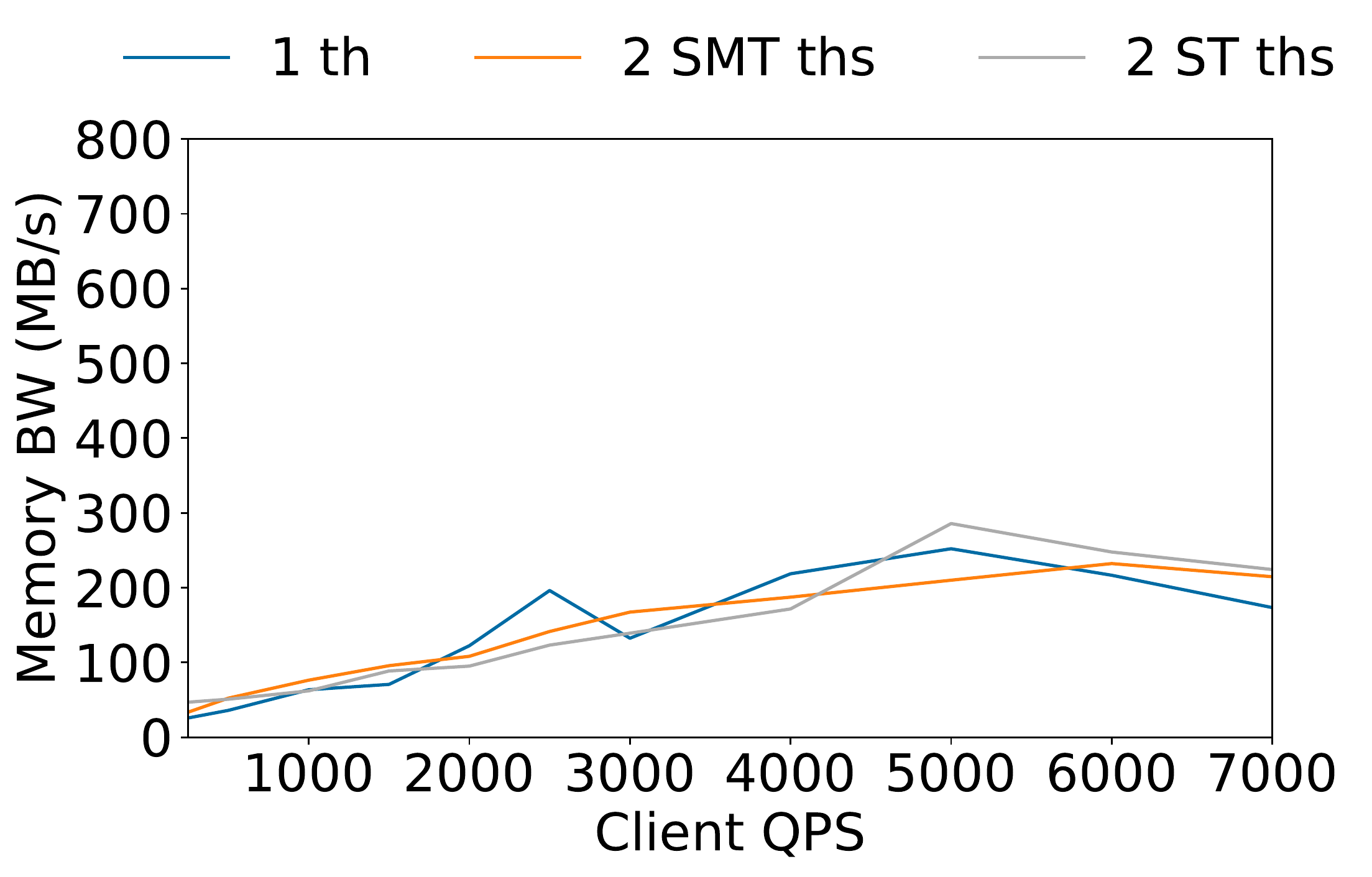}
       \label{fig:specjbb_mem}}
    \subfloat[LLC occupancy]{%
       \includegraphics[width=0.32\textwidth]{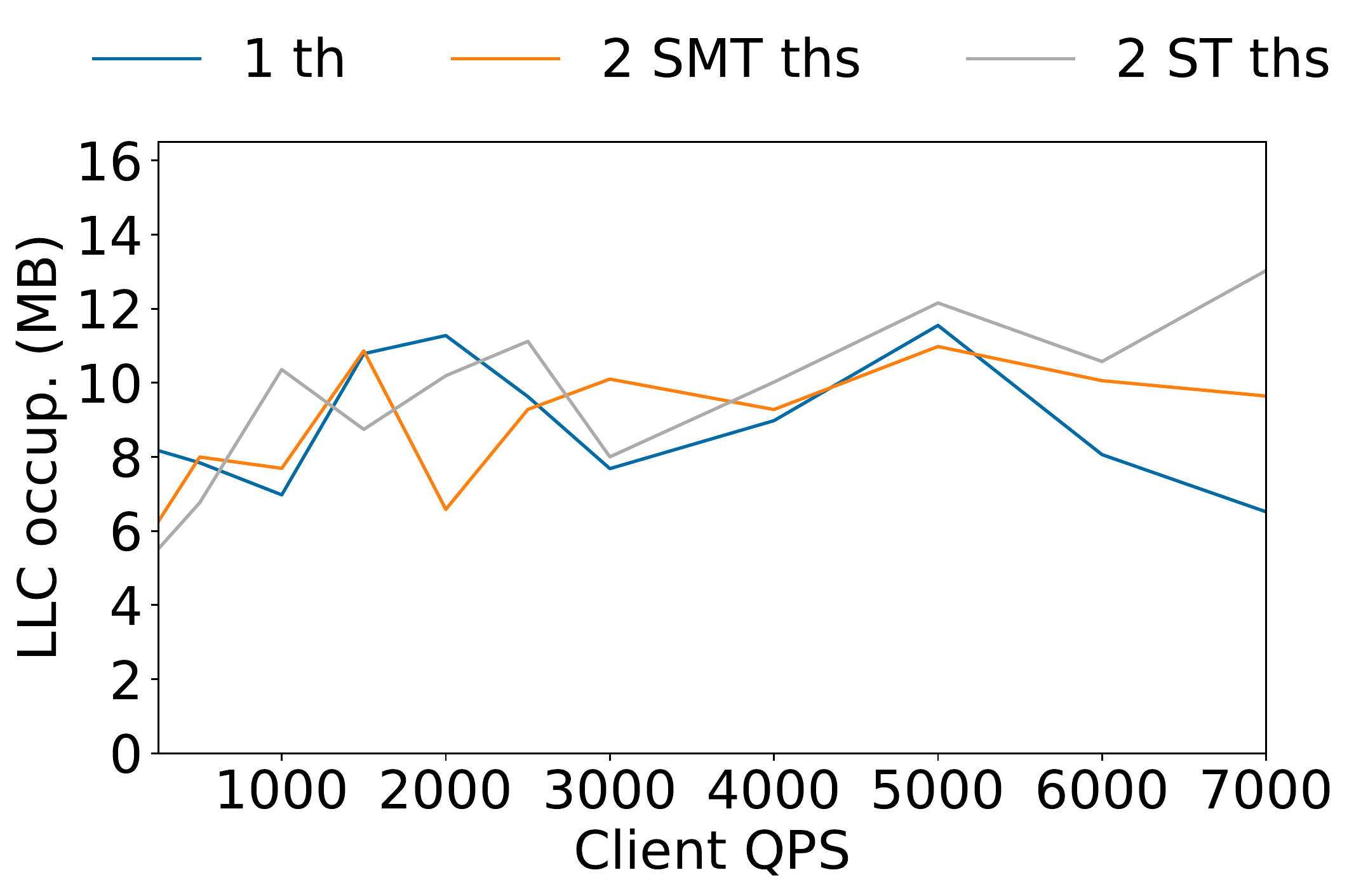}
       \label{fig:img-dnn_llc}}
    \caption{Specjbb characterization.}
    \label{fig:specjbb_noconstrains} 
\end{figure*}

\begin{figure*}[t!]
    \centering
    \subfloat[$95^{th}$ tail latency (ms)]{%
       \includegraphics[width=0.32\textwidth]{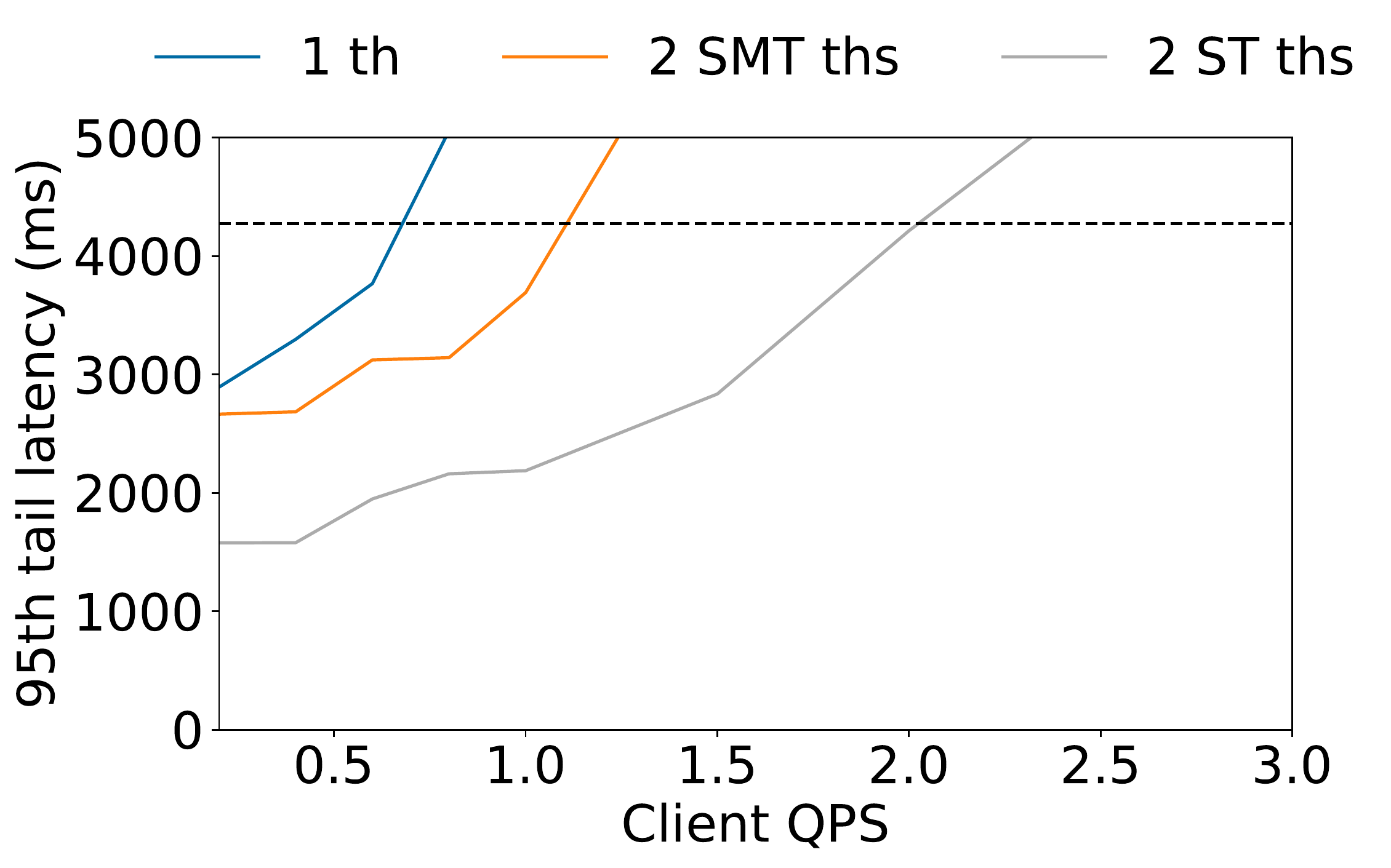}
       \label{fig:sphinx_lat}}
    \subfloat[CPU utilization]{%
       \includegraphics[width=0.32\textwidth]{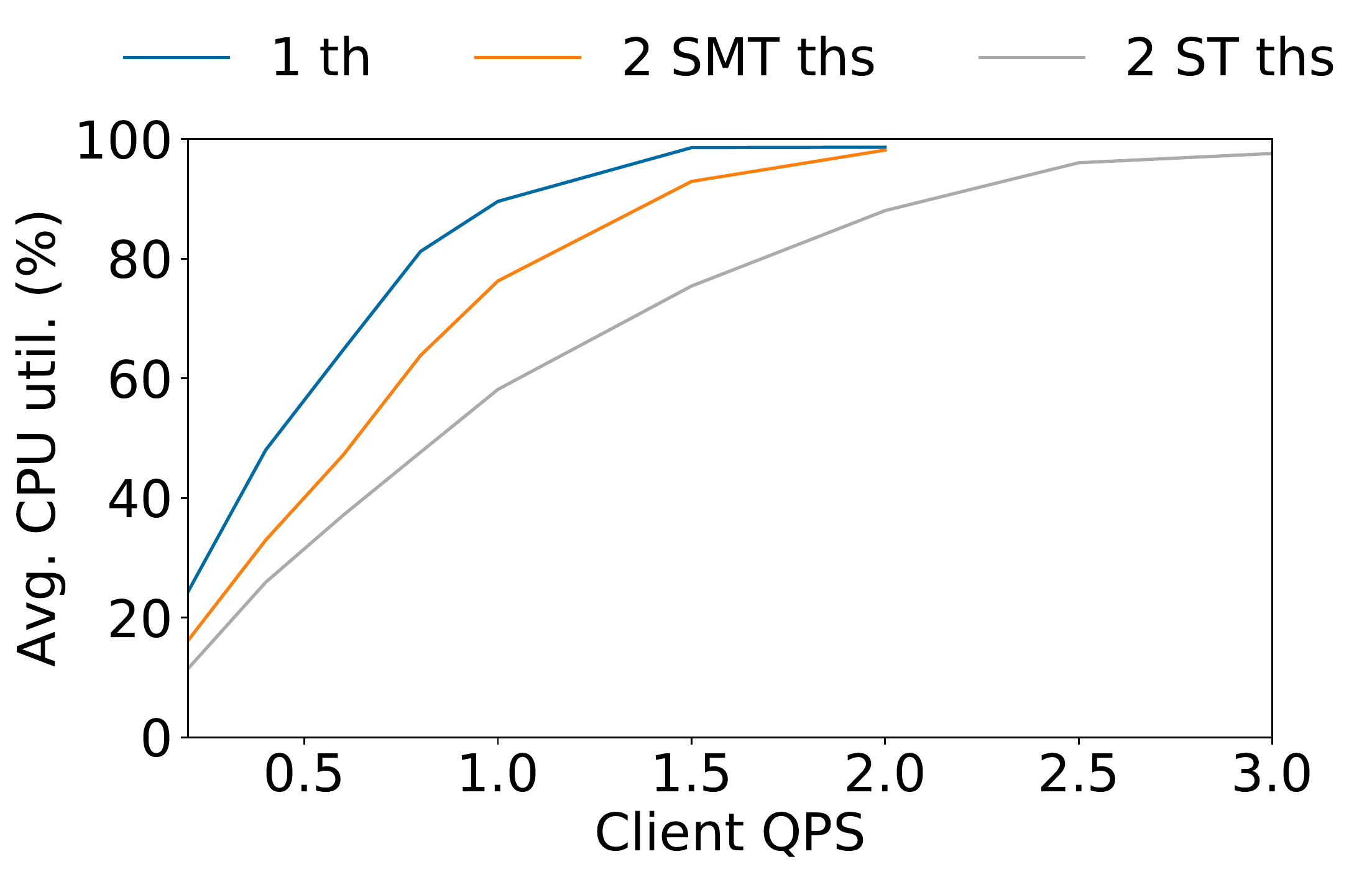}
       \label{fig:sphinx_util}}
    \subfloat[Network transmit bandwidth]{%
       \includegraphics[width=0.32\textwidth]{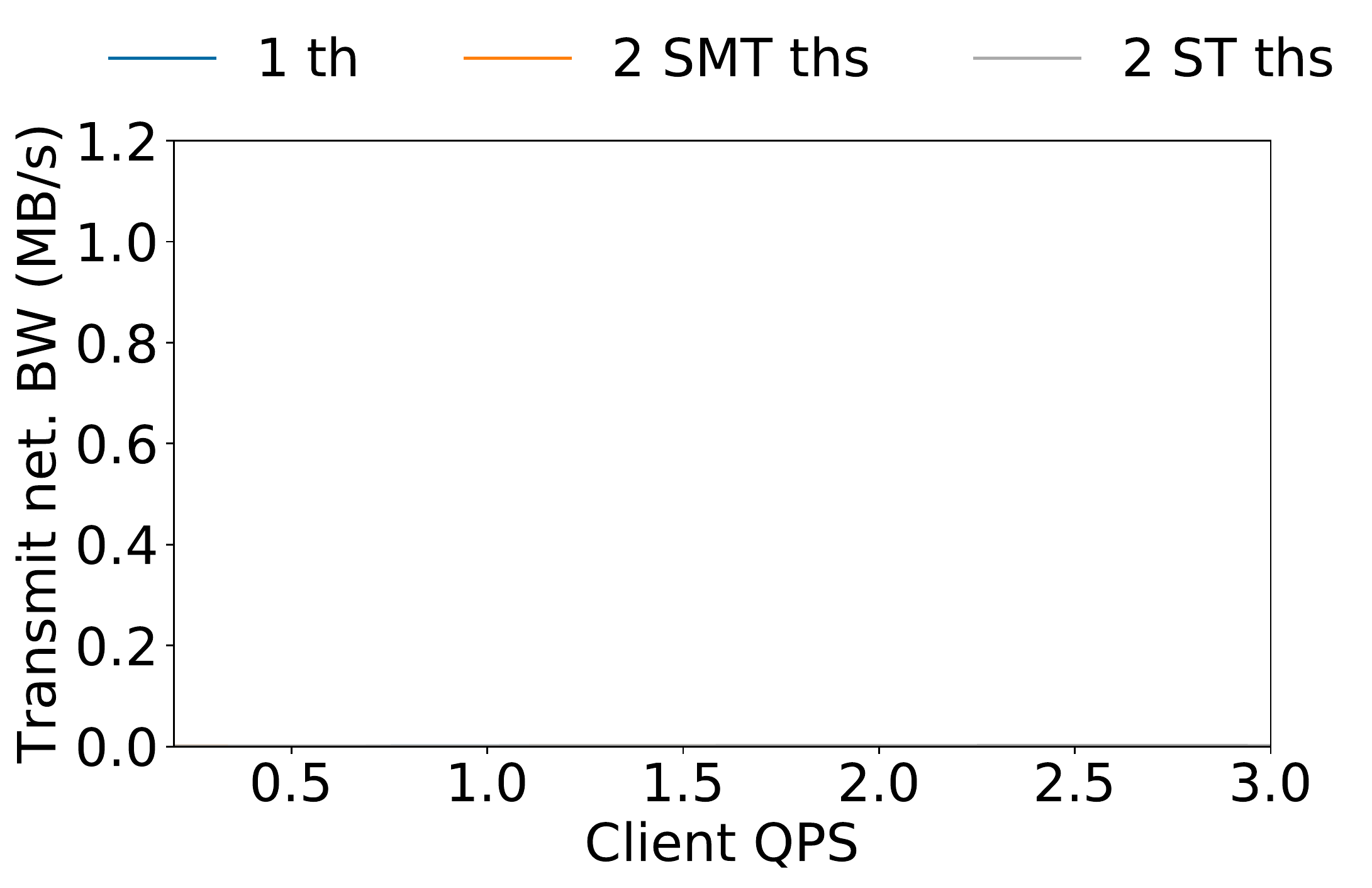}
       \label{fig:sphinx_tx}}
       \\
       \vspace{0.2cm}
    \subfloat[Disk bandwidth]{%
       \includegraphics[width=0.32\textwidth]{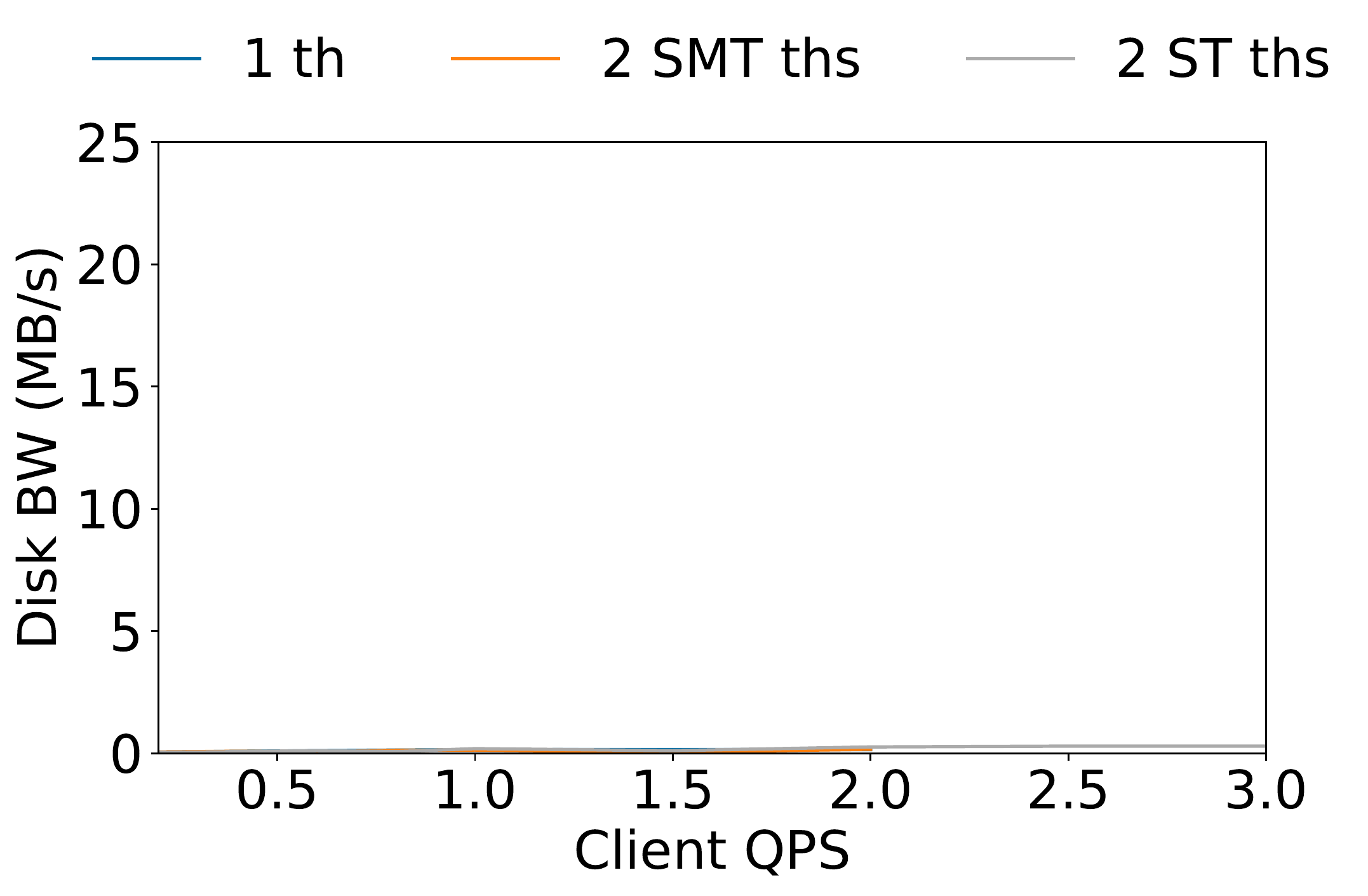}
       \label{fig:sphinx_disk}}
    \subfloat[Main memory bandwidth]{%
       \includegraphics[width=0.32\textwidth]{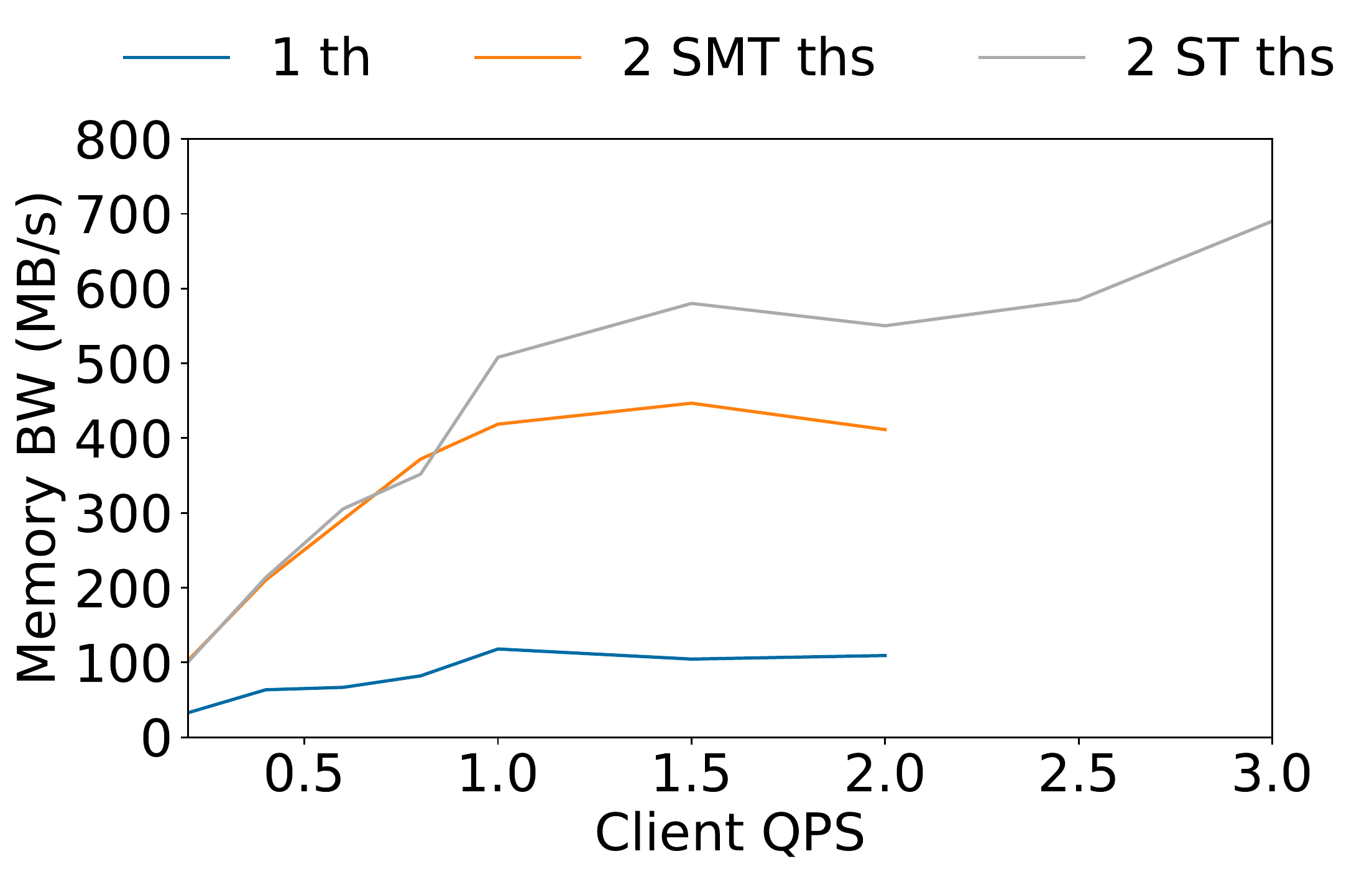}
       \label{fig:sphinx_mem}}
    \subfloat[LLC occupancy]{%
       \includegraphics[width=0.32\textwidth]{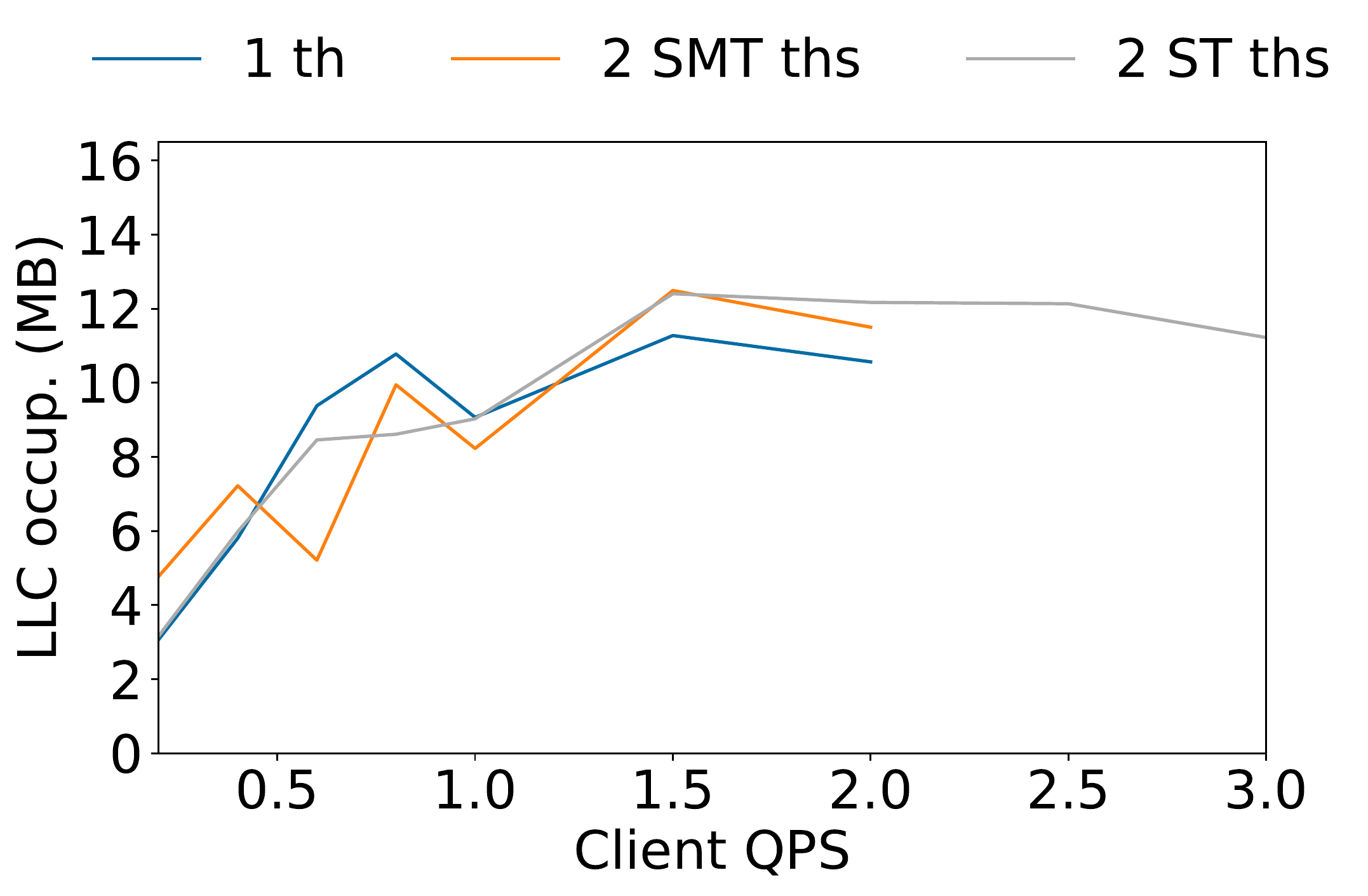}
       \label{fig:sphinx_llc}}
    \caption{Sphinx characterization.}
    \label{fig:sphinx_noconstrains} 
\end{figure*}

\begin{figure*}[t!]
    \centering
    \subfloat[$95^{th}$ tail latency (ms)]{%
       \includegraphics[width=0.32\textwidth]{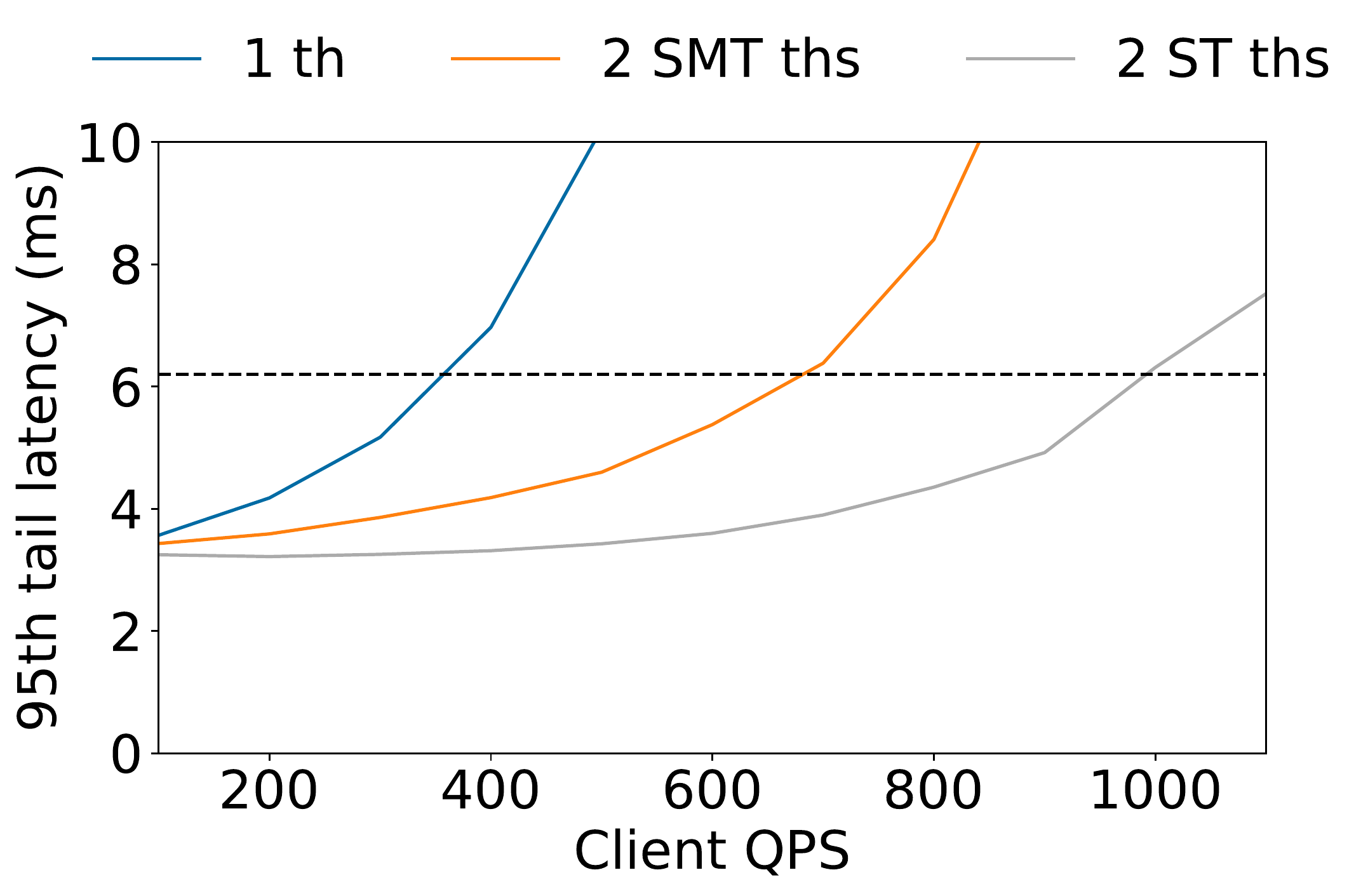}
       \label{fig:xapian_lat}}
    \subfloat[CPU utilization]{%
       \includegraphics[width=0.32\textwidth]{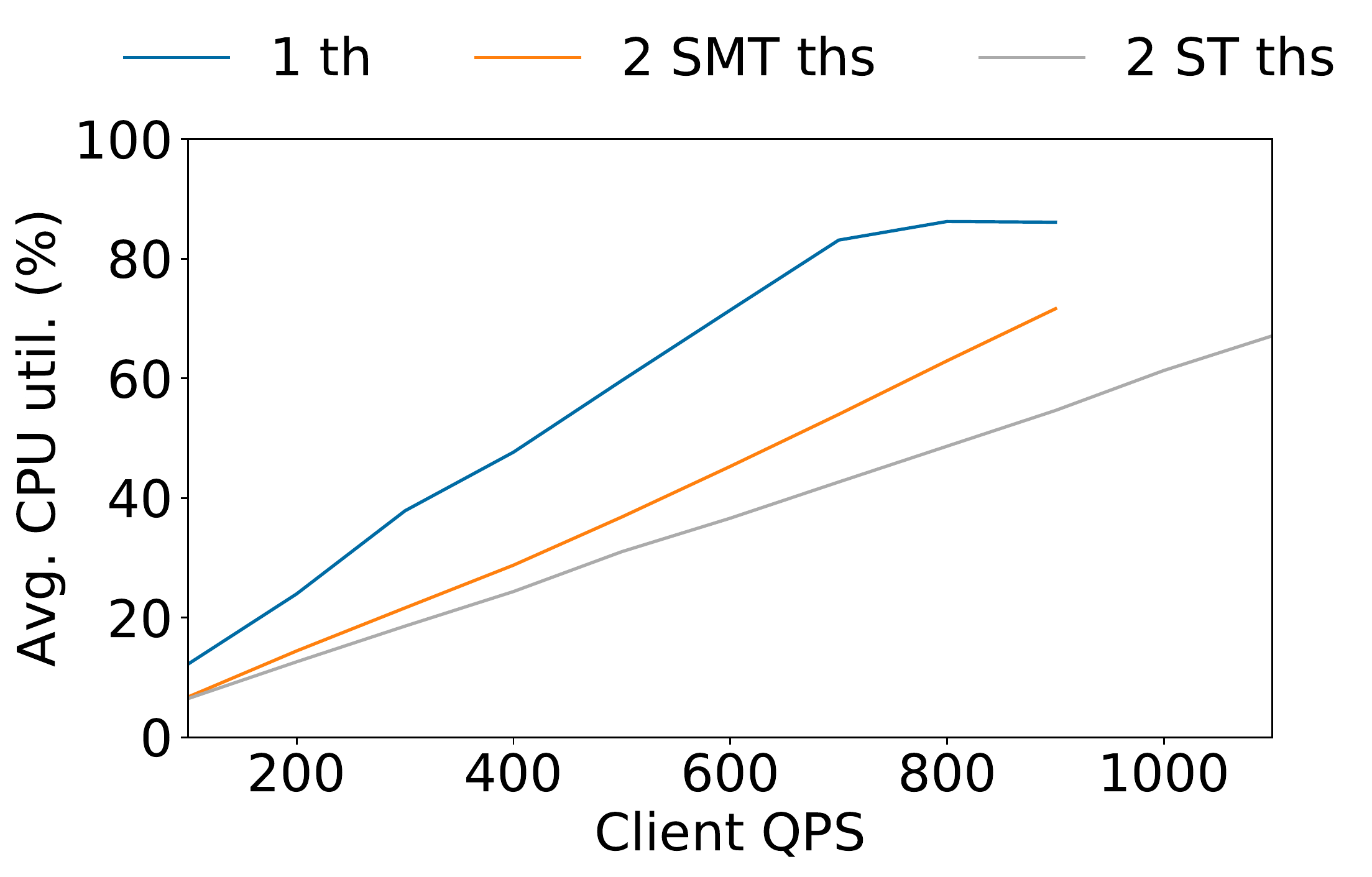}
       \label{fig:xapian_util}}
    \subfloat[Network transmit bandwidth]{%
       \includegraphics[width=0.32\textwidth]{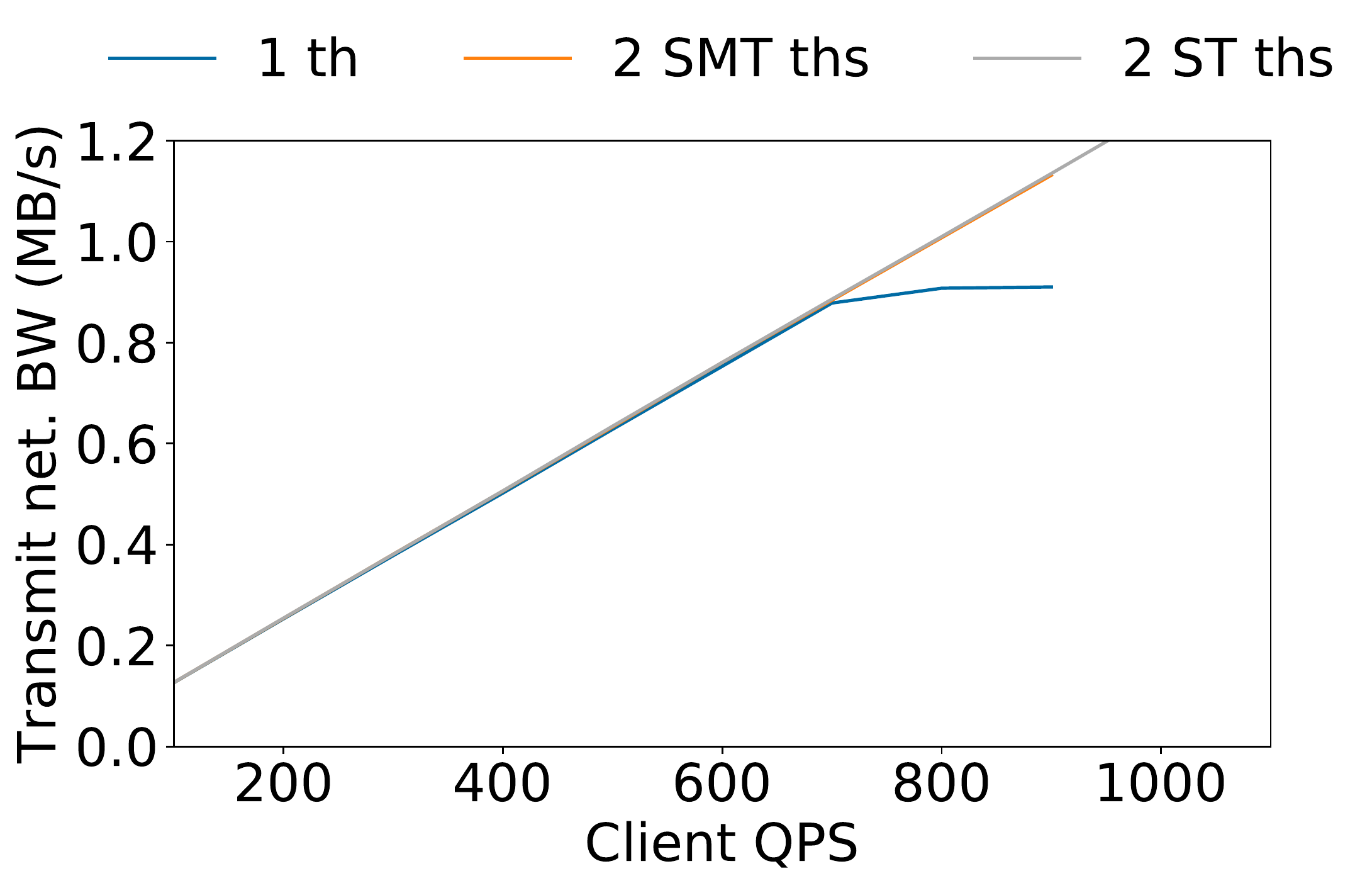}
       \label{fig:xapian_tx}}
       \\
       \vspace{0.2cm}
    \subfloat[Disk bandwidth]{%
       \includegraphics[width=0.32\textwidth]{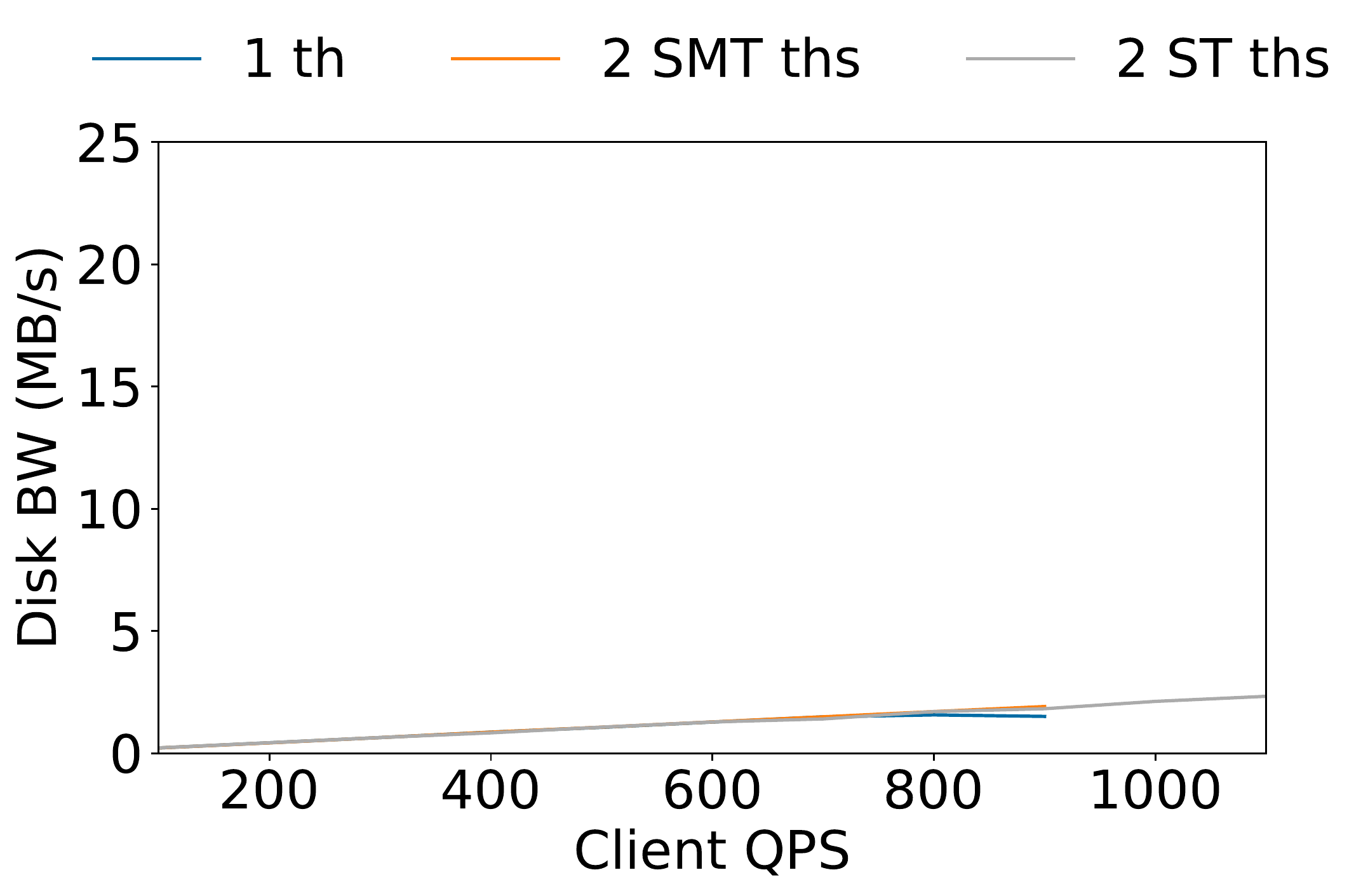}
       \label{fig:xapian_disk}}
    \subfloat[Main memory bandwidth]{%
       \includegraphics[width=0.32\textwidth]{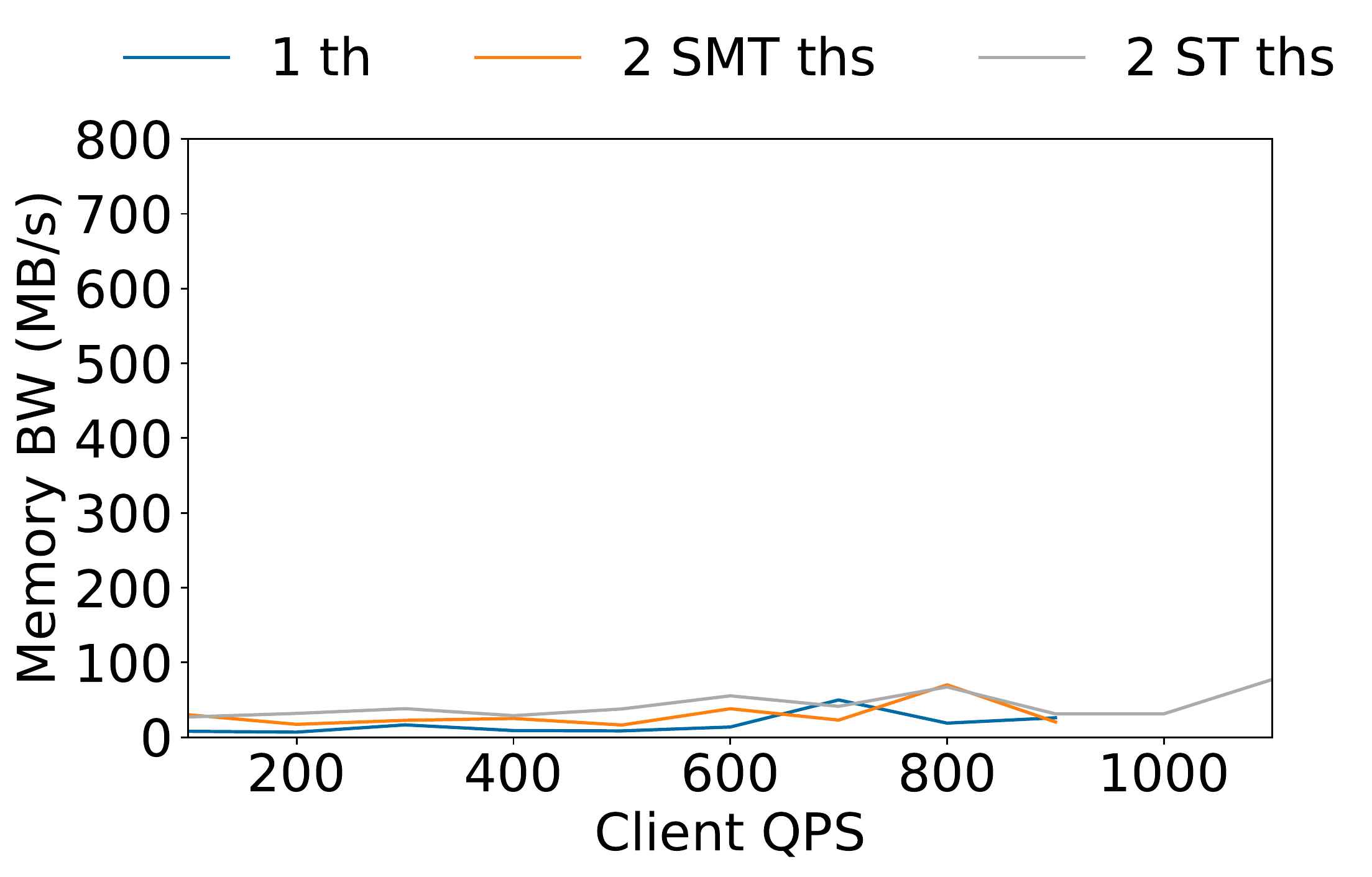}
       \label{fig:xapian_mem}}
    \subfloat[LLC occupancy]{%
       \includegraphics[width=0.32\textwidth]{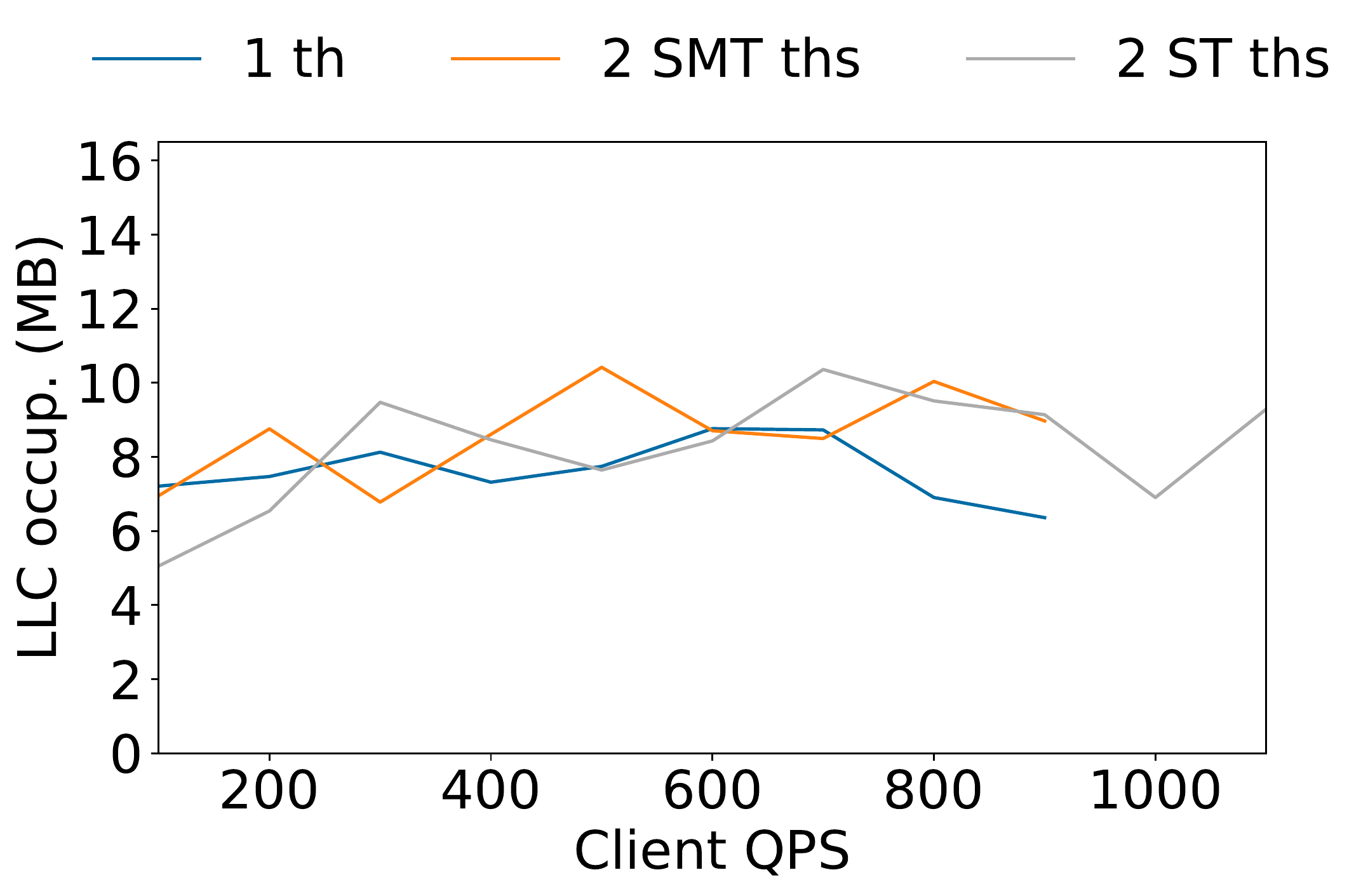}
       \label{fig:xapian_llc}}
    \caption{Xapian characterization.}
    \label{fig:xapian_noconstrains} 
\end{figure*}

\begin{figure*}[t!]
    \centering
    \subfloat[Transfer plus Response Time (ms)]{%
       \includegraphics[width=0.32\textwidth]{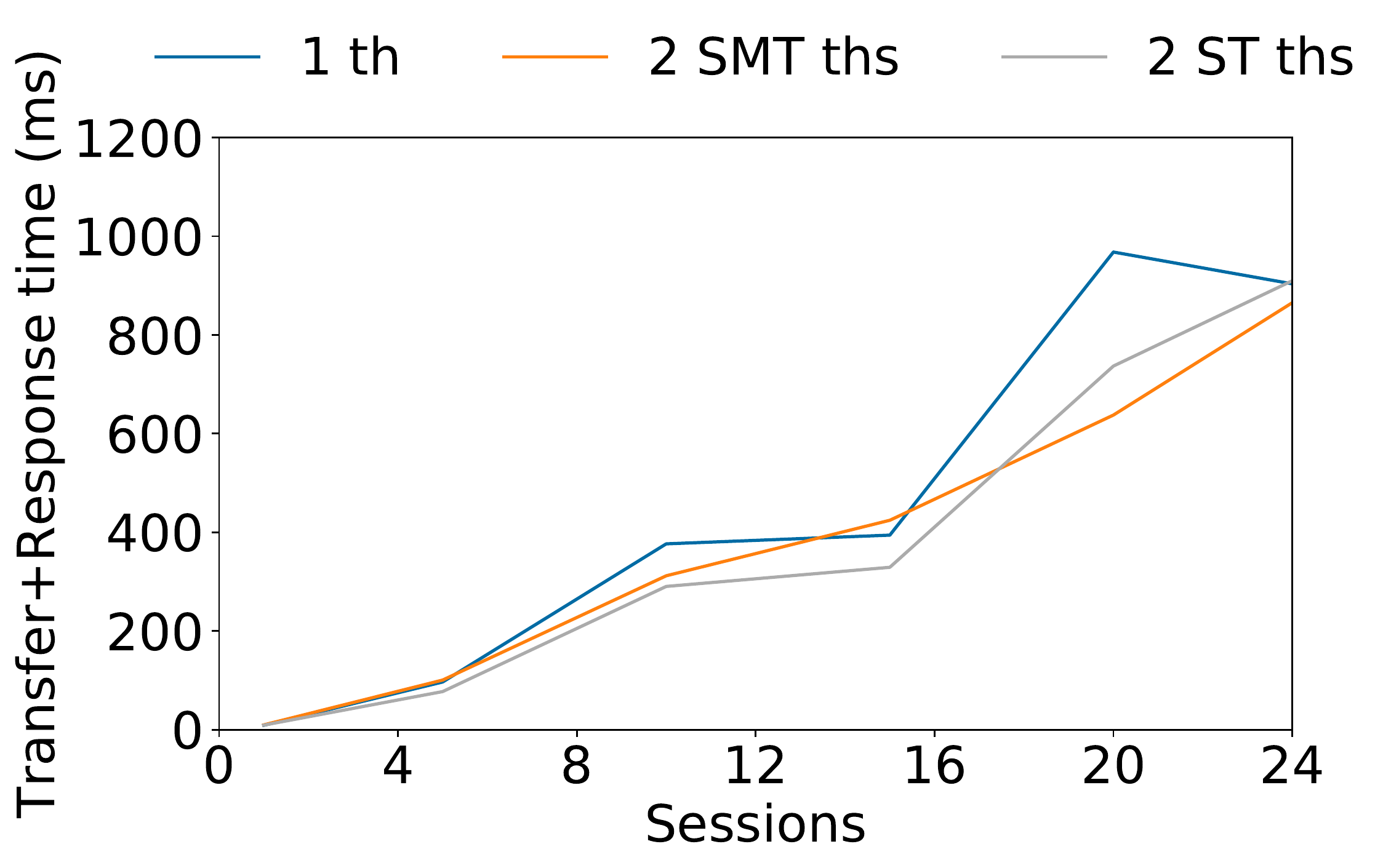}
       \label{fig:media_lat}}
    \subfloat[CPU utilization]{%
       \includegraphics[width=0.32\textwidth]{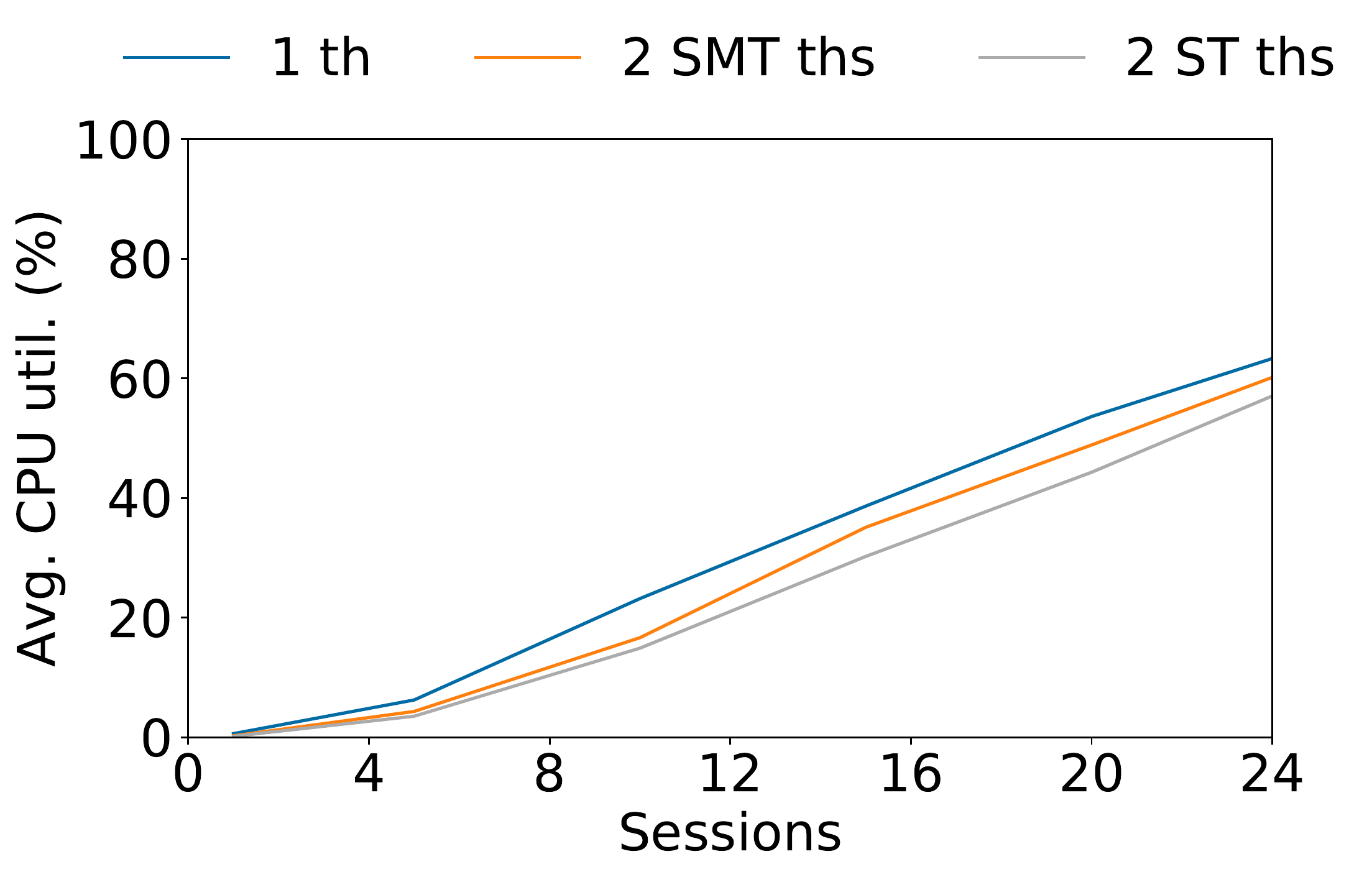}
       \label{fig:media_util}}
    \subfloat[Network transmit bandwidth]{%
       \includegraphics[width=0.32\textwidth]{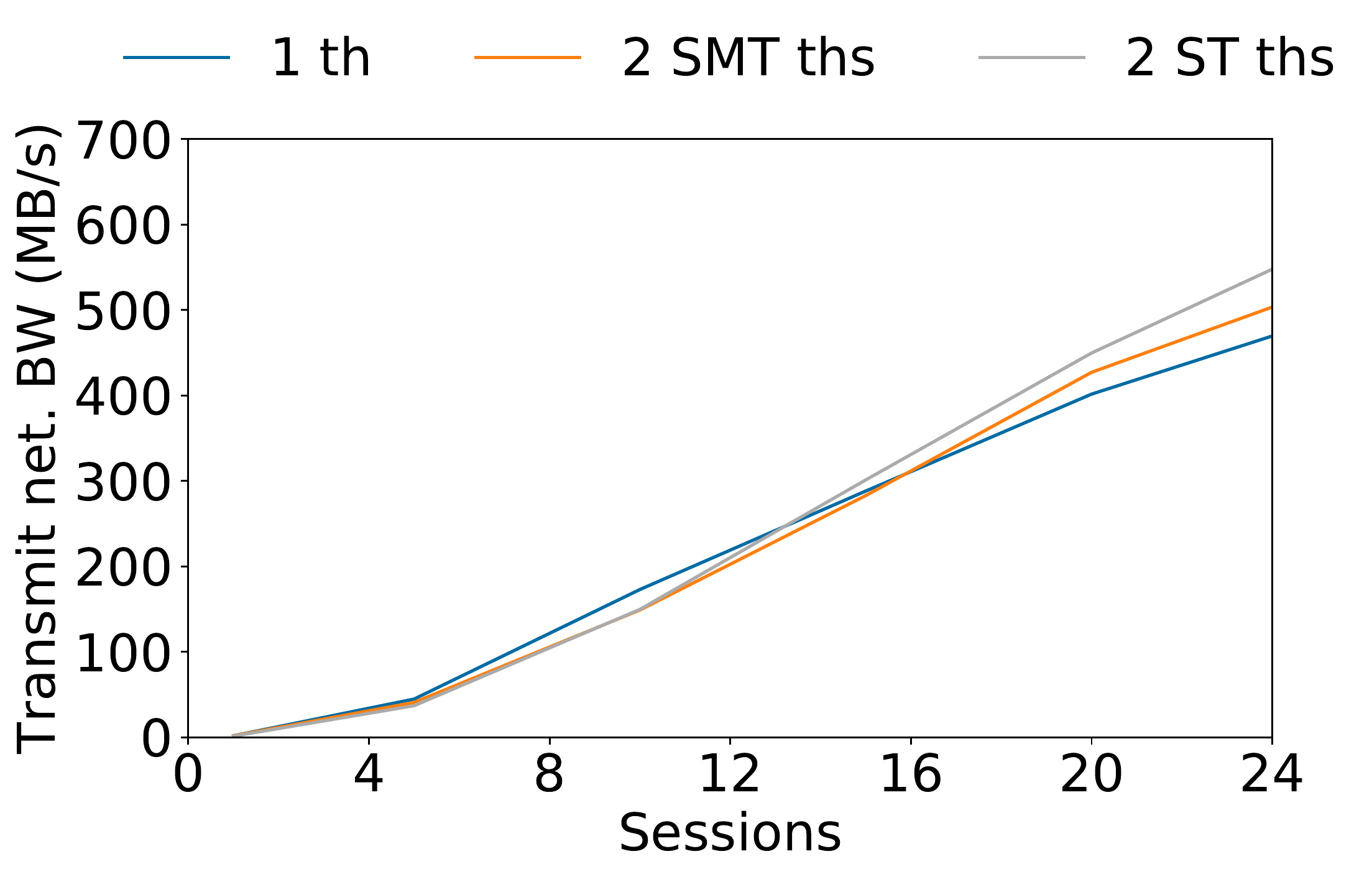}
       \label{fig:media_tx}}
       \\
       \vspace{0.2cm}
    \subfloat[Disk bandwidth]{%
       \includegraphics[width=0.32\textwidth]{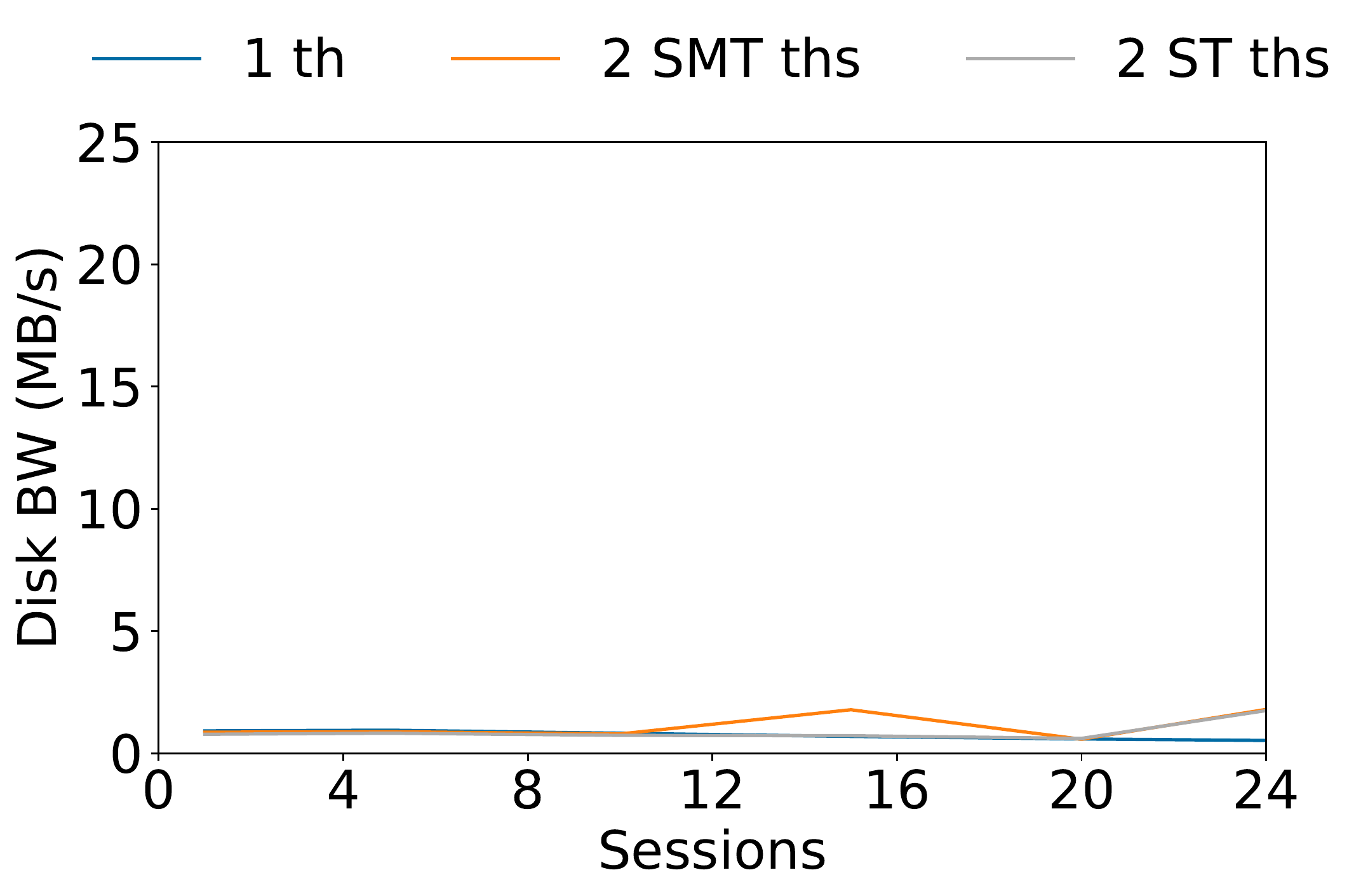}
       \label{fig:media_disk}}
    \subfloat[Main memory bandwidth]{%
       \includegraphics[width=0.32\textwidth]{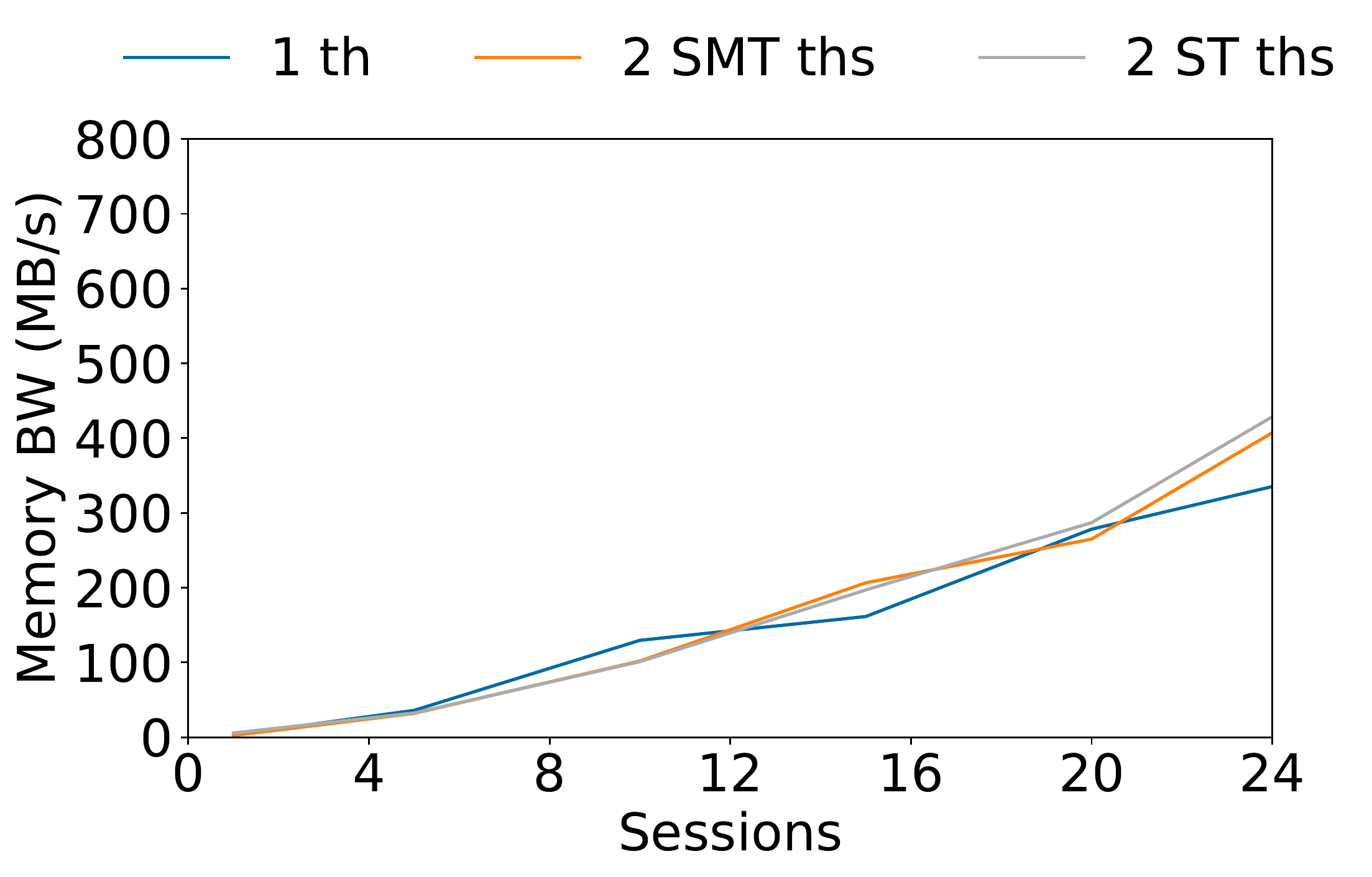}
       \label{fig:media_mem}}
    \subfloat[LLC occupancy]{%
       \includegraphics[width=0.32\textwidth]{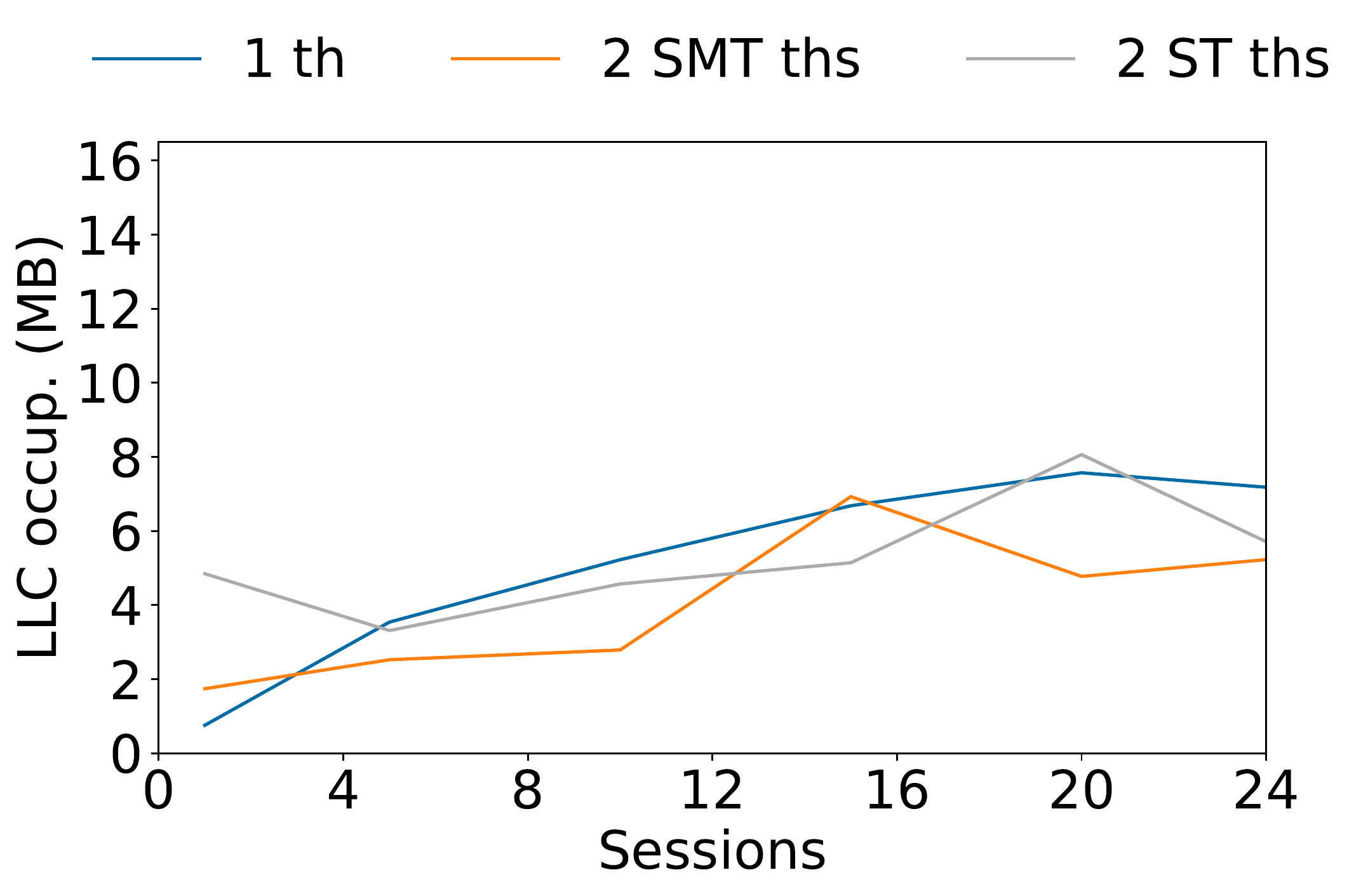}
       \label{fig:media_llc}}
    \caption{Media-streaming characterization.}
    \label{fig:media_noconstrains} 
\end{figure*}

Figures~\ref{fig:img_noconstrains} to \ref{fig:media_noconstrains} present the results under no resource constraint in the three aforementioned scenarios for the eight TailBench applications and \emph{media-streaming}.
In each figure there is a plot for each of the analyzed metrics and each plot presents how the considered metric (Y-axis) evolves varying the QPS ratio (X-axis). 
Six plots are included for each figure, depicting from a to f the values of the following average metrics:

\begin{itemize}

\item Plot (a) depicts the $95^{th}$ tail latency, i.e., the 95\% percentile (only 5\% of the responses are longer than that value). The dotted line of these plots represents the
QoS latency of each workload (discussed in Section~\ref{sec:charac:subsec:latency}). In case of \emph{media-streaming}, it shows the transfer time plus response time. 

\item Plot (b) depicts the CPU utilization of the cores, e.g., for the 2-thread scenarios, we report the average utilization of the two CPUs where the application is running.

\item Plot (c) shows the received and transmitted network bandwidth, respectively, measured at the server node.

\item Plot (d) presents the disk bandwidth consumed by the server node.

\item Plots (e) and (f) show the main memory bandwidth and LLC occupancy metrics, respectively.  

\end{itemize}

Next we discuss the results obtained, focusing on each system resource. 

\vspace{0.2cm}
\textbf{LLC Occupancy}
\vspace{0.2cm}

The LLC occupancy results, presented in plot (f), show that this metric usually ranges between 6 and 12 MB for the studied workloads.
The LLC is a critical resource since it can become one of the main sources for contention and performance degradation when multiple workloads run concurrently. 
Given that LLC occupancy ranges between 6 and 12 for each individual workload and that, in our environment, the LLC capacity for the 12 cores is 16.5 MB, contention will certainly appear when co-running multiple workloads.
Moreover, contention in the LLC also increases main memory bandwidth, which can become an additional performance bottleneck.

\vspace{0.2cm}
\textbf{Main memory Bandwidth}
\vspace{0.2cm}

Main memory bandwidth consumption of the TailBench applications ranges from 45~MB/s (\emph{shore}) to 700~MB/s (\emph{sphinx}).
This is a low bandwidth utilization taking into account that the available main memory bandwidth of the system amounts to 115~GB/s (9.5~GB/s per core).
However, these results are obtained when applications run in isolation, so it is expected that bandwidth utilization will grow when the effective LLC capacity is reduced by the effect of the interference.

Regarding the differences observed in the memory bandwidth for the three studied scenarios, the two-threaded configurations follow two different trends:
for \emph{moses} and \emph{sphinx}, the bandwidth consumption with respect to the single-threaded configuration is significantly higher (up to $5\times$ for the 2-ST configuration in \emph{sphinx};
for the remaining applications, the differences are small with respect to the main memory bandwidth of the single-threaded configuration.

\vspace{0.2cm}
\textbf{Network Bandwidth}
\vspace{0.2cm}

Among the studied workloads, \emph{media-streaming} is clearly the application that achieves the highest network bandwidth consumption.
Its network bandwidth utilization grows with the number of sessions, reaching a maximum transmit network bandwidth close to 550~MB/s. 
On the other hand, TailBench workloads present much lower network bandwidth requirements.

As it could be expected, the network traffic grows linearly with the QPS for the TailBench applications and with the number of sessions for \emph{media-streaming}.
In summary, the bandwidth consumption of the studied workloads is low compared to the network capabilities, which suggests that network bandwidth should not be a major performance bottleneck for the studied applications even in over-subscription scenarios.

\vspace{0.2cm}
\textbf{Disk Bandwidth}
\vspace{0.2cm}

Regarding the disk bandwidth results, the highest values are achieved by \emph{moses} and \emph{shore}, with 9 and 4~MB/s, respectively.
\emph{Xapian} presents a significantly lower but non-negligible disk bandwidth consumption, around 2~MB/s.
The remaining applications show a disk bandwidth consumption close to zero.

This is an expected fact for applications like \emph{masstree} and \emph{silo}, which are in-memory key-value store and in-memory database applications, respectively, and they present almost no disk accesses.
As it happened with the network bandwidth, disk bandwidth consumption also grows linearly with the QPS ratio, with the only exception of \emph{media-streaming}. Its bandwidth consumption starts at 0.9~MB/s and is reduced as the number of sessions grows.
This is because \emph{media-streaming} loads the videos from the disk once, and then they are served from the main memory. 
Therefore, when running with a few sessions, the disk accesses are averaged through a shorter time, which results in a higher bandwidth with respect to executions with more sessions.

The disk bandwidth is a resource that can potentially become a performance bottleneck when multiple workloads run concurrently, either on different cores or in the same core.
Although the SSD installed in the storage server allows up to 550~MB/s in large sequential reads, the bandwidth is significantly reduced when operations become small, random, and reads and writes are combined.
Consequently, disk bandwidth is a metric that should always be monitored to detect performance degradation when co-running multiple workloads.

\subsection{Analysis of CPU Utilization}\label{sec:charac:subsec:util}

This section discusses the results achieved by the three server configurations from a resource consumption perspective.
In other words, we study how inter-thread interference of the VMs at the system resources impacts on the studied metrics.
To do so, we perform two main comparisons: i) 1-ST vs 2-SMT server configuration; and ii) 2-SMT vs 2-ST server configuration.

\vspace{0.2cm}
\textbf{1-ST vs 2-SMT Threads}
\vspace{0.2cm}

This comparison study is useful in the analysis of inter-thread compute interference within the core.
That is, by comparing this two configurations, we can analyze how having both application threads competing among the core resources impacts on performance.
Results show that, if we look at 20\% utilization, the 2-SMT server doubles the QPS in four out of the seven applications compared to the 1-ST server.
In 50\% utilization, however, the QPS achieved by the 2-SMT server never doubles the QPS achieved by the 1-SMT configuration.
This happens because of the higher intra-core interference, which negatively impacts on hte performance that each thread achieves.
Therefore, results confirm the claim that the higher utilization values, the higher potential interference threads can incur. 

\vspace{0.2cm}
\textbf{2-SMT vs 2-ST Threads}
\vspace{0.2cm}

This comparison allows us to study how the QPS scales with the number of threads, as well as how the performance is improved when there can not be intra-core interference.
The achieved results show that, at 20\% utilization, there are small differences in the QPS supported by each server configuration. 
\emph{silo} increases the supported QPS by $1.23\times$ when the thread execute in different cores. The differences for the other workloads are negligible. This behavior could be expected since, with low CPU utilization, the 2-SMT threads configuration suffer little interference in the intra-core shared resources.Thus, it performs similarly to the dual-thread server configuration that pins the two threads to two different cores.
When the CPU utilization grows, intra-core interference also increases in the 2-SMT threads configuration. In this scenario, the 2-ST thread configuration, which does not suffer intra-core interference, achieves higher performance, which translates into more queries replied per second. This is the case of all workloads except \emph{mastree}, \emph{moses}, and \emph{shore}. The latter two are explained by the fact that they are limited by the disk and thus do not improve performance with the 2-ST  configuration. The case of \emph{masstree} is slightly surprising but it indicates that the two server threads do not suffer from intra-core interference but contention is mainly caused by main memory bandwidth and network bandwidth.

\vspace{0.2cm}
\textbf{To Take Away} \emph{As it is known, tenant applications cannot be instrumented in the public cloud, thus they need to be studied as a black box. This means that identifying a proxy of performance from the metrics we are able to collect is a must.
From the analysis of the results we can argue that different server applications present widely different compute demands that strongly affect their performance. In other words, the server can support a CPU utilization higher than 50\% in some applications (e.g., masstree) while still satisfying the client demands. Therefore, we cannot consider only the CPU utilization as a proxy of performance.}

\vspace{0.2cm}

\subsection{Analysis of Tail Latency and QoS Results}\label{sec:charac:subsec:latency}

\begin{table}[]
\centering
\begin{tabular}{|l|c|c|}
\hline
\textbf{Workload}   & \textbf{Tail latency QoS} & \textbf{QPS single-thread}   \\ \hline
img-dnn             & 3.6 ms       &        650   \\
masstree            & 1.4 ms       &        1000    \\
moses               & 7.1 ms     &          30 \\
shore               & 25 ms*      &        100 \\
silo                & 0.5 ms       &            1000\\
specjbb             & 0.7 ms       &            1500\\
sphinx              & 4275.4 ms**       &           0.7 \\
xapian              & 6.2 ms       &          350  \\\hline
\end{tabular}
\caption{Tail latency QoS requirements for the Tailbench workloads in our experimental platform.}
\label{tab:qos}
\end{table}

In this section, we analyze the characterization results achieved by the three studied configurations from a QoS perspective~\footnote{This analysis focuses in the Tailbench applications since the QoS metric is based on the latency results and \emph{media-streaming} does not report this metric.
}. 

Nowadays, many online service workloads present tail latency QoS constraints. 
These constrains are part of the Service Level Agreement (SLA) in some cases. 
In other cases, users should make sure of hiring enough resources to meet their target QoS. To determine realistic tail latency QoS constraints in our experimental setup and evaluate when workloads meet the QoS requirements and when they do not, we define the QoS requirement for each workload as a function of its average service time. 
This approach is based on the one proposed by Delimitrou et al~\cite{Delimitrou18}.
The QoS target for each workload is defined as $5\times$ the average service time that our system achieves with a CPU utilization of 20\%. 
We will refer to this value as LQoS . 
That is, to meet the QoS constraints, the $95^{th}$ tail latency should be lower than the LQoS. 

Table~\ref{tab:qos} presents the tail latency QoS requirements we defined and the QPS supported by the single-threaded server of each workload that meets the QoS latency constraint.
Tail latency requirements range from 0.5~ms for \emph{silo} to 4275.4~ms for \emph{sphinx}. 
We cannot set a tail latency requirement for \emph{shore} (*) based on this approach since its CPU utilization remains below 20\% even after it saturated, so we set the LQoS for \emph{shore} to 25~ms, point before its tail latency saturates.  Notice that tail latency QoS of \emph{sphinx} (**) is over 4~s, which seems rather high in comparison with other speech recognition services. However, we found that these values are in line with the results of this application presented in \cite{tailbench}.

Once we have defined the QoS requirements for each application, we can study the effect of adding another server thread from a QoS perspective.
To do so, we analyze the impact of adding more threads on the queries per second when the QoS requirements are met. 
That is, when the $95^{th}$ tail latency is below the QoS latency presented in Table~\ref{tab:qos}. 
The analysis is done by focusing on the QPS achieved by the different server configurations in Figures~\ref{fig:img_noconstrains}-\ref{fig:media_noconstrains}. In these figures, the QoS requirement of each TailBench application is represented by an horizontal dotted line.

Regarding the performance of the 2-ST server, most applications show great scalability, since the QPS they achieve while meeting the QoS requirements with the 2-ST server is by 2 times (or even higher) that of the achieved with the single-threaded server. This can be clearly observed in, for instance, the plots from \emph{img-dnn}, \emph{moses}, \emph{sphinx} and \emph{xapian}.
However, in some workloads the 2-ST server provides minor benefits in terms of QoS.
This is the case of \emph{specjbb}, which supports up to 2000~QPS regardless of the server configuration. 
Nevertheless, the \emph{specjbb} tail latency values achieved by the 2-ST configuration are still much lower than those of the other servers.

Next, we analyze the performance of the 2-SMT server. 
Significant performance scalability, similar to that of the 2-ST configuration in some cases, can be observed.
For example, in \emph{sphinx} and \emph{xapian}, the 2-SMT server provides by 1.4 and 1.9 times, respectively, the QPS supported by the single-threaded server, while also meeting the QoS latency. However, the performance scalability in the 2-ST server is even higher than the 2-SMT server.
Other workloads, like \emph{silo} and \emph{specjbb}, do not present significant benefits when running in the 2-SMT server.
For these applications, a 2-ST configuration is preferable to improve performance.

Notice that an unexpected behavior occurs with \emph{shore}. This application supports, the highest QPS with the 1-ST configuration. That is, increasing the amount of threads from one to two reduces considerably the saturation QPS (i.e., point at which the LQoS is achieved).
This issue has been also pointed out in \cite{tailbench}, where authors realized of this unexpected behavior and demonstrated by simulation that this poor scalability is due to both the overhead of adding threads (i.e., service times were constant instead of widely vary) and the contention at the shared resources.

\vspace{0.2cm}
\textbf{To Take Away} \emph{The analysis from the QoS perspective shows that the performance provided (in terms of QPS) by a 2-ST server while supporting the target QoS latency can be matched with a 2-SMT server; a configuration that involves a much lower amount of compute resources. 
Thus, a key challenge is to differentiate, as black boxes, those applications that behave in this way to take the appropriate thread-to-core allocation measures. 
The appropriate thread-to-core allocation is the one that devotes a lower number of resources to support the target QoS latency at the current QPS load. 
For instance, in \emph{specjbb}, for the considered QoS latency, a single thread is enough. 
In contrast, in \emph{xapian}, the 2-SMT configuration is the most efficient for a 500 QPS load, while a 2-ST server is required to support more than 700 QPS.}

\vspace{0.2cm}

\section{Workload Categories: A Resource Oriented Taxonomy}\label{sec:taxonomy}

Based on the results of the previous analysis,
we define the \emph{resource oriented taxonomy}. 
The taxonomy establishes four main categories based on the analysis of five major system resources and the level of load: 
i) CPU utilization, which evaluates both core computation and the time the core is waiting for main memory accesses, 
ii) main memory bandwidth consumption,
iii) $95^{th}$ tail latency,
iv) QPS, this metric defines the level of load for Tailbench workloads (the number of sessions is used for \emph{media-streaming}),
v) network bandwidth, 
and vi) disk bandwidth.
Next we discuss the devised workload categories.

\subsection{Category 1: High Processor (core and/or DRAM) Demands per Request} 
\label{ssec:category1}

This behavior is exhibited by compute-intensive applications with high requirements of CPU resources (both compute and DRAM memory resources) like \emph{sphinx} (Figure~\ref{fig:sphinx_noconstrains}), which shows:

\begin{itemize}
    \item {\bf CPU utilization.} In these applications the CPU utilization grows up to saturating (i.e., 100\%) with the single-threaded server. 
    \item {\bf Memory Bandwidth.} Significant memory bandwidth (over 600~MB/s) consumption is reached. The fact that CPU saturates means that memory access time is fully overlapped with CPU time.
    \item {\bf$95^{th}$ tail latency.} This latency is very high, in the order of thousands of milliseconds (in the plot it ranges from few to tens of seconds).
    \item {\bf QPS.} This metric is dramatically low, e.g., below 2, since requests take long time to be processed.
    \item {\bf Network bandwidth.} Not an issue.
    \item {\bf Disk bandwidth.} Not an issue.
\end{itemize}

\subsection{Category 2: High Disk Bandwidth Demands} 
\label{ssec:category2}

The behavior of these applications is mainly defined by the disk utilization, although main memory or CPU utilization also contribute to achieve good performance although in a lower extend. That is, they present low to moderate compute and DRAM memory resources demands.
This behavior is shown by \emph{moses}, which stores the language model and phrase table on disk to reduce its main memory utilization (Figure~\ref{fig:moses_noconstrains}), 
and by \emph{shore} (Figure~\ref{fig:shore_noconstrains}), whose queries access large databases stored in the disk. A summary of their system resource demands are:

\begin{itemize}
    \item {\bf CPU utilization.} These applications require low to moderate CPU demands regardless of the QPS for the target QoS.
    \item {\bf Memory Bandwidth.} Around 100~MB/s for 200 QPS and 10~MB/s for 100 QPS before violating the QoS latency (dotted horizontal line), in \emph{moses} and \emph{shore}, respectively.
    \item {\bf $95^{th}$ tail latency.} Relatively large, ranging from 3 to 20~ms in the single-threaded server configuration before QoS violation. That is, much smaller than the one obtained in applications of category 1 but much larger than applications in category 3 discussed below.
    \item {\bf QPS.} Medium, ranging from 10 to 250 before QoS violation. That is, much larger than applications in category 1 but much smaller than applications in category 3.
    \item {\bf Network bandwidth.} Almost null, not an issue.
    \item {\bf Disk bandwidth.} This is the key resource that characterizes applications in this category. This value is by 5~MB/s for 100~QPS in \emph{moses}, and around 3.4~MB/s for 100~QPS in \emph{shore}.
\end{itemize}

\subsection{Category 3: Fast} \label{ssec:category3}

This behavior is shown in most of the applications like 
\emph{img-dnn} (Figure~\ref{fig:img_noconstrains}), 
\emph{masstree} (Figure~\ref{fig:masstree_noconstrains}), 
\emph{silo} (Figure~\ref{fig:silo_noconstrains}),
\emph{specjbb} (Figure~\ref{fig:specjbb_noconstrains}), and
\emph{xapian} (Figure~\ref{fig:xapian_noconstrains}), 
whose requests require little amount of resources and so,
they are served very fast --at most a few milliseconds-- and the processor is able to attend a significant number of request per second (i.e., over one thousand). 
Even though medium CPU resources (both compute and DRAM memory resources) are needed per request, the aggregate values can be noticeable.
A summary of their system resource demands is exposed next:

\begin{itemize}
    \item {\bf CPU utilization.} In these applications CPU utilization grows with the QPS up to nearly 100\% in some cases, despite QoS being already damaged.
    \item {\bf Memory Bandwidth.} Main memory bandwidth consumption is moderate and depends on the workload. For instance, \emph{silo}'s memory bandwidth utilization is around 70~MB/s and \emph{specjbb}'s utilization is around 200~MB/s for the QPS for which their tail latency violates the QoS latency.
    \item {\bf$95^{th}$ tail latency.} Latency is very low (sometimes below one millisecond).
    \item {\bf QPS.} This metric is very high, in the order of thousands.
    \item {\bf Network bandwidth \& Disk bandwidth.} These applications present different combinations of I/O operations to the network and disk.\emph{Img-dnn} and \emph{mastree} have negligible disk bandwidth but significant network bandwidth. The opposite occurs with \emph{xapian}, which presents significant disk bandwidth but negligible network bandwidth.
    \emph{Silo} and \emph{specjbb} show negligible utilization of both resources.
    This occurs due to the high QPS ratio they can attend and their low response time, the requirements of each of these resources \emph{per request} is low; however little contention on these resources might impact on their very short latency, so harming their performance.
\end{itemize}

\subsection{Category 4: Streaming Workloads} \label{ssec:category4}

The behavior of applications in this category is the one typically shown by servers that stream multimedia contents. \emph{Media-streaming}, from Cloudsuite, is the application that presents this behavior among the studied workloads.  Next, we summarize its main system resource demands:

\begin{itemize}
\item {\bf CPU utilization.} In general, CPU utilization is moderate since this type of workloads spend a significant fraction of their execution performing I/O operations. CPU utilization could grow if the server needed to perform some processing to the content to be streamed. 

\item {\bf Memory Bandwidth.} Main memory bandwidth utilization is usually high as the streamed contents have low cache locality. Thus, the server needs to frequently access to the main memory to retrieve the content to be delivered to the clients. \emph{Media-streaming} consumes close to 400MB/s of main memory bandwidth, on average, along its entire execution.

\item {\bf Network bandwidth} These workloads have a high utilization of the network bandwidth to transmit the streamed content to the clients. For instance, \emph{media-streaming} transmits up to 600 MB/s of video to the clients (see Figure~\ref{fig:media_tx}). 
    
\item {\bf Disk bandwidth.} Depending on the amount of main memory, the size and locality of the content the workload is streaming, this kind of workloads can have a modest average disk bandwidth utilization such as \emph{media-streaming} in our experimental setup) when the streamed content is stored in main memory, to a very high disk bandwidth requirements when the server has to perform a noticeable number of accesses to the disk to look for the requested content.
\end{itemize}

\section{Effect of Constraining the Main Shared Resources}
\label{sec:constrain_results}

\begin{figure*}[t!]
    \centering
    \subfloat[$95^{th}$ tail latency (ms)]{%
       \includegraphics[width=0.32\textwidth]{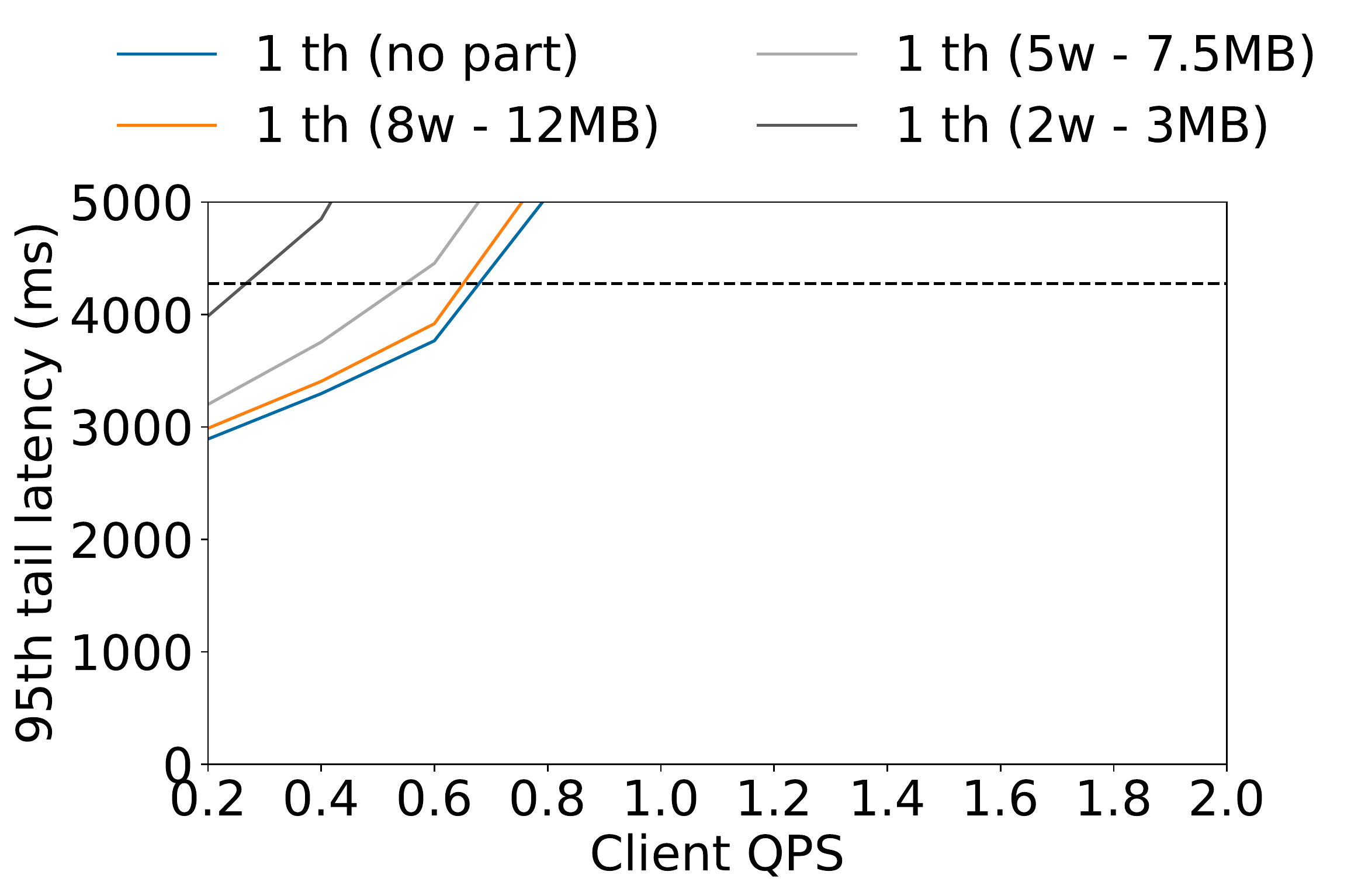}
       \label{fig:sphinx_lat_part}}
    \subfloat[LLC occupancy]{%
       \includegraphics[width=0.32\textwidth]{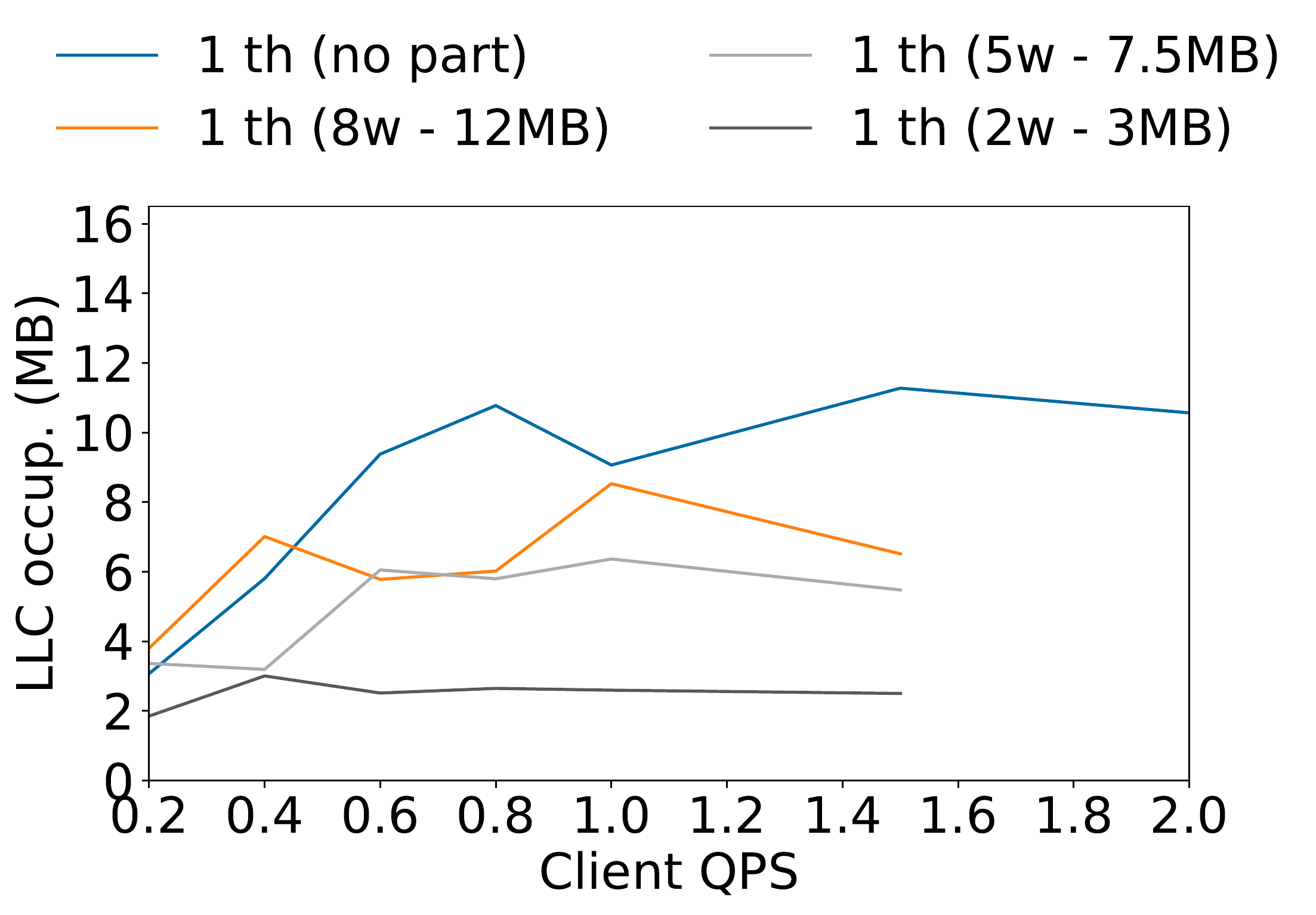}
       \label{fig:sphinx_llc_part}}
    \subfloat[Main memory bandwidth]{%
       \includegraphics[width=0.32\textwidth]{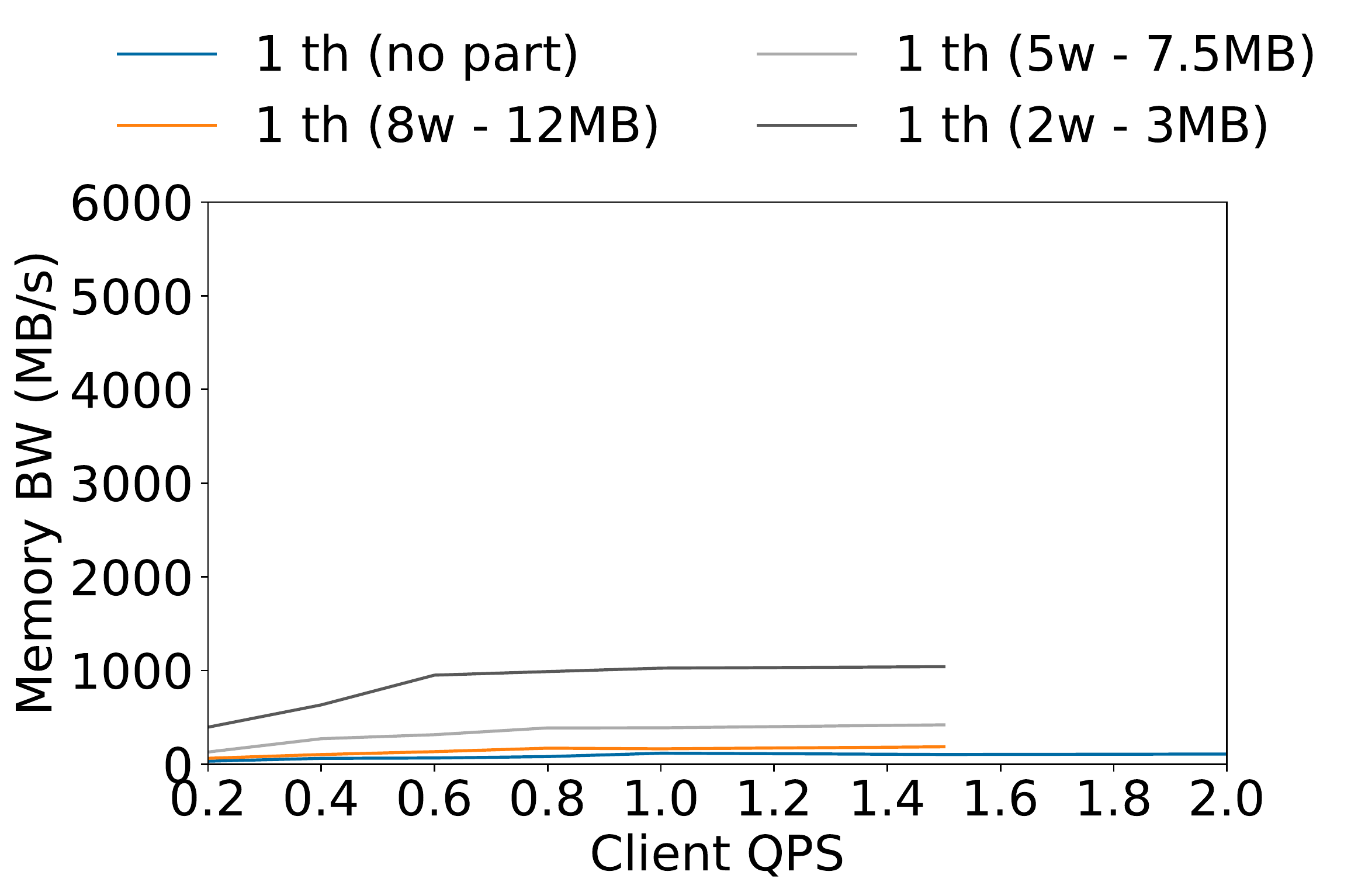}
       \label{fig:sphinx_mem_part}}
    \caption{Sphinx characterization with LLC partitioning.}
    \label{fig:sphinx_llc_constrains} 
\end{figure*}

\begin{figure*}[t!]
    \centering
    \subfloat[$95^{th}$ tail latency (ms)]{%
       \includegraphics[width=0.32\textwidth]{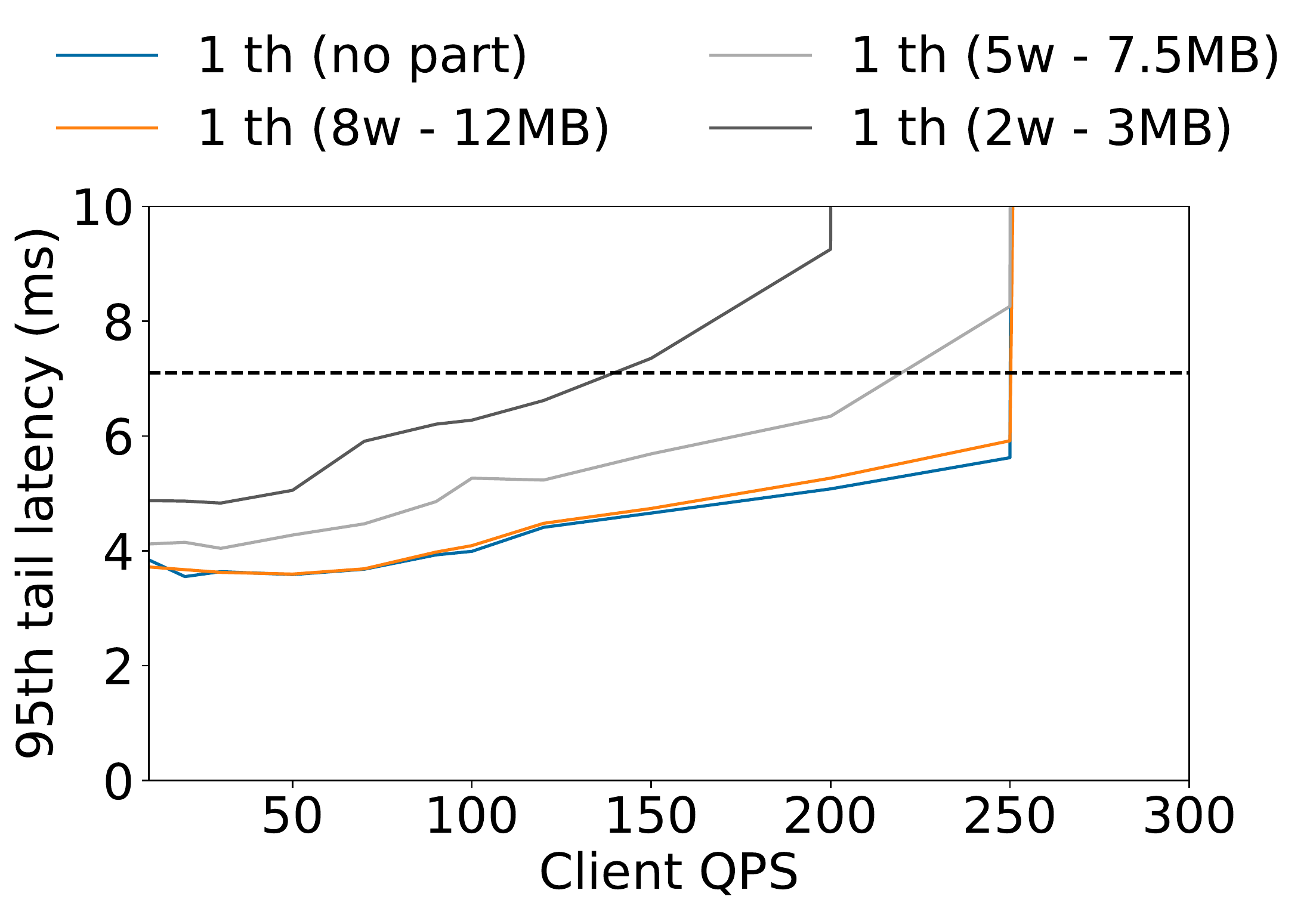}
       \label{fig:moses_lat_part}}
    \subfloat[LLC occupancy]{%
       \includegraphics[width=0.32\textwidth]{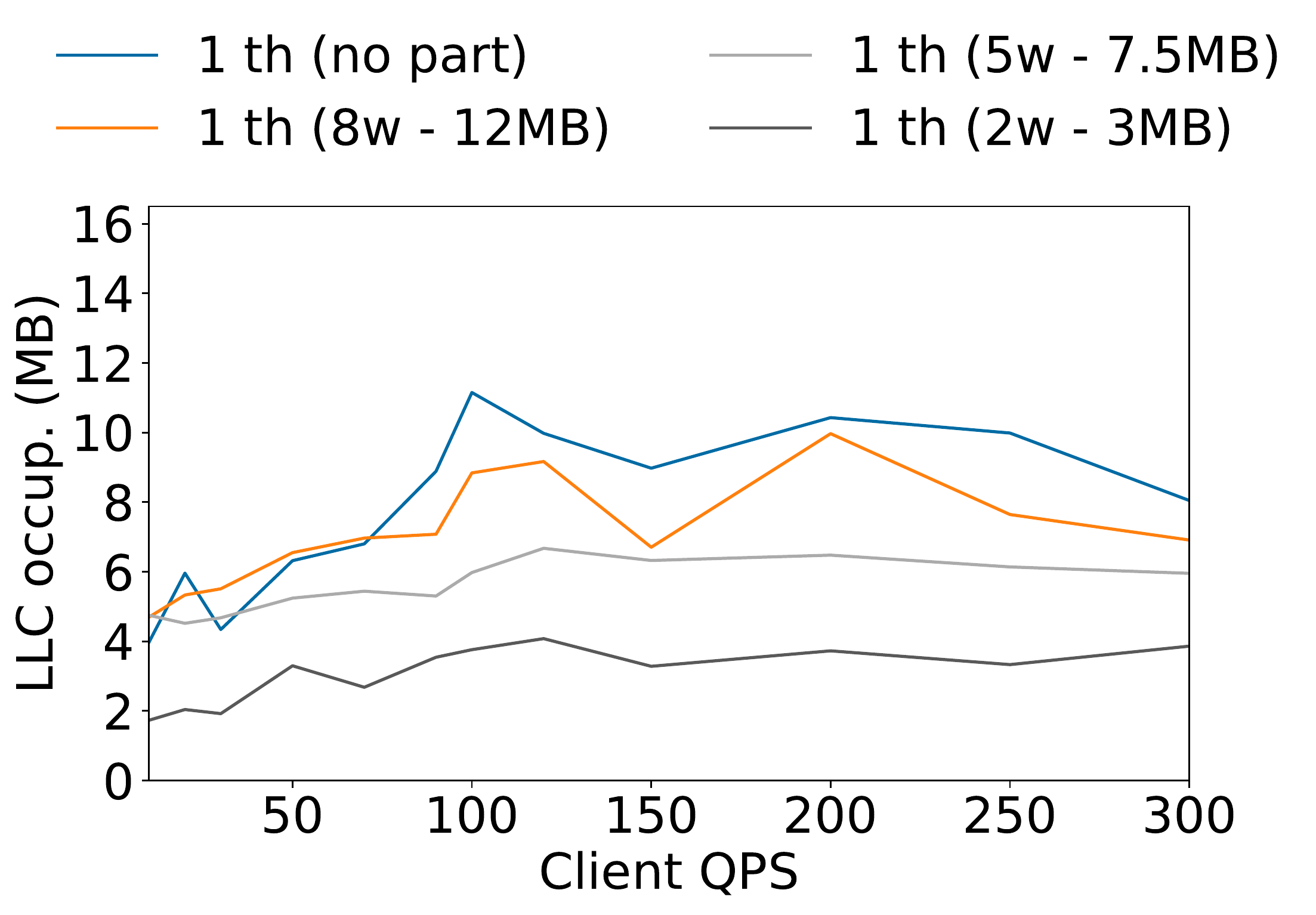}
       \label{fig:moses_llc_part}}
    \subfloat[Main memory bandwidth]{%
       \includegraphics[width=0.32\textwidth]{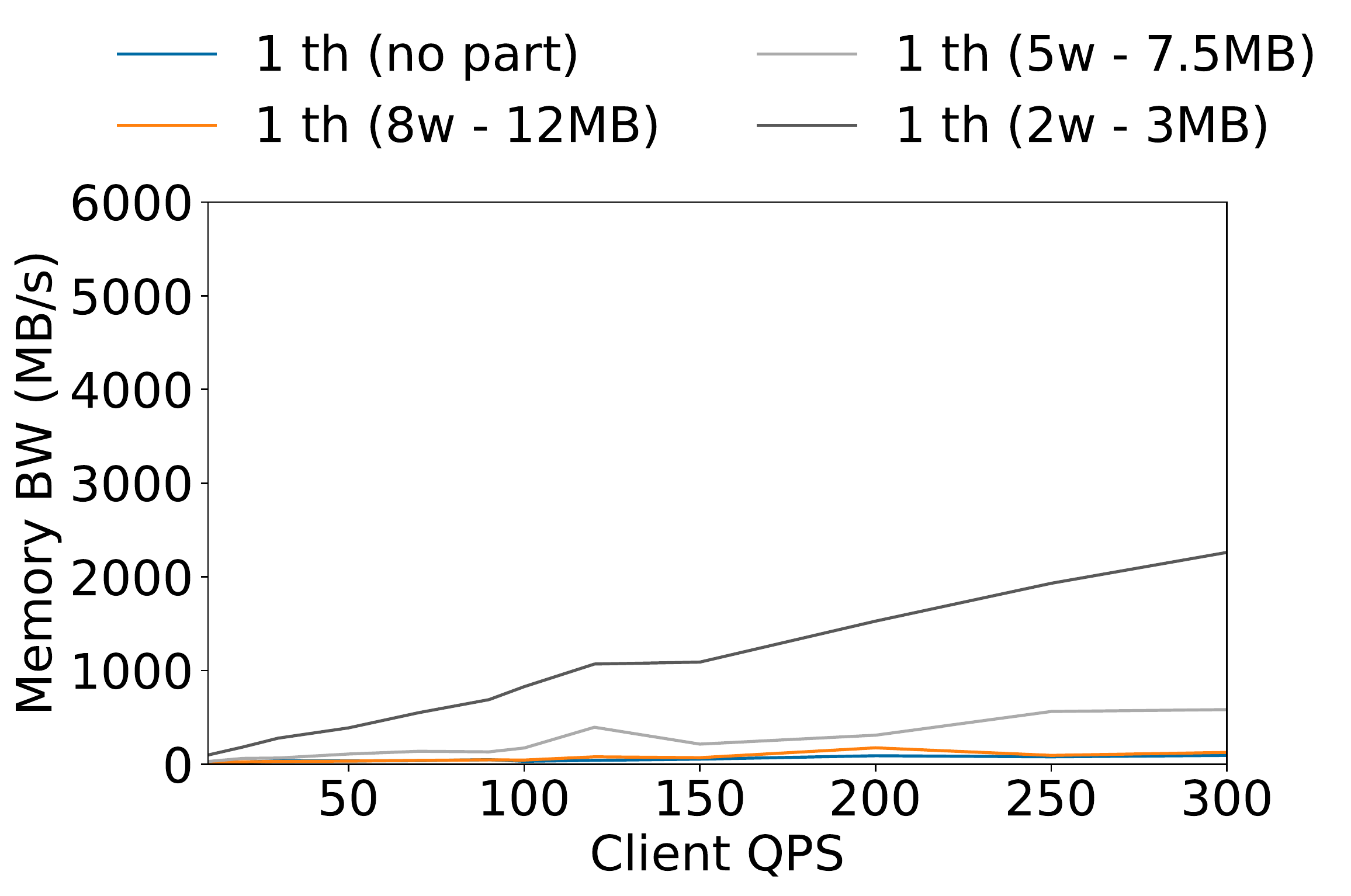}
       \label{fig:moses_mem_part}}
    \caption{Moses characterization with LLC partitioning.}
    \label{fig:moses_llc_constrains} 
\end{figure*}

\begin{figure*}[t!]
    \centering
    \subfloat[$95^{th}$ tail latency (ms)]{%
       \includegraphics[width=0.32\textwidth]{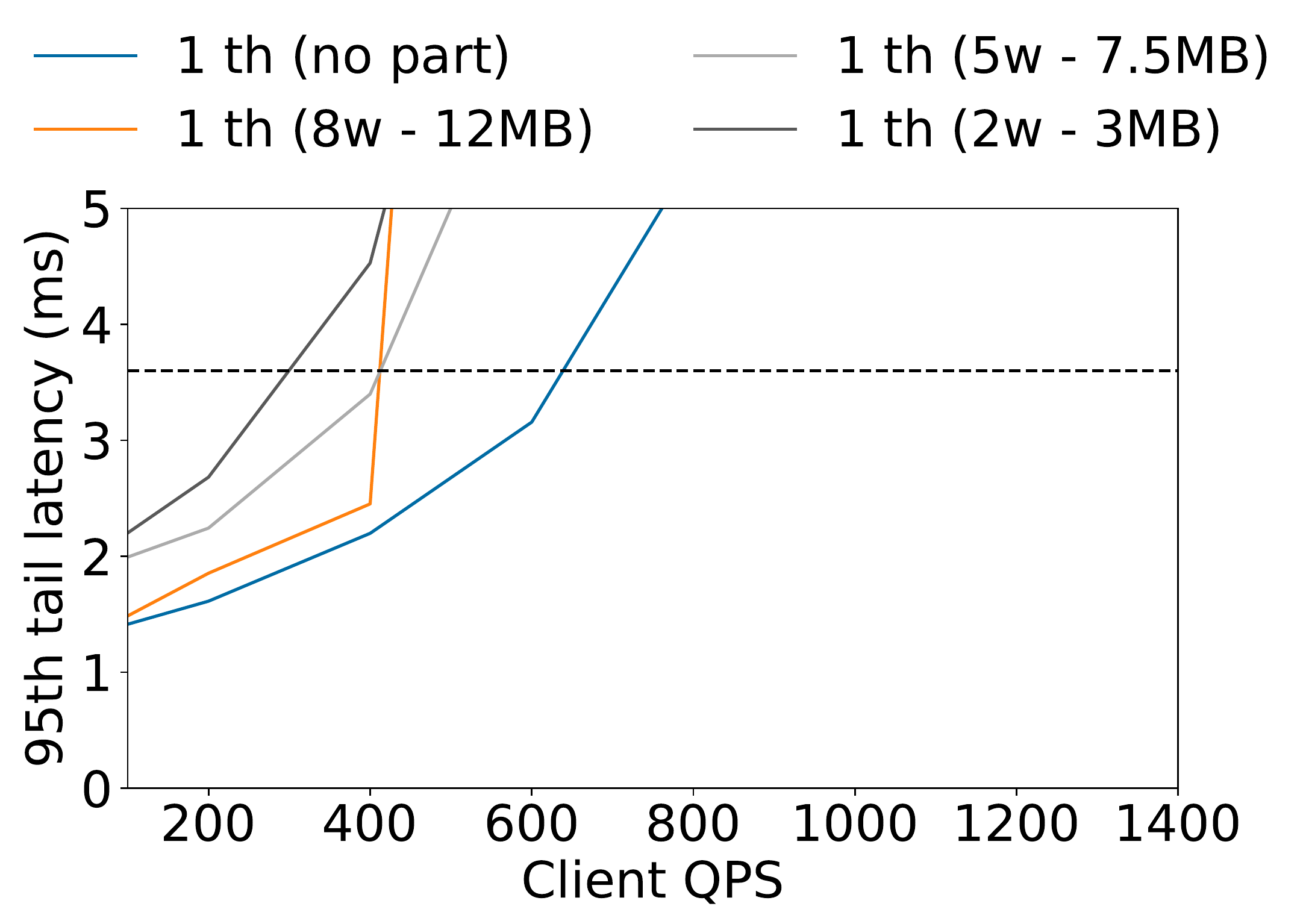}
       \label{fig:img_lat_part}}
    \subfloat[LLC occupancy]{%
       \includegraphics[width=0.32\textwidth]{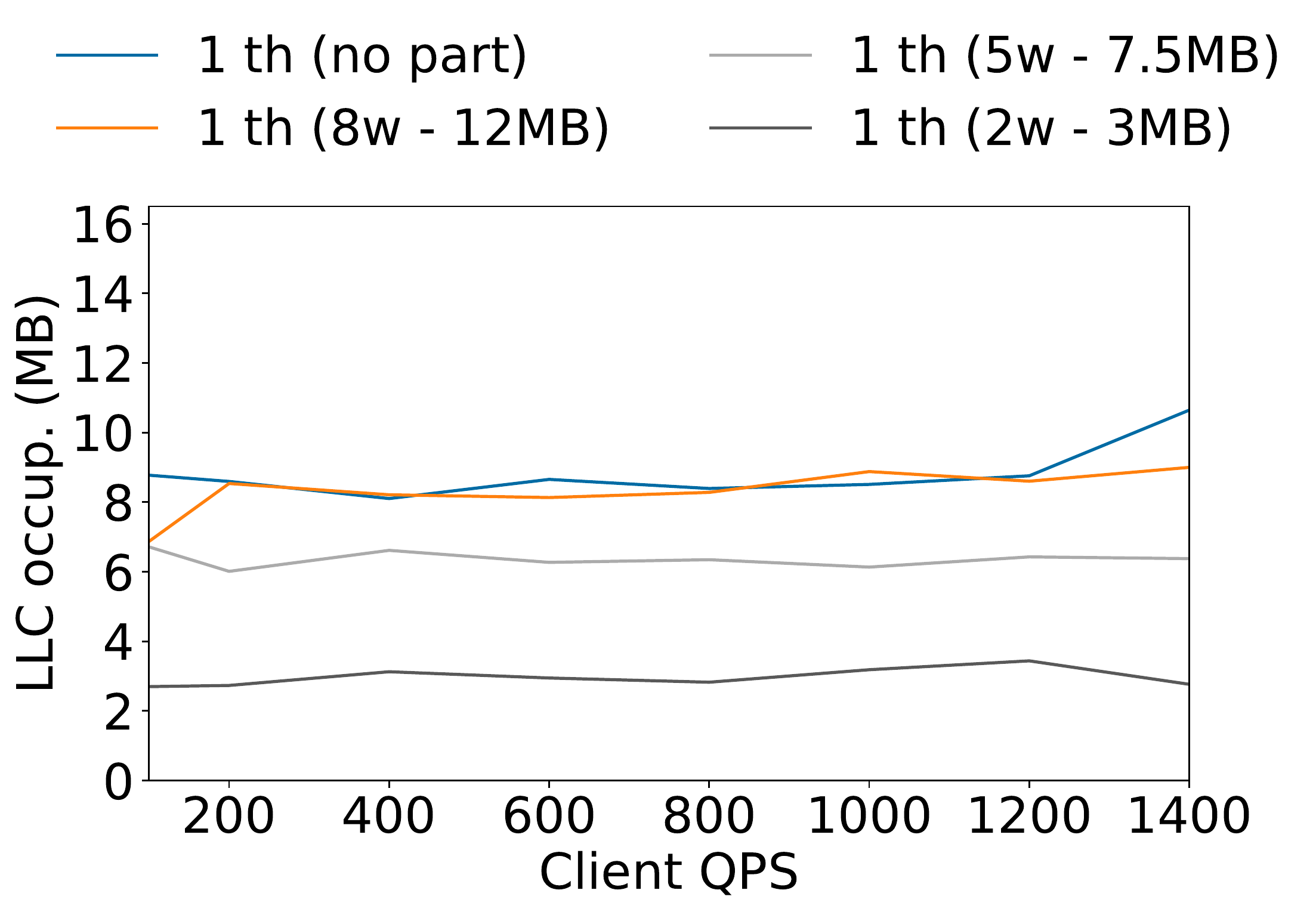}
       \label{fig:img_llc_part}}
    \subfloat[Main memory bandwidth]{%
       \includegraphics[width=0.32\textwidth]{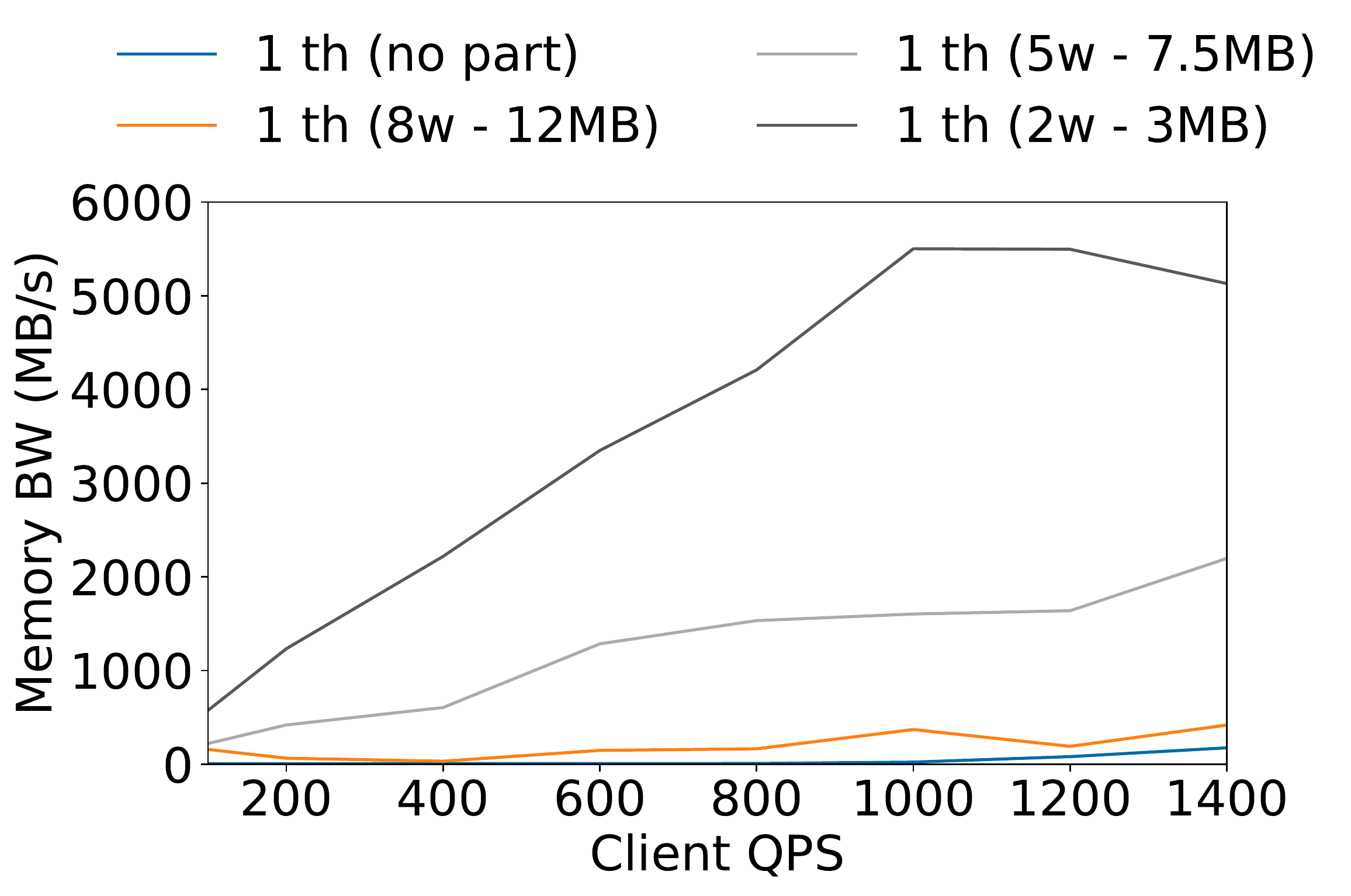}
       \label{fig:img_mem_part}}
    \caption{Img-dnn characterization with LLC partitioning.}
    \label{fig:img-dnn_llc_constrains} 
\end{figure*}

\begin{figure*}[t!]
    \centering
    \subfloat[$95^{th}$ tail latency (ms)]{%
       \includegraphics[width=0.32\textwidth]{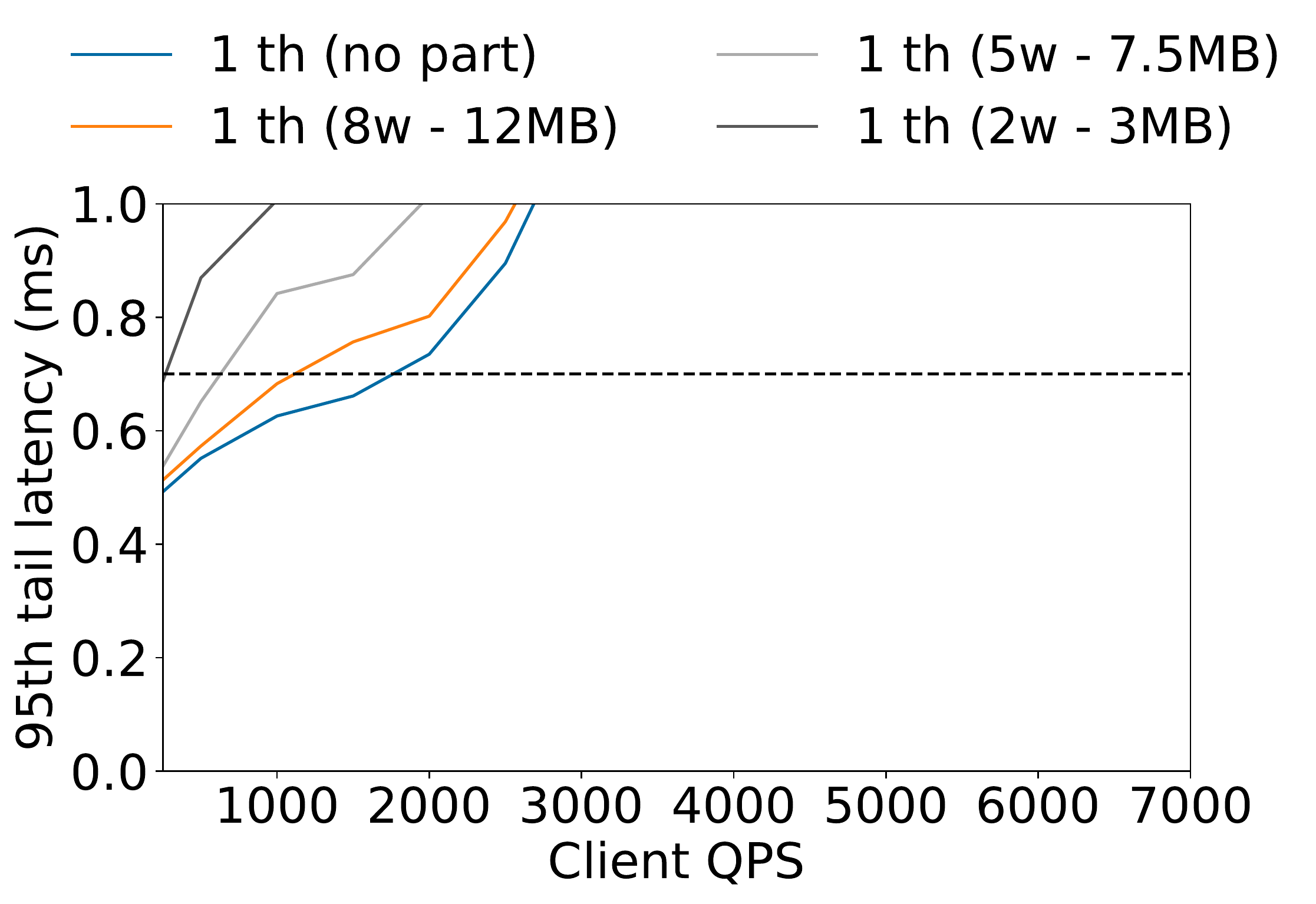}
       \label{fig:specjbb_lat_part}}
    \subfloat[LLC occupancy]{%
       \includegraphics[width=0.32\textwidth]{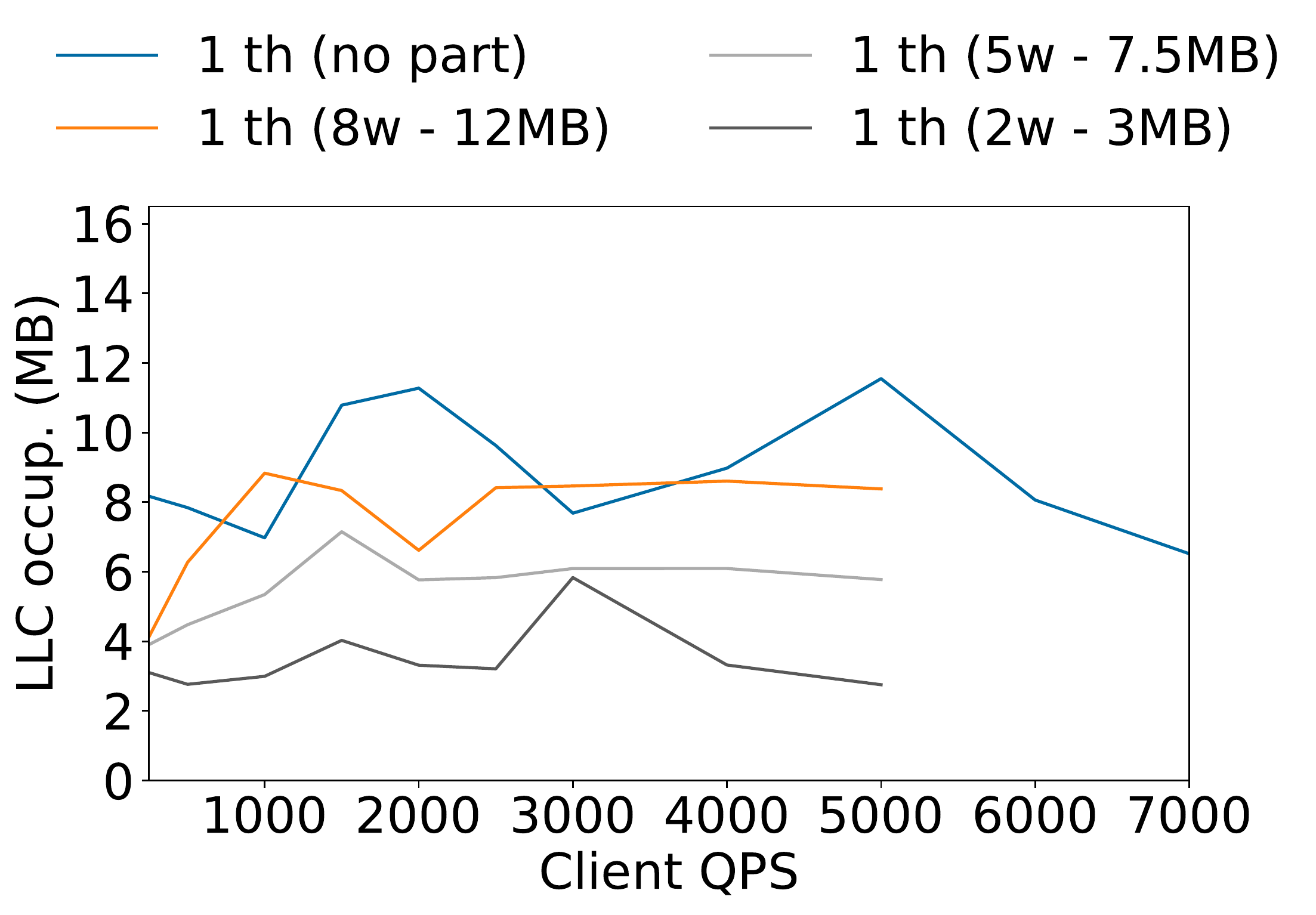}
       \label{fig:specjbb_llc_part}}
    \subfloat[Main memory bandwidth]{%
       \includegraphics[width=0.32\textwidth]{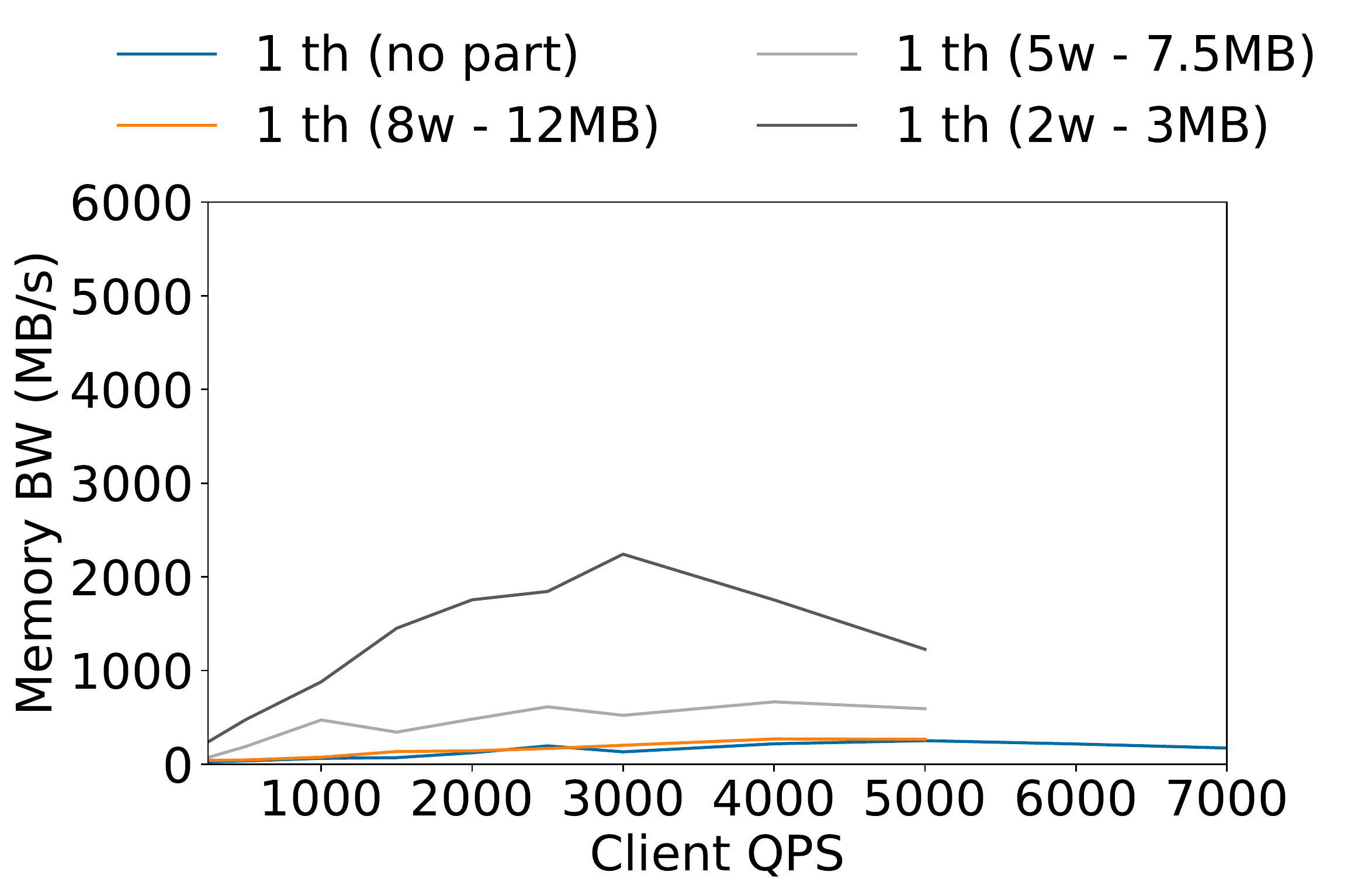}
       \label{fig:specjbb_mem_part}}
    \caption{Specjbb characterization with LLC partitioning.}
    \label{fig:specjbb_llc_constrains} 
\end{figure*}

\begin{figure*}[t!]
    \centering
    \subfloat[Transfer + Response times (ms)]{%
       \includegraphics[width=0.32\textwidth]{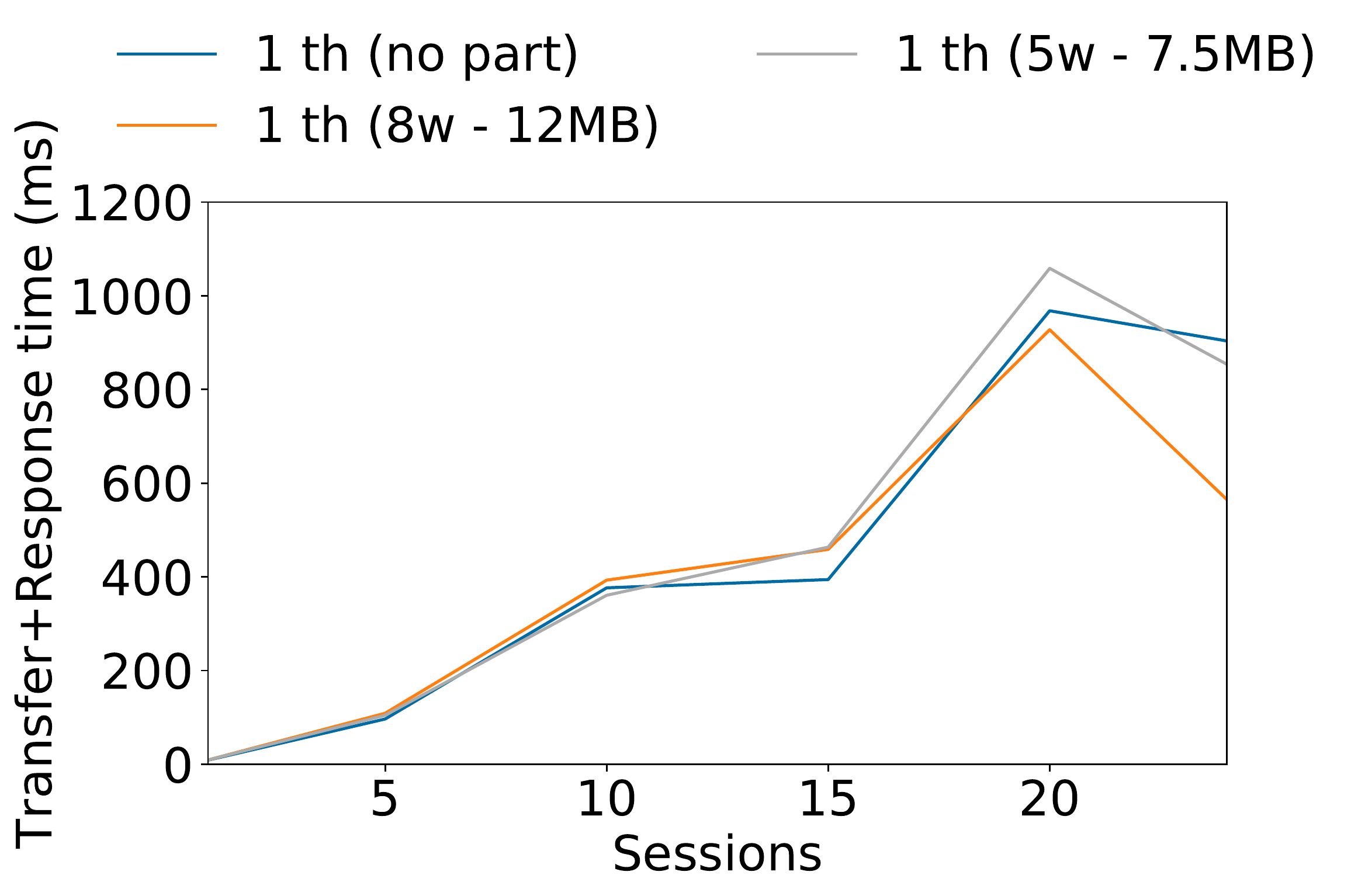}
       \label{fig:media_lat_part}}
    \subfloat[LLC occupancy]{%
       \includegraphics[width=0.32\textwidth]{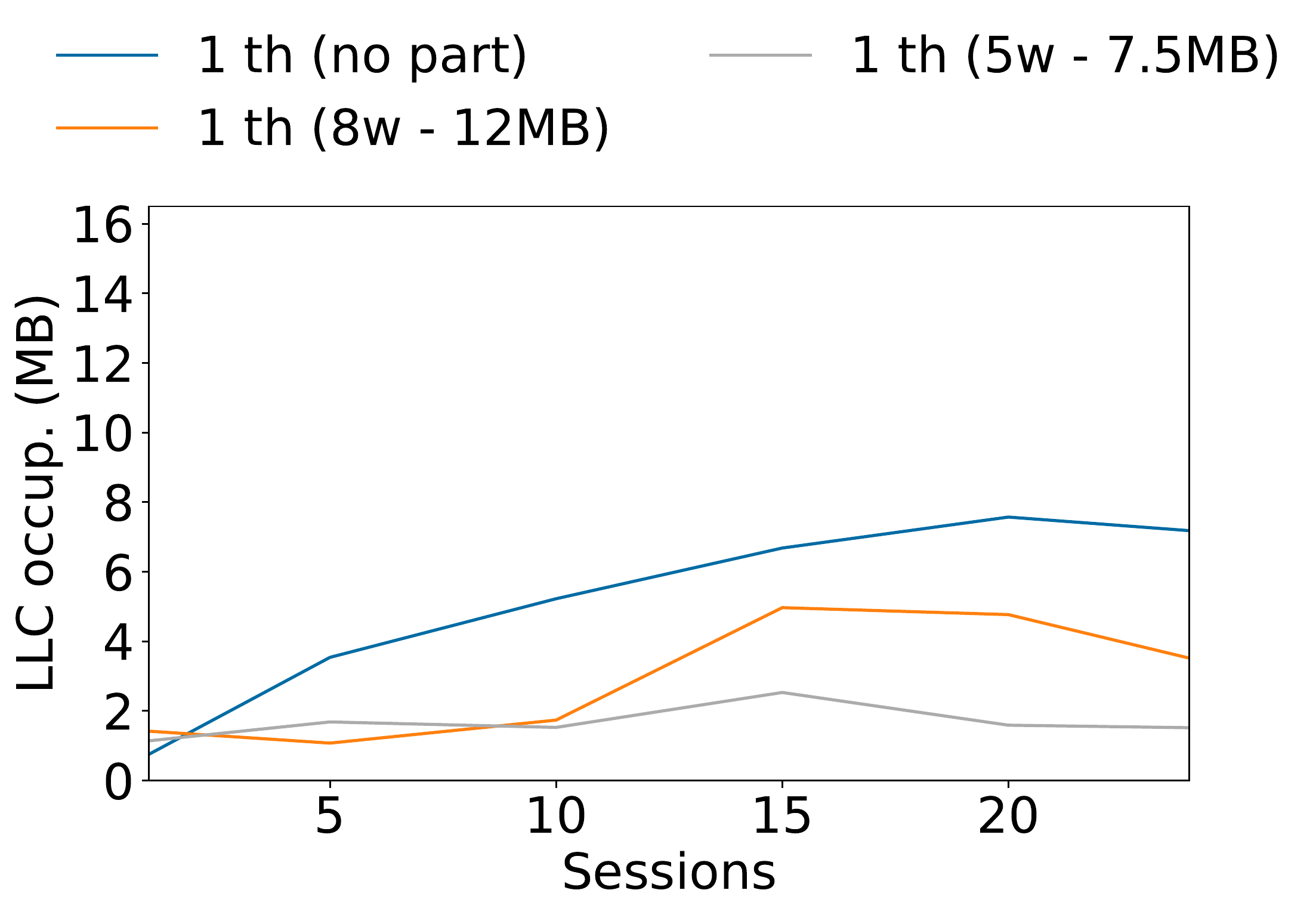}
       \label{fig:media_llc_part}}
    \subfloat[Main memory bandwidth]{%
       \includegraphics[width=0.32\textwidth]{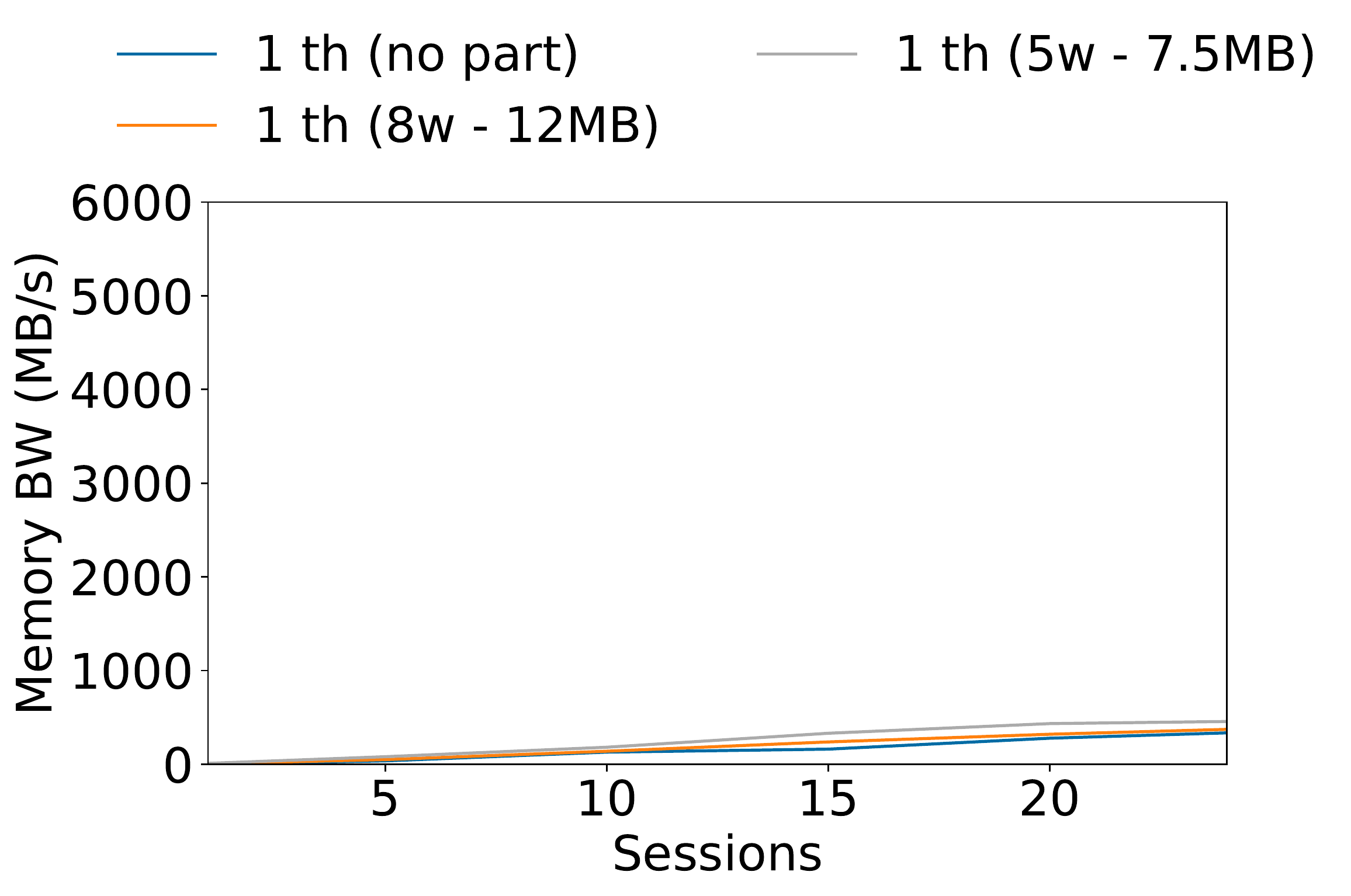}
       \label{fig:media_mem_part}}
    \caption{Media-streaming characterization with LLC partitioning.}
    \label{fig:media_llc_constrains} 
\end{figure*}

This section analyzes how constraining the main system shared resources affects the performance of the studied applications. 
We focus our study on the LLC and main memory bandwidth, since the disk and network bandwidth of the studied applications is far below the maximum supported consumption.

To carry out this study, we use the Intel Resource Director Technology (RDT)~\cite{intel_rdt}, which implements the Cache Monitoring Technology (CMT) and Cache Allocation Technology (CAT) that allow monitoring the LLC occupancy and partitioning the LLC, respectively. 

Partitioning is performed using Classes of Service (CLOS), which have associated a (set of) application(s) and a bitmask which indicates the LLC ways applications can use. 
Cache blocks accessed by the application assigned to a CLOS can only be stored in the cache ways indicated in the CLOS's bitmask.
We also use Intel Memory Bandwidth Allocation (MBA), which works similarly to Intel CAT; applications are assigned to CLOSes and so we can limit the amount of memory bandwidth that the CLOSes can use.

\subsection{LLC Partitioning Analysis}
\label{sec:constrain:subsec:llc}

This section analyzes the effect of constraining the amount of LLC storage (i.e., LLC ways) that the server workload can use on the client perceived performance (i.e., latency), CPU utilization, and other resource utilization metrics. As mentioned previously, based on the resource utilization of the server workloads, the available LLC storage shows up as one of the most critical resources that VMs should share when running concurrently. 

The Intel Xeon Silver 4116 processor, used as server in the experimental platform, has a 11-way 16.5~MB LLC, hence each cache way provides 1.5~MB storage capacity.
We study the impact of constraining the LLC capacity (that is, limiting the cache ways) for the server workloads from the original 16.5~MB (no partitioning) to 12~MB, 7.5~MB, and 3~MB. More precisely, we reduce the number of ways assigned to the application's CLOS from 11 to 8, 5, and 2, respectively. These scenarios emulate different levels of LLC storage capacity for a given VM when running concurrently with other VMs on the server. 
The analysis focuses on the single-threaded server since the results are applicable to a multi-threaded server.

Figures \ref{fig:sphinx_llc_constrains} to \ref{fig:media_llc_constrains} present the $95^{th}$ tail latency (transfer+response times in case of \emph{media-streaming}), LLC occupancy, and main memory bandwidth for a representative workload of each devised category when limiting the LLC storage the application can use to 12~MB, 7.5~MB, and 3~MB.
Notice that two workloads (\emph{img-dnn} and \emph{specjbb}) have been shown for the Fast category, since two behaviors have been identified.

We found that the tail latency of the workloads is affected differently by LLC partitioning. On the one hand, \emph{img-dnn} and \emph{sphinx} gradually reduce the supported QPS while guaranteeing the QoS latency as the available LLC space reduces. For instance, \emph{img-dnn} supports around 600 QPS with the entire LLC available, but this value reduces to 400, and 300 when the LLC occupancy is limited. In the case of \emph{specjbb}, only when the maximum LLC occupancy is set to 3~MB, the supported QPS reduces. A similar situation occurs with \emph{moses}.
On the other hand, \emph{media-streaming} its performance is not affected by the amount of LLC assigned. Notice that this application on has bee reduced to 5 and 2 ways since its normal LLC occupancy is low, below 8MB.

To sum up, high processor demands and fast applications are the most likely candidates to be affected by LLC space reductions. However, some fast applications like \emph{specjbb} may only be affected by a high reduction of space. Disk applications like \emph{moses} have a similar behavior. Finally, network applications like \emph{media-streaming} are not affected by the LLC space. Therefore, applications with no or low sensitivity to the LLC are eligible candidates to have their LLC space reduced in multi-program execution.

\subsection{Main Memory Bandwidth Analysis}
\label{sec:constrain:subsec:mbw}

\begin{figure*}[t!]
    \centering
    \subfloat[$95^{th}$ tail latency (ms)]{%
       \includegraphics[width=0.32\textwidth]{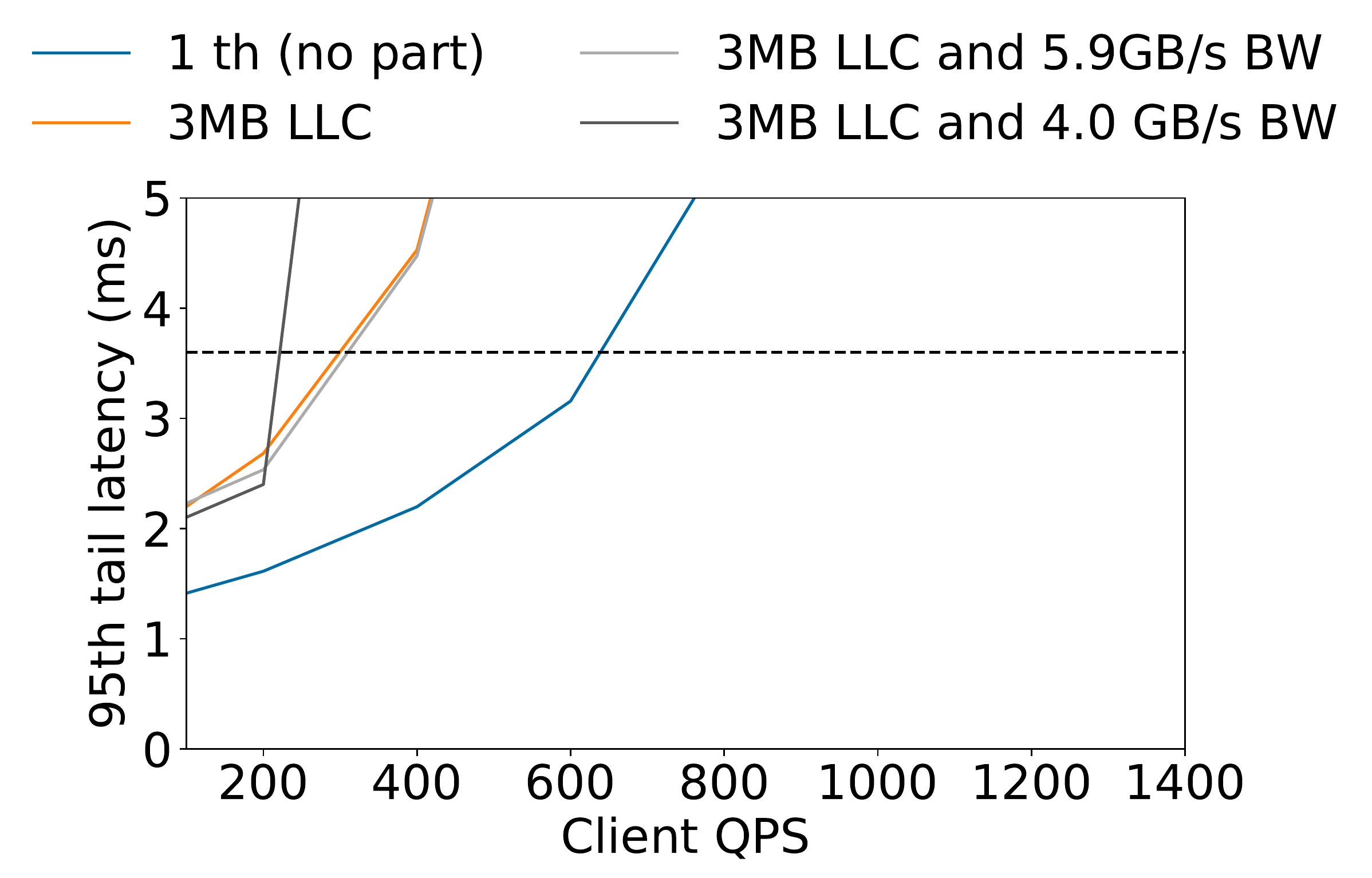}
       \label{fig:img_lat_part_mbw}}
    \subfloat[CPU utilization]{%
       \includegraphics[width=0.32\textwidth]{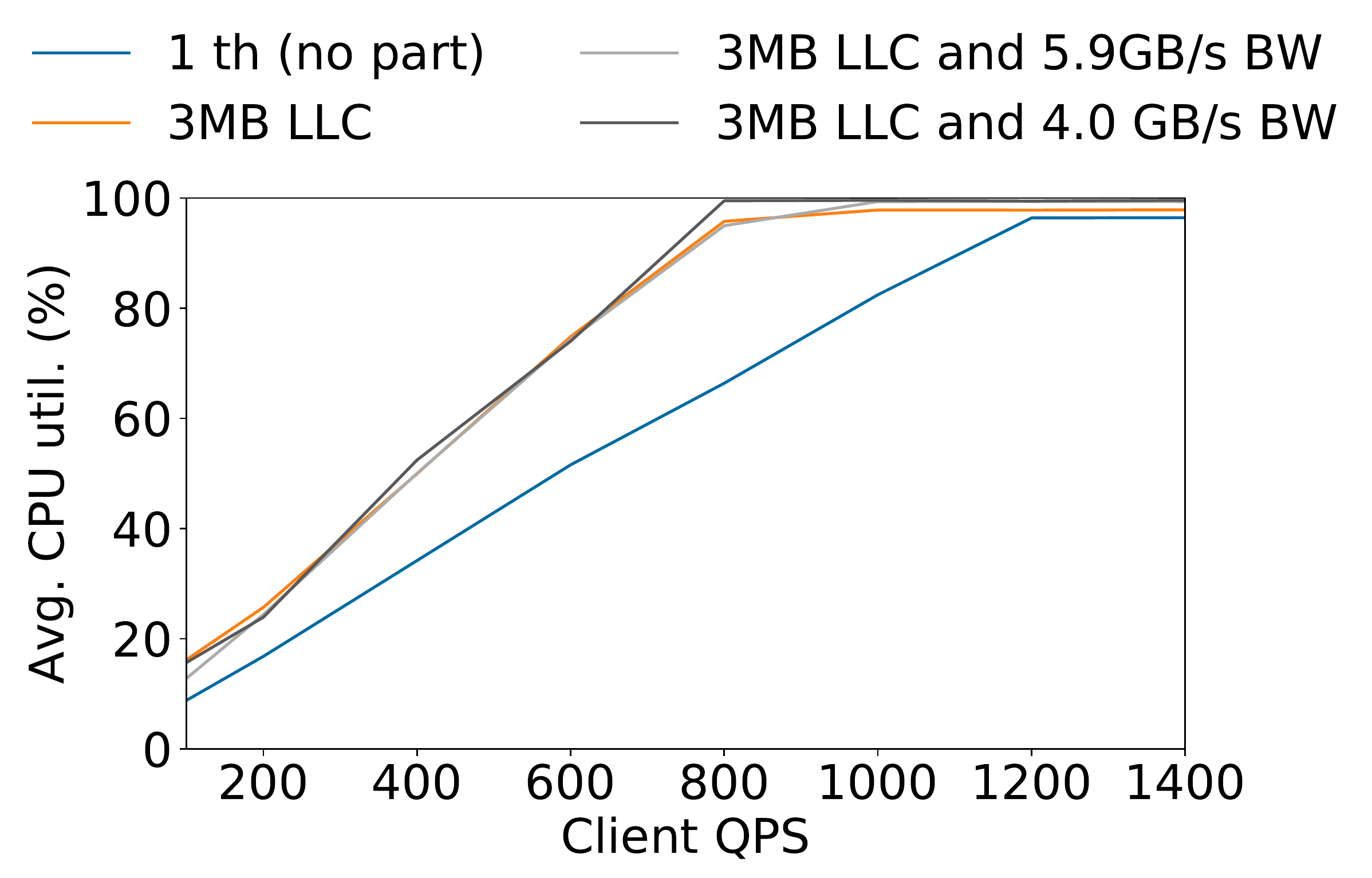}
       \label{fig:img_util_part_mbw}}
    \subfloat[Main memory bandwidth]{%
       \includegraphics[width=0.32\textwidth]{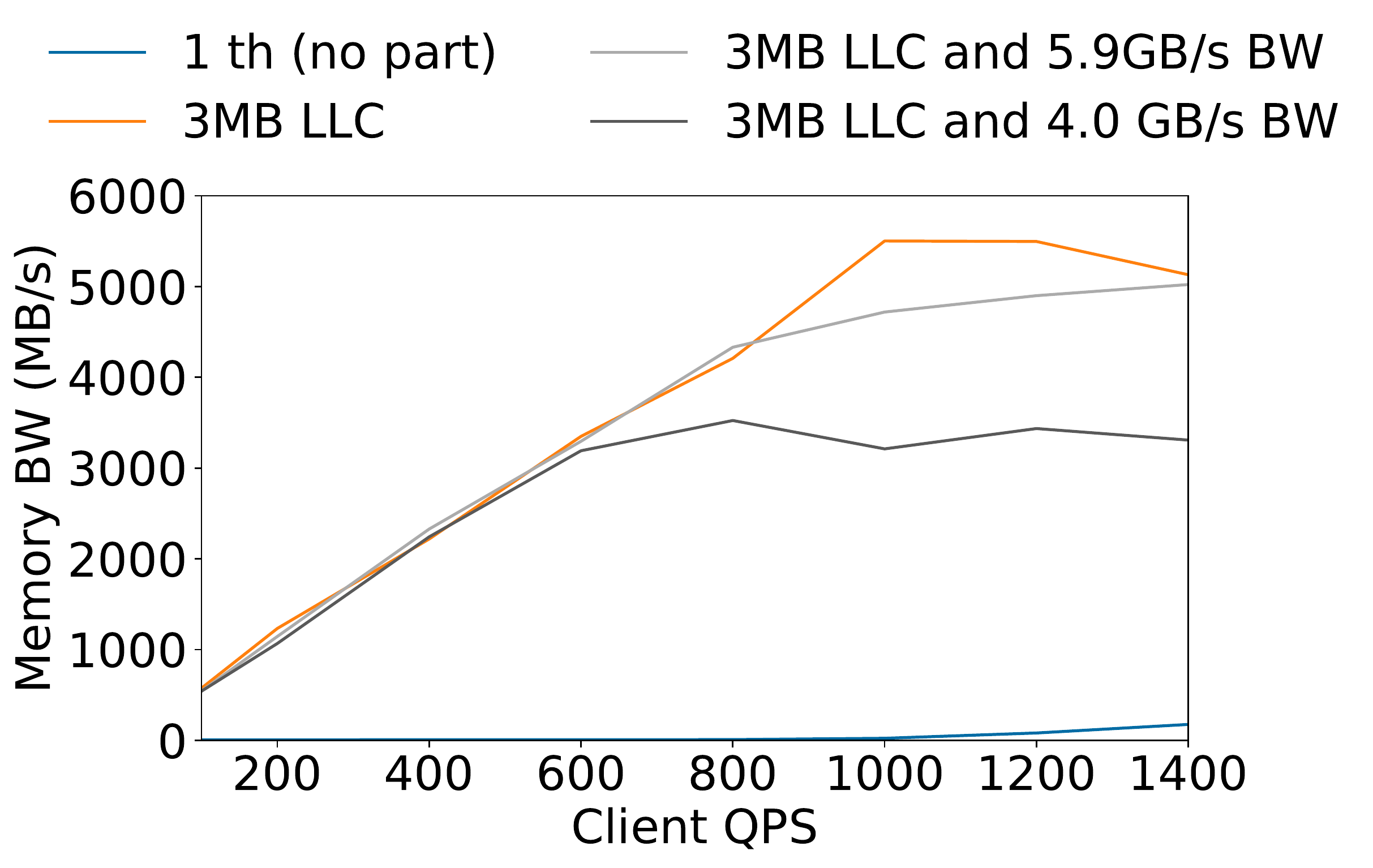}
       \label{fig:img_mem_part_mbw}}
    \caption{Img-dnn metrics varying the LLC and the memory bandwidth limits.}
    \label{fig:img-dnn_llc_constrains_mbw} 
\end{figure*}

From the characterization results shown in Section~\ref{sec:charac:subsec:results_individual}, it can be concluded that the studied applications barely consume memory bandwidth compared to the existing system bandwidth.
Moreover, taking into account that the \emph{real} memory bandwidth of the experimental platform is around 111~GB/s\footnote{To find out the memory bandwidth supported, we have carried out experiments with the STREAM benchmark~\cite{STREAM-benchmark}.} the TailBench workloads present far lower values than the maximum bandwidth that a core can consume (around 9 GB/s).
For instance, \emph{specjbb} is the TailBench application that achieves higher values, but it only reaches around 250~MB/s (with one server thread and 5000 QPS).
The \emph{media-streaming} workload consumes more memory bandwidth (over 330~MB/s in our experiments) than TailBench applications, but it is still far from the maximum.
Therefore, according to these results, it makes no sense to limit the memory bandwidth to these applications, since this metric does not impose a constraint to their latency or QoS when running alone in the experimental platform.

However, as mentioned above, limiting the LLC space of applications presents, in most cases, the side effect of rising the memory bandwidth consumption since it results in more accesses to the main memory. 
This can be observed in \emph{img-dnn}, whose memory bandwidth rises up to 5500~MB/s when its LLC space is limited to 2 cache ways (3~MB).
To study this situation, in this section we include results obtained applying both memory bandwidth and LLC partitioning.

Figure~\ref{fig:img-dnn_llc_constrains_mbw} shows the results of limiting both the memory bandwidth of \emph{img-dnn} to 4000 MB/s and the LLC to 2 cache ways, which corresponds to approximately one third of the maximum memory bandwidth achievable with one core. 
As it can observed, limiting the memory bandwidth has a noticeable negative effect. The amount of supported QPS is lower since the saturation point is reached before. %
Additionally, the amount of CPU utilization rises up to 90\% for a QPS of 700, the maximum QPS this application is able to cope while guaranteeing a 97,5\% timely. 
The CPU utilization rises because, as mentioned above, more memory accesses are performed thus the CPU requires more time to execute the application. 
Therefore we can conclude that main memory bandwidth plays an important role regarding performance and latency in these applications.

\section{Conclusions}\label{sec:conclusions}
\label{sec:conclus}

This paper is aimed at helping cloud system administrators detect the sources of potential performance losses.
To this end, we have analyzed three major concerns for cloud providers in order to improve the system utilization and provide service to a higher level of load:
 i) impact on performance of the level of load,
 ii) impact on performance of hyper-threading, 
 and iii) effect of constraining major system resources. All of them have been carried out in 
an experimental hardware/software platform with a configuration closely resembling to a real public cloud provider system.

As a first step, we have characterized TailBench workloads and CloudSuite's \emph{media-streaming} application varying the level of load and the number of logical cores occupied by the different threads of each application (to study the hyper-threading effect).
All the major system resources at the server (CPU, memory, disk) and the network have been evaluated, proving that CPU utilization alone cannot be considered as a proxy to estimate performance for running applications but other metrics are required.
Regarding the effect of hyper-threading, results have shown that sharing intra-core resources among the running threads usually has a negative impact on the supported QPS. However, this is not always the case since some applications achieve the same performance with both 2-SMT and 2-ST configurations.

Taking into consideration the load level, the perceived latency and the resource consumption, we have devised four major workload categories. By sizing the resources accordingly to each category, system utilization, and therefore, the overall system performance, can be improved while avoiding QoS violations.  
These results can help cloud providers improve the overall performance by defining allocation and partitioning policies.

Finally, we have used Intel RDT to study the impact of limiting the LLC space and the main memory bandwidth on the overall system performance for each  application.
Results have shown that the applications' performance is affected differently by LLC partitioning, being \emph{high processor} applications the most sensitive to reductions in the LLC capacity.
Besides, we found out that constraining both LLC and memory bandwidth can have a noticeable negative effect, despite Tailbench applications present small memory bandwidth needs.
This means that cloud providers can increase performance by devising suitable partitioning policies for their systems.

\section*{Acknowledgements}
\label{sec:acks}

This work has been supported by Huawei Cloud.

\bibliographystyle{elsarticle-num} 
\bibliography{bibliography}

\end{document}